\documentclass[journal,twocolumn]{IEEEtran}

\ifCLASSINFOpdf
\else
\fi

\usepackage{url,amsmath,amsfonts,amssymb,bm}
\usepackage{pifont}
\usepackage{cite}
\newcommand{\cmark}{\ding{51}}%
\newcommand{\xmark}{\ding{55}}%
\usepackage{enumitem}
\usepackage{graphicx}
\usepackage{algorithm}
\usepackage{algorithmicx, algpseudocode }
\usepackage{multirow}
\usepackage{color}
\usepackage{mathtools}
\usepackage{tikz}

\usetikzlibrary{calc}

\usepackage{caption}
\usepackage{subcaption}
\usepackage{bm}

\long\def\comment#1{}

\newfont{\bbb}{msbm10 scaled 700}

\newcommand{\av}{{\bf a}}
\newcommand{\bv}{{\bf b}}
\newcommand{\cv}{{\bf c}}
\newcommand{\dv}{{\bf d}}

\newcommand{\fv}{{\bf f}}

\newcommand{\uv}{{\bf u}}
\newcommand{\wv}{{\bf w}}
\newcommand{\vv}{{\bf v}}
\newcommand{\xv}{{\bf x}}
\newcommand{\yv}{{\bf y}}
\newcommand{\zv}{{\bf z}}

\newcommand{\Am}{{\bf A}}

\newcommand{\Cm}{{\bf C}}
\newcommand{\Dm}{{\bf D}}

\newcommand{\Gm}{{\bf G}}
\newcommand{\Hm}{{\bf H}}
\newcommand{\Id}{{\bf I}}
\newcommand{\Jm}{{\bf J}}

\newcommand{\Lm}{{\bf L}}
\newcommand{\Mm}{{\bf M}}

\newcommand{\Pm}{{\bf P}}
\newcommand{\Qm}{{\bf Q}}
\newcommand{\Rm}{{\bf R}}
\newcommand{\Sm}{{\bf S}}
\newcommand{\Tm}{{\bf T}}
\newcommand{\Um}{{\bf U}}
\newcommand{\Wm}{{\bf W}}
\newcommand{\Vm}{{\bf V}}

\newcommand{\Zm}{{\bf Z}}
\newcommand{\Lam}{{\bf \Lambda}}

\newcommand{\Ac}{{\cal A}}
\newcommand{\Bc}{{\cal B}}

\newcommand{\Ec}{{\cal E}}

\newcommand{\Gc}{{\cal G}}

\newcommand{\Oc}{{\cal O}}

\newcommand{\Vc}{{\cal V}}

\newcommand{\Lcb}{{\bm {\mathcal L}}}

\newcommand{\Lambdam}{\hbox{\boldmath$\Lambda$}}

\newcommand{\Sigmam}{\hbox{\boldmath$\Sigma$}}
\newcommand{\Phim}{\hbox{\boldmath$\Phi$}}

\newcommand{\Psim}{\hbox{\boldmath$\Psi$}}
\newcommand{\Thetam}{\hbox{\boldmath$\Theta$}}

\newcommand{\diag}{{\hbox{diag}}}

\newcommand{\sign}{{\hbox{sign}}}

\usepackage{amsthm}

\usepackage[colorlinks=true, linkcolor=blue, citecolor=blue, urlcolor=black]{hyperref}
\hypersetup{pdfcreator  = {LaTeX with hyperref package},
	pdfproducer = {dvips + ps2pdf} }
\newtheorem{definition}{Definition}
\newtheorem{remark}{Remark}

\newtheorem{theorem}{Theorem}
\newtheorem{lemma}{Lemma}

\newtheorem{proposition}{Proposition}

\newcommand{\R}{{\mathbb R}}

\newcommand{\N}{{\mathbb N}}
\newcommand{\E}{{\mathbb E}}

\newcommand{\Lblock}{{\bf L}^{bd}}
\newcommand{\Lbip}{{\bf L}^{bi}}
\newcommand{\Wblock}{{\bf W}^{bd}}
\newcommand{\Wbip}{{\bf W}^{bi}}
\newcommand{\Dblock}{{\bf D}^{bd}}
\newcommand{\Dbip}{{\bf D}^{bi}}

\newcommand{\Pbip}{{\bf P}^{bi}}
\newcommand{\Pblock}{{\bf P}^{bd}}

\newcommand{\Tcb}{{\bm {\mathcal T}}}

 \DeclareMathOperator*{\argmax}{arg\,max}
 \DeclareMathOperator*{\argmin}{arg\,min}
  
  \DeclareMathOperator*{\rank}{{rank}}

\def\equationautorefname~#1\null{(#1)\null} 
\def\itemautorefname~#1\null{#1\null}

\def\NEW#1{\textcolor[rgb]{0.00,0.0,0.0}{#1}}%

\begin{document}
%
%

\title{Two Channel Filter Banks on  Arbitrary Graphs with Positive Semi Definite Variation Operators }
%
%

\author{ Eduardo~Pavez,~Benjamin~Girault,~Antonio~Ortega, ~Philip A. Chou
  %
 \thanks{E. Pavez and A. Ortega  are with the Department of Electrical and Computer Engineering,
 University of Southern California, Los Angeles,
 CA, USA. (e-mail: pavezcar@usc.edu, ortega@sipi.usc.edu).
B. Girault is with Université de Rennes, ENSAI, CNRS, CREST-UMR 9194, Rennes, France. (e-mail: benjamin.girault@ensai.fr).
P. A. Chou is with Google Research, Seattle, Washington, USA. (e-mail:philchou@google.com).
  }
}

%
%

%


\maketitle
%
\begin{abstract}
We propose novel two-channel  filter banks for signals on graphs. Our designs   can be applied to  arbitrary graphs,  given a positive semi definite variation operator, while  using arbitrary vertex partitions for downsampling.  
The  proposed generalized filter banks (GFBs) also satisfy several desirable properties including perfect reconstruction  and  critical sampling, while having efficient implementations.  Our results generalize previous approaches that were only valid for the normalized Laplacian of bipartite graphs. 
Our approach is based on novel  graph Fourier transforms (GFTs) given by the generalized eigenvectors of the variation operator. These  GFTs are orthogonal in an alternative inner product space which depends on the   downsampling  and  variation operators. 
Our key theoretical contribution is showing that the \emph{spectral folding  property} of the normalized Laplacian of bipartite graphs, at the core of bipartite filter bank theory, can be generalized for the proposed  GFT if the inner product matrix is chosen properly. 
In addition, we study vertex domain and spectral domain properties of GFBs and  illustrate their probabilistic interpretation    using  Gaussian graphical models. While GFBs can be defined given any choice of a vertex partition for downsampling,  we propose an algorithm to optimize these partitions with a criterion that favors balanced partitions with large graph cuts, which are shown to lead to efficient and stable GFB implementations. Our numerical experiments show that partition-optimized GFBs can be implemented efficiently on  3D point clouds with hundreds of thousands of points (nodes), while also improving the color signal representation quality over competing state-of-the-art approaches. 
\end{abstract}
%
%
%
%
%
%
%
\IEEEpeerreviewmaketitle

\section{Introduction}
%
%
%
 \IEEEPARstart{G}{raphs} are powerful tools to model unstructured data. On a graph, nodes and edges represent objects of interest and  their similarity relations, respectively. A function on the nodes is called a graph signal, with examples including color attributes in images and  3D point clouds,  measurements obtained from   sensor networks,  or biomedical brain signals  \cite{ortega2018graph}. Graph signal processing (GSP)  develops new theories and algorithms for restoration, compression, and analysis of graph signals \cite{shuman2013emerging,sandryhaila2013discrete,ortega2021book}.
Filter banks and other multiresolution representations (MRR)    have been widely applied to signal processing  problems on regular grids \cite{vetterli1995wavelets,daubechies1992ten,vaidyanathan2006multirate,mallat1999wavelet}, which  has motivated their  generalization to graphs \cite{shuman2013emerging, shuman2020localized}. 
%
%
%
%
%
%
%

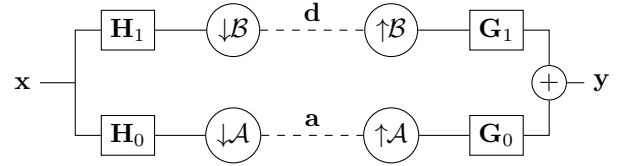
\begin{figure}[t]
	\centering
\def\hpdeltay{1}
\def\lpdeltay{-1}
\def\inputtosplitx{1}
\def\splittofilterx{1}
\def\filtertosamplingx{2}
\def\minspacedownupx{3}
\def\filtertooutputsplitx{1}
\begin{tikzpicture}[scale=0.7]
    \node (input) at (0, 0) {$\xv$};
    \coordinate (input_split) at (\inputtosplitx,0);

	\node[draw,rectangle] (H1) at ($(\inputtosplitx+\splittofilterx,\hpdeltay)$) {$\Hm_1$};
	\node[draw,circle,inner sep=2pt] (downB) at ($(H1)+(\filtertosamplingx,0)$) {$\downarrow\mkern-6mu \Bc$};

	\node[draw,rectangle] (H0) at ($(\inputtosplitx+\splittofilterx,\lpdeltay)$) {$\Hm_0$};
	\node[draw,circle,inner sep=2pt] (downA) at ($(H0)+(\filtertosamplingx,0)$) {$\downarrow\mkern-6mu\Ac$};

	\node[draw,circle,inner sep=2pt] (upB) at ($(downB)+(\minspacedownupx,0)$) {$\uparrow\mkern-6mu\Bc$};
	\node[draw,rectangle] (G1) at ($(upB)+(\filtertosamplingx,0)$) {$\Gm_1$};

	\node[draw,circle,inner sep=2pt] (upA) at ($(downA)+(\minspacedownupx,0)$) {$\uparrow\mkern-6mu\Ac$};
	\node[draw,rectangle] (G0) at ($(upA)+(\filtertosamplingx,0)$) {$\Gm_0$};

    \node[draw,circle, inner sep=1pt](add) at ($(G1)+(\filtertooutputsplitx,\lpdeltay)$) {$+$};


    \node (output) at ($(add)+(\inputtosplitx,0)$) {${\yv}$};

    \draw (input) -- (input_split);
	\draw (input_split) |- (H1) -- (downB);
	\draw (input_split) |- (H0) -- (downA);

	\draw[dashed] (downB) -- node[above] {$\dv$} (upB);
	\draw[dashed] (downA) -- node[above] {$\av$} (upA);

    \draw (upB) -- (G1) -| (add);
    \draw (upA) -- (G0) -| (add);

	\draw (add) -- (output);
\end{tikzpicture}
	\caption{Two-channel filter bank implemented with spectral graph filters. Low- and high-pass analysis filters are denoted by $\Hm_0$ and $\Hm_1$ respectively.  $\Gm_0$ and $\Gm_1$ correspond to low- and high-pass synthesis filters.  $\Ac$ and $\Bc$ are the sampling sets of the low- and high-pass channels respectively.  $\xv$ and ${\yv}$ denote input and output signals respectively. $\av$ is the vector of approximation 
 (low-pass) coefficients, and $\dv$ is the vector of detail (high-pass) coefficients.  }
	\label{fig:2chanfb}
\end{figure}

\emph{Bipartite Filter Banks} (BFBs) \cite{narang2012perfect,narang2013compact} are critically sampled two-channel filter banks on bipartite graphs. 
They are 
constructed with low- and high-pass spectral 
graph filters of the normalized graph Laplacian, along with downsampling and upsampling operators (see  \autoref{fig:2chanfb}).   
BFBs satisfy several desirable properties: 
(i)  perfect reconstruction, (ii)  critical sampling (non redundant), (iii)  compact support (polynomial  filters) and  (iv)  energy preservation (orthogonality).  
BFBs are closely related to filter bank designs for signals on regular grids since both satisfy the same design conditions, (i)-(iv), and  specific design techniques  (e.g., Meyer and  Daubechies filters \cite{vetterli1995wavelets, daubechies1992ten}) 
can be adapted to construct BFBs  \cite{narang2012perfect,narang2013compact,sakiyama2016spectral,tay2015techniques}. Given their efficient implementations, close relation with  filter banks on regular grids, and their 
well-understood theoretical properties, BFBs   have found  numerous applications in  compression \cite{nguyen2014compression,anis2016compression,zeng2017hyperspectral,tzamarias2019compression}, denoising \cite{iizuka2014depth} 
and signal analysis \cite{levorato2012reduced,sharma2018efficient,qiao2019target}.  
However, a major limitation of  BFBs  is that they are only valid for bipartite graphs, using either the normalized Laplacian \cite{narang2012perfect,narang2013compact} or the adjacency matrix \cite{tay2015techniques,tay2017bipartite}. 
In practice, the graph in a given application is rarely bipartite, so that bipartite graph learning or bipartite approximation are required \cite{narang2010local,zeng2017bipartite,jiang2019admm,pavez2018learning,kumar2020unified}, which increases computational complexity significantly, and may hinder the  signal representation quality.  
Another limitation of BFBs comes from their use of the normalized Laplacian   
(with lowest frequency eigenvector entries  proportional to the node degree) instead of the combinatorial Laplacian (constant lowest frequency eigenvector 
entries), which is usually a better choice for signal representations and 
results in better energy compaction \cite{tzamarias2021graphBior,narang2013compact,girault2018irregularity}.

For discrete time signals, many filter bank designs obey properties (i)--(iv) \cite{vetterli1995wavelets}, while others obeying  (i)--(iii) are also used (e.g., near orthogonal separable filter banks  \cite{Taubman2002JPEG2000I}).  
These filter banks can be  
implemented efficiently by exploiting lifting structures, separable filtering and  the  \emph{Fast Fourier Transform} (FFT).
In contrast,    fast \emph{Graph Fourier Transform} (GFT) algorithms  are only available for very limited families of graphs  \cite{lu2019fast}, while separable and lifting structures are only known for bipartite graphs \cite{narang2012perfect,narang2013compact, tay2017bipartite},  thus preventing the use of these techniques for reduced complexity implementations on arbitrary graphs. 
We are  motivated by 3D point cloud applications, where graphs can have  millions of nodes \cite{d20178i} making it difficult to use higher complexity methods, e.g., filter banks that require eigendecomposition or  bipartite graph learning.  As a result, we seek   designs that can  leverage graph sparsity for efficient implementation, for instance by using low degree polynomial graph filters.

We generalize the critically sampled filter bank design problem so that solutions can be found for non-bipartite graphs, as well as  with operators other than the normalized Laplacian. 
Instead, \emph{Generalized Filter Bank} (GFB) solutions are obtained for \textit{arbitrary graphs}, for \textit{any positive semi-definite graph variation operator}  and using \textit{arbitrary 
 vertex partitions for downsampling}, 
while still satisfying properties (i)--(iv).  We show that  for sparse graphs,  GFBs have scalable and eigendecomposition-free  implementations,  relying only on sparse 
matrix-vector products and sparse linear system solvers.
%

%

The key  innovation enabling GFBs is the adoption of a new inner product, which allows us to depart from the traditional Hilbert space (induced by the dot product) underlying  the majority of GSP methods\footnote{The use of non traditional Hilbert spaces has proven effective in various  studies involving irregularly structured data, such as   graph  sparsification \cite{spielman2011graph}, machine learning \cite{hein2007graph},  compression of 3D point clouds \cite{chou2019volumetric,krivokuca2019volumetric},  graph signal sampling \cite{girault2020graph}, and perceptual coding \cite{lu2020perceptually}.}. We build upon \cite{girault2018irregularity} where graphs are represented by a positive semi definite  variation operator $\Mm \succeq 0$,   which measures signal smoothness, and an  inner product   $\langle \xv, \yv \rangle_{\Qm} = \yv^{\top} \Qm \xv$, with $\Qm \succ 0$. The $(\Mm,\Qm)$ \emph{Graph Fourier Transform} ($(\Mm,\Qm)$-GFT) is defined as the generalized eigenvectors of $\Mm$,  which form a $\Qm$-orthonormal basis. 

\NEW{In our solution,  $\Qm$ is chosen as a function of $\Mm$ and the downsampling operator (determined by a vertex partition). 
More precisely, for a given a variation operator $\Mm$ our GFB theory can find a valid $\Qm$ for  \textit{any}  vertex partition, so that GFBs can be constructed using spectral graph filters of the $(\Mm,\Qm)$-GFT. Moreover, we show that it is important to optimize these  partitions so that they are balanced (\autoref{sec_properties_examples}) and result in  $\Qm$ matrices that are sparse and  close to  diagonal  (\autoref{sec_vertex_partitioning}). }
%
%
Our main contributions are summarized next.

%
\textbf{Theory of GFBs (\autoref{sec_prel_2chanfb}):} 
We introduce a new {\em spectral folding property} for the $(\Mm,\Qm)$-GFT, analogous to that satisfied by the eigenvectors and eigenvalues of the normalized Laplacian  of bipartite graphs \cite{chung1997spectral}.   
For a given variation operator $\Mm$ and a vertex partition for downsampling, we show that there exists a unique inner product matrix $\Qm$ such that the $(\Mm,\Qm)$-GFT obeys the spectral folding property.  
Based on this result,  we propose    perfect reconstruction and   $\Qm$-orthogonal  filter banks using spectral graph filters  of the $(\Mm,\Qm)$-GFT. 
Our conditions in the graph frequency domain are exactly  those developed in  \cite{narang2012perfect,narang2013compact} for the normalized Laplacian of bipartite graphs and therefore we can reuse any of the previously proposed filter designs, including  those in \cite{narang2012perfect,narang2013compact} and improved solutions such as  \cite{sakiyama2016spectral,tay2015techniques}.  
When $\Mm$ is the normalized Laplacian and the graph is bipartite, we recover the BFB framework.  

In our preliminary version of this work \cite{pavez2020spectral} we introduced 
the spectral folding property and two-channel filter banks on arbitrary graphs (Sections \ref{ssec_folding} and \ref{ssec_filterbanks_arbitrary}), while 
an application of GFBs to  image compression has been recently published \cite{tzamarias2021graphBior}.
In this paper, we further develop the GFB theory, 
providing all proofs not given in \cite{pavez2020spectral}, discussing the results in more depth and introducing the following novel contributions.

\begin{figure}[t]
	\centering
	\includegraphics[width=0.4\textwidth]{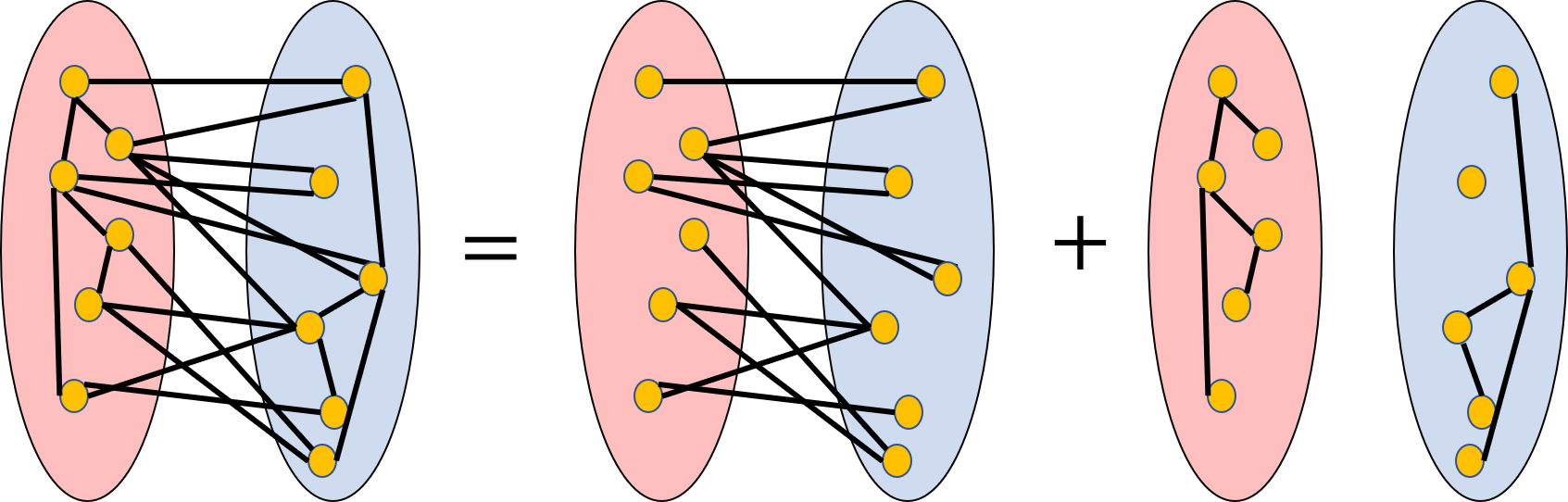}
\caption{ Representation of a graph (left) as the sum of a bipartite graph (center) and a disconnected graph with $2$ or more connected components (right). Red and blue circles contain node sets $\Ac$ and $\Bc$, respectively.}	\label{fig:bip_block_graph_decomp}
\end{figure}

\NEW{\textbf{Properties, interpretations and examples (\autoref{sec_properties_examples}):}
Since $(\Mm,\Qm)$-GFTs are relatively new, especially when $\Qm$ is not diagonal, \autoref{sec_properties_examples} is devoted to studying their properties. 
We show that some spectral properties of the normalized and random walk Laplacians can be extended to the proposed $(\Mm,\Qm)$-GFTs. We also provide examples  of $(\Mm,\Qm)$-GFTs with and without the spectral folding property (\autoref{fig:path_graphs_MQ}).
Representing an arbitrary graph as the sum of a bipartite and a disconnected graph (\autoref{fig:bip_block_graph_decomp}) leads to a vertex domain interpretation of the proposed spectral graph filters.  Finally,   we use Gaussian graphical models to give a probabilistic  interpretation of GFBs   and the spectral folding property.}

\NEW{ \textbf{Vertex partitioning (\autoref{sec_vertex_partitioning}):} While the GFB theory is valid for arbitrary vertex partitions, not all vertex partitions are desirable for downsampling. As an example, in \autoref{sec_properties_examples} we prove that unbalanced  vertex partitions lead to GFBs with poor frequency selectivity, and thus balanced partitions should be used. In \autoref{sec_vertex_partitioning} we propose  a numerical stability criterion for vertex set partition along with a computationally efficient  algorithm based on an approximate solution to a maximum cut (max-cut) problem. Essentially, our algorithm finds vertex partitions leading to a graph decomposition (\autoref{fig:bip_block_graph_decomp}) 
where the bipartite part has the largest cut, while the disconnected graph is sparse and has  small weights, which is also helpful to reduce the complexity of GFB implementations. Note that bipartite approximation algorithms for BFBs have been formulated as max-cut problems \cite{narang2010local}. However, these approach result in the removal of 
non-bipartite edges, while in our case edge  removal is unnecessary. 
}

\NEW{In this paper, the inner product defined by $\Qm$ plays a fundamental role: It not only leads to a definition of the $(\Mm,\Qm)$-GFT and to conditions for $\Qm$-orthogonal filter banks, but it is also needed to understand biorthogonal designs (i.e., good biorthogonal filters are such that they are approximately $\Qm$-orthogonal). 
Since the choice $\Qm=\Id$ is the dominant one in both signal processing and GSP, it is important to understand the implications of not using the standard ($\Qm = \Id$) inner product. 
To address this point, note that relaxing the $\Id$-orthogonality requirement is often an option if it is necessary to achieve some desirable properties. As an example JPEG2000 uses biorthogonal filters that are nearly $\Id$-orthogonal because they have finite support and are symmetric, which is desirable for imaging applications and would not be possible with orthogonal filters. Similarly, biorthogonal filter banks on bipartite graphs are preferred over $\Id$-orthogonal ones  because they have polynomial implementations \cite{narang2013compact}. 
Thus, in practice a non $\Id$-orthogonal solution can be useful as long as it is sufficient close to being $\Id$-orthogonal. 
In this work, 
we use biorthogonal filters (obtained from \cite{narang2013compact}) that lead to polynomial filters and can be designed to be nearly $\Qm$-orthogonal (see \autoref{sec_prel_2chanfb}).  
Additionally, in \autoref{sec_vertex_partitioning}  we show that the downsampling operator can be optimized 
so that the $\Qm$ matrices have favorable properties (sparsity, close to diagonal and with small condition number in operator norm) so that the $\Qm$ inner product approximates the conventional dot product. How closely the best $\Qm$ matrix for a given graph approximates the dot product  depends on the graph itself (degree irregularity, sparsity, etc), but we show examples for specific graphs (sensor network, spatial graphs, and 3D point cloud graphs) showing that good approximations can be achieved. Further analysis of this question for more general scenarios is left for future work. }

\NEW{The rest of the paper is organized as follows. 
In Section \autoref{sec_related} we review related work. 
In  \autoref{sec_gsp_MQ}  we introduce the fundamentals of GSP on general Hilbert spaces, while \autoref{sec_prel_2chanfb} is dedicated to bipartite and generalized filter banks. \autoref{sec_properties_examples} focuses on  properties of $(\Mm,\Qm)$-GFTs and GFBs.
\autoref{sec_vertex_partitioning} studies optimal vertex partitioning for downsampling. 
We end this paper with  numerical results,    and conclusions in  Sections \ref{sec_exp}   and \ref{sec_conc} respectively. Additional proofs   can be found in the Appendix.}

\section{Related  work}
\label{sec_related}
%
%
%
\begin{table*}[t]
	\centering
	\scalebox{1}{
		\begin{tabular}{|l|l|l|l|l | l|l|l|l|}
			\hline
			Ref.					 		& PR 		& O 		& Poly.  	& Crit. Samp. & AO 	& SO 	&Graph Type 	& Graph Matrix \\ \hline 
			Laplacian BFB  \cite{narang2012perfect} 		& \cmark  	& \cmark ($\Id$)	& \xmark& \cmark 	& SGF 	& SGF 	& Bipartite 	& $\Lcb$ \\ 
			Biorthogonal Laplacian BFB\cite{narang2013compact} 		& \cmark  	& \xmark	& \cmark	& \cmark 	& SGF 	& SGF 	& Bipartite 	& $\Lcb$ \\ 
			M channel FB \cite{teke2016extending_1,teke2016extending_2}& \cmark  	& \xmark	& \cmark 	& \cmark 	& SGF 	& SGF 	& M block cyclic 	& $\Wm$ \\ 
			Polyphase adjacency BFB\cite{tay2015techniques,tay2017bipartite}		& \cmark  	& \cmark ($\Id$) 	& \cmark  	& \cmark 	& SGF 	& SGF 	& Bipartite 	& $\Wm$ \\ 	\hline 
			Oversampled FB \cite{sakiyama2014oversampled} 	& \cmark  	& \xmark	& \cmark 	& \xmark 	& SGF 	& SGF 	& Any 	& $\Lcb$ \\ 	
			Graph Pyramid\cite{shuman2015multiscale} 	& \cmark  	& \xmark	& \cmark 	& \xmark 	& SGF 	& pinv 	& Any 	& $\Mm$ \\ 
			Ideal filter FB\cite{chen2015discrete} & \cmark & \cmark ($\Id$) & \xmark & \cmark &SGF & Interp & Any & Any  \\
			Approximate ideal filter FB\cite{li2019scalable} 			& \xmark  	& \xmark	& \cmark 	& \cmark 	& SGF 	& Interp 	& Any 	& $\Mm$ \\ 	
			Spectral Sampling FB\cite{sakiyama2019two}		& \cmark  	& \cmark ($\Id$)	& \xmark 	& \cmark 	& SGF 	& SGF 	& Any 	& Any \\ 	%
			\hline
			Generalized Filter Banks (GFB)										& \cmark  	& \cmark ($\Qm$)	& \cmark ($\Zm$) & \cmark 	& SGF 	& SGF 	& Any 	& $\Mm$ \\ \hline 
		\end{tabular}
	}
	\caption{Comparison of multiresolution representations for graph signals. Transform properties: perfect reconstruction (PR), orthogonal (O) in $\Qm$ or $\Id$ inner product, polynomial filter (Poly), critical sampling (CS), analysis operator (AO), synthesis operator (SO), graph type (GT), and graph operator (GO). Analysis operators are implemented with spectral graph filters (SGF), while synthesis operators may be implemented via pseudo inverse (pinv), interpolation (Interp), or a SGF.
	}
	\label{tab_relatedwork}
\end{table*}
We discuss related filter banks and multiresolution representations (MRR) on graphs from three perspectives:  graph topology, graph  matrix, and  downsampling sets.

{\bf Graph topology.} 
Several existing   filter bank theories are applicable only to graphs with certain types of  topology such as:  bipartite  \cite{narang2012perfect,narang2013compact}, $M$-block cyclic  \cite{teke2016extending_1,teke2016extending_2}, circulant  \cite{kotzagiannidis2019splines}, and  acyclic \cite{ekambaram2015spline}.  
While  some of these graph structures appear naturally in some applications \cite{cheung2018graph,kao2014graph}, 
 many cases graphs of interest do  not belong to any of these categories. 
To apply BFBs to arbitrary graphs one can decompose any graph as a sum of bipartite graphs and apply the filter bank in a separable manner \cite{narang2012graph}. Other approaches include bipartite approximation \cite{narang2010local},  bipartite graph learning methods \cite{zeng2017bipartite,jiang2019admm,pavez2018learning,lu2018learning,kumar2020unified}, graph oversampling \cite{sakiyama2014oversampled}, and vertex partition optimization \cite{anis2017critical}.
%
Earlier filter bank designs for  arbitrary graphs  are difficult to invert (e.g., least squares reconstruction is needed) \cite{crovella2003graph,coifman2006diffusion,hammond2011wavelets}. 
More recent MRRs  fail to be simultaneously perfect reconstruction and orthogonal \cite{shuman2015multiscale}, while others, require full eigendecomposition \cite{chen2015discrete,cloninger2020natural}.
Existing approaches that are valid for arbitrary graphs have several disadvantages. On the one hand, approximation-based methods may reduce signal representation quality while also requiring additional computational resources (to select the best bipartite approximation) \cite{narang2010local,narang2012graph,zeng2017bipartite,jiang2019admm}. 
On the other hand, methods that allow the original graph to be used can do so at 
the expense of other desirable features, such as low complexity \cite{chen2015discrete,cloninger2020natural}, perfect reconstruction  or orthogonality (in $\Id$ or other $\Qm$ inner product) \cite{anis2017critical,shuman2015multiscale} (see \autoref{tab_relatedwork} for a comparison of some of these approaches).  Therefore, 
a theoretical formulation leading to critically sampled filter banks for arbitrary graphs, without the aforementioned disadvantages,  can be an attractive alternative. 

{\bf Graph matrix.}
The choice of graph matrix,   such as the graph Laplacian or the adjacency matrix, whose non-zero pattern encodes the graph structure, is an important design decision  \cite{dong2019learning,mateos2019connecting}. Existing filter bank frameworks are built for specific types of graph matrices because of their special algebraic or spectral properties \cite{narang2012perfect,teke2016extending_1}, but these choices may not  be suitable for a particular application. As an example, BFBs  use the normalized Laplacian \cite{narang2012perfect}, while the random walk Laplacian has been shown to achieve better coding performance \cite{tzamarias2021graphBior}. More recent filter bank approaches \cite{li2019scalable,sakiyama2019two} can be applied to the larger class of positive semi-definite variation operators.  However,  they lack perfect reconstruction   \cite{li2019scalable} or  require computing a full eigendecomposition of the graph operator   \cite{sakiyama2019two}, which significantly limits their  application to large graphs. 
The proposed GFBs   are valid for any positive semi definite graph matrix (i.e., a variation operator), including commonly used graph Laplacians,  without compromising on other  properties.  

{\bf Downsampling.}
For   discrete-time signals,  a downsampling by $2$ operator  keeps ``every other sample'' and discards the rest.  In graphs, there is no obvious notion of ``every other vertex'' unless the graph is bipartite \cite{narang2012graph}.  For graph filter banks,  vertex partitions of the node set can be chosen under various criteria \cite{li2019scalable,anis2017critical,shuman2020localized}.
Recently, \cite{sakiyama2019two,miraki2021spectral} used spectral domain sampling, and while this approach leads to an attractive theory, spectral sampling requires computing all eigenvectors and eigenvalues of the graph matrix, which can have significant computation complexity. 
\cite{li2019scalable} extended sampling theory of graph signals to filter banks but  practical implementations of this framework cannot achieve perfect reconstruction. In this work,  we show that  any partition of the vertex set is a valid downsampling operator. We also propose strategies for optimally choosing these partitions in Section \ref{sec_vertex_partitioning}. 
\section{GSP in general Hilbert spaces}
\label{sec_gsp_MQ}
\subsection{Notation}
Scalars, vectors and matrices are written in lower case regular, lower case bold and upper case bold respectively (e.g.,  $a$, $\bv$, $\Cm$). Positive definite and semi-definite matrices are denoted by $\Am \succ 0$,  and $\Am \succeq 0$ respectively. We will denote by $\Cm_{\Ac \Bc}$, the sub-matrix of $\Cm$ whose rows and columns are indexed by the sets $\Ac$ and $\Bc$, respectively.  The spectral norm or largest singular value of a matrix $\Am$ is denoted by  $\Vert \Am \Vert$.
\subsection{Graph signal processing}
Consider a  graph $\Gc = (\Vc, \Ec)$ with vertex set $\Vc = \lbrace 1,\cdots,n \rbrace $, and edge set $\Ec \subset \Vc \times \Vc$. 
A graph signal is a function $x: \Vc \rightarrow \mathbb{R}$, which can be represented by a vector $\xv = [x_1,\cdots,x_n]^{\top}$, and $x_i$ is the signal value at  vertex $i \in \Vc$.
\subsubsection{Graph signal variation}
The graph is equipped with a symmetric positive semi definite variation matrix $\Mm = (m_{ij})$, with sparsity pattern  determined by the edge set, that is, $m_{ij} = m_{ji}  = 0$ when $(i,j) \notin \Ec$ and $i \neq j$.
Throughout the paper we will assume that the graph is connected\footnote{\NEW{If the graph is disconnected,  each connected component can be treated independently.} } and thus  $\Mm$ is an irreducible matrix \cite{horn2012matrixbook,biyikoglu2007laplacian}.
The graph signal variation is
\begin{equation}
    \Delta(\xv) = \xv^{\top} \Mm \xv.
\end{equation}
Intuitively, given two signals of equal energy, the one with larger variation is the one with higher energy in the higher frequencies of the spectrum.
Popular choices of variation operator are the normalized,  combinatorial and generalized Laplacian matrices \cite{biyikoglu2007laplacian}, which we define next.  

The adjacency matrix is a non negative symmetric matrix $\Wm = (w_{ij})$, where $w_{ij} =0$, whenever $(i,j) \notin \Ec$. The degree of node $i$ is $d_i = \sum_{j \in \Vc }w_{ij}$, and the degree matrix is $\Dm = \diag(d_1,\cdots, d_n)$. 
The \emph{combinatorial graph Laplacian} (CGL) is $\Lm = \Dm - \Wm$, while the \emph{normalized graph Laplacian} (NGL) is $\Lcb = \Dm^{-1/2}\Lm \Dm^{-1/2} = \Id - \Dm^{-1/2} \Wm \Dm^{-1/2}$.  A \emph{generalized graph Laplacian} (GGL) is any positive semidefinite matrix with non positive off diagonal entries, which includes the normalized and combinatorial graph Laplacian matrices \cite{biyikoglu2007laplacian, Kurras.ICML.2014,egilmez2017graph,pavez2020ragft}. Other variation operators are listed in \cite{girault2018irregularity,anis2016efficient}.
\subsubsection{The $(\Mm,\Qm)$-GFT \cite{girault2018irregularity}}
Given a  positive definite matrix  $\Qm$, we define the  $\Qm$ inner  product between  $\xv$ and $\yv$, as
\begin{equation}
    \langle \xv, \yv \rangle_{\Qm} = \yv^{\top} \Qm \xv,
\end{equation}
with   induced  $\Qm$-norm   $\Vert \xv \Vert_{\Qm} = \sqrt{\langle \xv, \xv \rangle_{\Qm}}$.
We say $\lbrace \uv_k \rbrace_{k=1}^{n}$ is a $\Qm$-orthonormal set   if it satisfies
\begin{equation}
  \langle \uv_i, \uv_j \rangle_{\Qm} = \left\{ \begin{array}{cc}
		1  & \mbox{if } i = j \\
		0 & \mbox{if } i \neq j.
	\end{array} \right.
\end{equation}
In matrix form this corresponds to $\Um^{\top}\Qm\Um = \Id$, where $\Um = [\uv_1,\cdots, \uv_n]$, which also implies that $\Um^{-1} = \Um^{\top}\Qm$. 
Following  \cite{girault2018irregularity}, the $(\Mm,\Qm)$ \emph{Graph Fourier Transform} ($(\Mm,\Qm)$-GFT)   
is defined as the $\Qm$-orthonormal set that minimizes the graph signal variation, \NEW{ that is 
\begin{equation}\label{eq_MQGFT_def1}
    \uv_1 = \argmin_{\uv: \uv \neq \mathbf{0},~ \Vert \uv \Vert_{\Qm}=1} \uv^{\top} \Mm \uv,
\end{equation}
and  for any $ 2 \leq k \leq n$
\begin{equation}\label{eq_MQGFT_def2}
    \uv_{k} = \argmin_{\uv: \uv \neq \mathbf{0},~ \Vert \uv \Vert_{\Qm}=1} \uv^{\top} \Mm \uv \textnormal{ s.t. }  \langle \uv, \uv_i \rangle_{\Qm} = 0, \forall i < k.
\end{equation}
Note that this  definition of the $(\Mm,\Qm)$-GFT basis is consistent with the traditional definition of GFT (i.e., the $(\Mm,\Id)$-GFT). In particular, when $\Mm=\Lm$, the variation of $\xv$ is $\xv^{\top}\Lm\xv = \sum_{(i,j)\in \Ec}w_{ij}(x_i - x_j)^2$, and as in that case,  the generalized eigenvectors  $\uv_k$ have increasing variation (quantified with the same operator $\Mm$) for larger $k$. The only difference is that now the basis vectors are   $\Qm$-orthonormal instead of $\Id$-orthonormal. }
The  $(\Mm,\Qm)$-GFT  is also the solution to the generalized eigendecomposition problem \cite{horn2012matrixbook,girault2018irregularity}:
\begin{equation}\label{eq_MQGFT_gen_eigen}
    \Mm \uv = \lambda \Qm \uv.
\end{equation}

The matrix of unit $\Qm$-norm generalized eigenvectors  is denoted by  $\Um = [\uv_1,\cdots, \uv_n]$, with generalized eigenvalues (or graph frequencies) forming a set $\sigma(\Mm,\Qm) = \{\lambda_1,\cdots, \lambda_n\}$ where $\lambda_k = \uv_k^{\top}\Mm \uv_k$ and 
 $\lambda_1 \leq \lambda_2 \cdots \leq \lambda_n$. 
 The generalized eigenvalue matrix is   $\Lambdam = \diag(\lambda_1,\cdots, \lambda_n)$.  
 The \emph{fundamental matrix} is  $\Zm = \Qm^{-1}\Mm$, which is diagonalized by $\Um$,  so
\begin{equation}
\Zm = \Qm^{-1}\Mm = \Um \Lambdam \Um^{-1} = \Um \Lambdam \Um^{\top}\Qm. 
\end{equation}
 Then we can factorize the variation operator $\Mm$ as follows:
 \begin{equation}
\Mm = \Qm \Um \Lambdam \Um^{\top} \Qm.
 \end{equation}
 
%
A graph signal $\xv$  has a $(\Mm,\Qm)$-GFT representation given by:
\begin{equation}\label{eq_MQGFT}
    \xv = \sum_{i=1}^n \langle \xv, \uv_i \rangle_{\Qm} \uv_i = \Um \hat{\xv}, 
\end{equation}
where   $\hat{\xv} = \Um^{\top} \Qm \xv$ is the $(\Mm,\Qm)$-GFT of $\xv$ \cite{girault2018irregularity}. 
Thus,  \eqref{eq_MQGFT} is the inverse $(\Mm,\Qm)$-GFT, where $\xv$ is reconstructed as a linear combination of the frequency components in $\hat{\xv}$. 
Note that the inverse transform is given by $\xv = \Um \hat{\xv}$, since $\Um \Um^{\top} \Qm = \Id$.
\subsubsection{Spectral graph filters}
Given a function (i.e., a filter kernel) $h: \sigma(\Mm,\Qm) \rightarrow\mathbb{R}$, then  a \emph{spectral graph filter}   (SGF) is any matrix $\Hm$  of the form
\begin{equation}\label{eq_SGF_Definition}
    \Hm = \Um h(\Lam) \Um^{\top} \Qm = h(\Zm),
\end{equation}
where  $h(\Lambdam) = \diag(h(\lambda_1), h(\lambda_2),\cdots, h(\lambda_n))$.  When $h(\lambda)=\lambda$ we have that 
 $\Hm=\Zm = \Qm^{-1} \Mm $ and we recover the fundamental matrix. \NEW{  A spectral graph filter is implemented by first applying the $(\Mm,\Qm)$-GFT to a signal  $\xv$,  then multiplying each transformed  coefficient $\hat{x}_i = \langle \xv, \uv_i \rangle_{\Qm}$  by the filter coefficient $h(\lambda_i)$. The resulting signal is transformed back to the vertex domain with the inverse $(\Mm,\Qm)$-GFT.
 Filtering $\xv$ using $h$  produces the graph signal $\Hm \xv$, since 
 \begin{equation}
     \Hm \xv = \Um h(\Lambdam) \hat{\xv} = \sum_{i=1}^n h(\lambda_i)\hat{x}_i \uv_{i}.
 \end{equation}
  When $h$ is a polynomial  $h(\lambda) = a_0 + a_1 \lambda + \cdots +a_d \lambda^d$ for  $d \in \N$, the resulting SGF is a polynomial of $\Zm$, thus $\Hm=h(\Zm) =a_0\Id + a_1 \Zm + \cdots +a_d \Zm^d$, 
   can be implemented efficiently without  eigendecomposition using  matrix vector products.
  Thus, if $\Zm$ is sparse and $d$ is 
  relatively small, 
  graph filtering with polynomial SGFs 
  can be applied efficiently on large graphs. Complexity of polynomial graph filters and filter banks is reviewed in more detail in \autoref{sec_vertex_partitioning}.
 }
 \subsection{Vertex and frequency domain Hilbert spaces}
 \label{sec_vertex_freq_hilbert}
Both a graph signal, $\xv$, and its $(\Mm,\Qm)$-GFT,  $\hat{\xv}$, can be represented by vectors in $\R^{n}$.  
However, in our formulation they belong to different Hilbert spaces, namely,  $(\R^n,\langle \cdot, \cdot \rangle_\Qm)$ for $\xv$ and $(\R^n,\langle \cdot, \cdot \rangle_{\Id})$ for $\hat{\xv}$, where $\langle \cdot, \cdot \rangle_{\Id}$ is the traditional dot product.  
 %
Thus, we can view the $(\Mm,\Qm)$-GFT, $\Um^{\top}\Qm$, as a mapping from  $(\R^n,\langle \cdot, \cdot \rangle_\Qm)$ to $(\R^n,\langle \cdot, \cdot \rangle_{\Id})$, while the inverse $(\Mm,\Qm)$-GFT $\Um$ is a mapping from $(\R^n,\langle \cdot, \cdot \rangle_{\Id})$ to $(\R^n,\langle \cdot, \cdot \rangle_{\Qm})$. 
Making explicit the Hilbert spaces involved in this mapping allows us to state the following property.
\begin{theorem}[Parseval \cite{girault2018irregularity}] 
\label{th_parseval} 
Let $\xv,\yv \in \R^{n}$, and let $\hat{\xv}$ and $\hat{\yv}$ be their respective frequency domain representations, then
\begin{equation}\label{eq_ortho_GFT}
    \langle \xv,\yv \rangle_{\Qm} = \langle \hat{\xv},\hat{\yv} \rangle_{\Id}.
\end{equation}
\end{theorem}
From    \eqref{eq_SGF_Definition} it is clear  that SGFs are functions from  $(\R^n,\langle \cdot, \cdot \rangle_{\Qm})$ to the same Hilbert space. 
For the rest of the paper we will use the $\Qm$ inner product for graph signals, and the $\Id$ inner product for their  frequency domain representations. 
%
%
%
\section{Two-channel filter banks on graphs}
\label{sec_prel_2chanfb}
\NEW{In this section we introduce the two-channel filter bank theory for arbitrary graphs. In  \autoref{subsec_vertex_cond_arbitrary}   we formulate the 
problem.     \autoref{subsec_BFB}  reviews the solution for bipartite graphs from    \cite{narang2012perfect}.   We illustrate how to go from bipartite to arbitrary graphs through an example in \autoref{ssec:lazy_example}. Our main theoretical result, the generalized spectral folding property is presented in \autoref{ssec_folding}.   Generalized filter banks are constructed in \autoref{ssec_filterbanks_arbitrary} and  \autoref{ssec_treeGFB}.}
\subsection{Problem formulation}
\label{subsec_vertex_cond_arbitrary}
A two-channel filter bank   (\autoref{fig:2chanfb})  is composed of downsampling and upsampling operators on the sets $\Ac$ and $\Bc$, and  two sets of $\vert \Vc \vert \times \vert \Vc \vert$ matrices, corresponding to the analysis flters ($\Hm_0$, $\Hm_1$) and the synthesis filters ($\Gm_0$,$\Gm_1$).  
Throughout the paper, we will consider critically sampled filter banks.
\begin{definition}
A two-channel filter bank is {\bf critically sampled} if the downsampling sets $\Ac$ and $\Bc$ form a partition of $\Vc$.
\end{definition}
 Without loss of generality we assume that $\Ac = \lbrace 1, \cdots, \vert \Ac \vert \rbrace$, and $\Bc = \Vc \setminus \Ac$. Downsampling   $\xv$ on  $\Ac$ keeps the entries $\lbrace x_i: i \in \Ac \rbrace$ and discards the rest, resulting in
\begin{equation}
    \xv_\Ac = \Sm_\Ac \xv,
\end{equation}
where $\Sm_\Ac = [\Id_\Ac, \mathbf{0}]$ is a $\vert \Ac \vert \times \vert \Vc \vert$ selection matrix.
The upsampling operator   $\Sm_\Ac^{\top}$ maps the signal back to $\Vc$ by filling the entries on $\Bc$ with zeroes. Downsampling followed by upsampling corresponds to
\begin{equation}
     \Sm_\Ac^{\top} \Sm_\Ac^{\phantom{\top}} \xv=
      \Sm_\Ac^{\top} \xv_{\Ac} =
     \begin{bmatrix}
     	\xv_\Ac^{\top} &
     	\mathbf{0}
     \end{bmatrix}^{\top}.
\end{equation}
%
The analysis operator $\Tm_a$ (i.e., filtering followed by downsampling) from \autoref{fig:2chanfb}  can be written  as
\begin{equation}\label{eq_Ta_analysis}
\Tm_a =  \Sm_\Ac^{\top} \Sm_\Ac^{\phantom{\top}} \Hm_0 + \Sm_\Bc^{\top} \Sm_\Bc^{\phantom{\top}} \Hm_1 = \begin{bmatrix}
\Sm_\Ac \Hm_0 \\ \Sm_\Bc \Hm_1
\end{bmatrix}.
\end{equation}
The outputs of the low pass and high pass channels, approximation, $\av$, and detail, $\dv$, coefficients, respectively, are 
\begin{equation}
\Tm_a \xv = \begin{bmatrix}
\av^{\top} &
\dv^{\top}
\end{bmatrix}^{\top}
= \begin{bmatrix}
(\Sm_\Ac \Hm_0\xv)^{\top} & (\Sm_\Bc \Hm_1 \xv)^{\top} 
\end{bmatrix}^{\top}.
\end{equation}
The synthesis operator   can be expressed as
\begin{equation}\label{eq_synthesis_operator}
    \Tm_s = \Gm_0 \Sm_\Ac^{\top} \Sm_\Ac^{\phantom{\top}}  + \Gm_1 \Sm_\Bc^{\top} \Sm_\Bc^{\phantom{\top}} = \begin{bmatrix}
    \Gm_0 \Sm_\Ac^{\top} & \Gm_1 \Sm_\Bc^{\top} 
    \end{bmatrix}.
\end{equation}
Both the analysis and synthesis operators  map signals  from  $(\R^n,\langle \cdot, \cdot \rangle_\Qm)$ to $(\R^n,\langle \cdot, \cdot \rangle_\Qm)$ (see also \autoref{sec_vertex_freq_hilbert}), thus  $\xv, \Tm_a\xv$, and $\Tm_s\Tm_a\xv$ are vertex domain signals.  
We are interested in  perfect reconstruction two-channel filter banks.
\begin{definition}A two-channel filter bank is {\bf perfect reconstruction} (PR) if $\Tm_s \Tm_a \xv = \xv$, for all $\xv$.
\end{definition}
Since we only consider critically sampled filter banks, the PR condition is equivalent  to $\Tm_s = \Tm_a^{-1}$, thus a necessary and sufficient condition for perfect reconstruction  is given by
\begin{equation}\label{eq_PR_condition_sampling}
 \Tm_s \Tm_a = \Gm_0 \Sm_\Ac^{\top} \Sm_\Ac \Hm_0  + \Gm_1 \Sm_\Bc^{\top} \Sm_\Bc  \Hm_1 = \Id.  
\end{equation}
We also introduce  $\Qm$-orthogonal filter banks. 
\begin{definition}\label{def_Qorth}
A  two-channel filter bank is $\Qm$-{\bf orthogonal}, if for every pair of graph signals  $\xv, \yv$, 
\begin{equation}\label{eq_Q_ortho_FB}
    \langle \xv, \yv \rangle_{\Qm} =  \langle \Tm_a \xv, \Tm_a \yv  \rangle_{\Qm}.
\end{equation}
In matrix form,  \eqref{eq_Q_ortho_FB}  is equivalent to $\Tm_a^{\top} \Qm \Tm_a = \Qm$.
\end{definition}
The traditional notion of orthogonal filter banks corresponds to the case $\Qm=\Id$.
Critically sampled $\Qm$-orthogonal filter banks are always perfect reconstruction with synthesis  operator
\begin{equation}\label{eq_synthesis_Qorth}
	\Tm_s = \Qm^{-1}\Tm_a^{\top} \Qm.
\end{equation}
It is  easy to verify that \eqref{eq_synthesis_Qorth} also obeys  $\Tm_s^{\top} \Qm \Tm_s = \Qm$.
\begin{remark}
Because the vertex domain and frequency domain Hilbert spaces are different (see \autoref{sec_vertex_freq_hilbert}),  $\Qm$-orthogonality for filter banks is different than $\Qm$-orthogonality of the $(\Mm,\Qm)$-GFT. This is because the $(\Mm,\Qm)$-GFT maps graph signals from their vertex domain representation to their frequency domain representation, that is, $\Um^{\top}\Qm:~(\R^n,\langle \cdot, \cdot \rangle_\Qm) \rightarrow (\R^n,\langle \cdot, \cdot \rangle_\Id)$, which results in the  relationship of  
\autoref{th_parseval} (see \cite{girault2018irregularity}). 
In contrast,  the analysis and synthesis operators of a PR critically sampled filter bank  stay in the same Hilbert space $(\R^n,\langle \cdot, \cdot \rangle_\Qm)$.
\end{remark}
\NEW{We seek two-channel filter banks that can provide  good signal representations,   
while having efficient   implementations on large (arbitrary) graphs. In practice, this can be achieved by designing SGFs $\Hm_i, \Gm_i$ that: (i) have good frequency selectivity and 
(ii) can be written as polynomials of   $\Zm$. We review bipartite graph solutions \cite{narang2012perfect,narang2013compact}  before introducing our proposed GFBs for arbitrary graphs. 
}
 %
\subsection{Bipartite filter banks \cite{narang2012perfect,narang2013compact}}
\label{subsec_BFB}
\NEW{The necessary and sufficient filter bank design conditions of \cite{narang2012perfect,narang2013compact} 
apply to  
bipartite graphs. 
}
\begin{definition}[Bipartite graph \cite{chung1997spectral}]
A graph $\Gc=(\Vc,\Ec)$  is  bipartite on  a partition of the node set $\Ac, \Bc$, if
for  all $(i,j)\in \Ec$, $i \in \Ac$ and $j \in \Bc$, or $i \in \Bc$ and $j \in \Ac$.
\end{definition}
Some examples of bipartite graphs include \autoref{fig:bip_block_graph_decomp} (middle graph), \autoref{fig:path} and \autoref{fig:image_graph_4con}.
In bipartite filter banks (BFB) the sets $\Ac$ and $\Bc$ are used for downsampling. In a bipartite graph  only edges between  $\Ac$ and $\Bc$ exist, so  we have: 
\begin{equation}\label{eq_bipartite_laplacians}
    \Lm = \begin{bmatrix}
    \Dm_\Ac & -\Wm_{ \Ac\Bc} \\ -\Wm_{\Bc\Ac} & \Dm_\Bc
    \end{bmatrix},~ \Lcb = \begin{bmatrix}
    \Id_\Ac & -\tilde{\Wm}_{\Ac\Bc} \\ -\tilde{\Wm}_{\Bc\Ac} & \Id_\Bc
    \end{bmatrix}.
\end{equation}
%
BFBs are implemented using spectral graph filters of  the $(\Lcb,\Id)$-GFT, 
that is $\Zm = \Lcb$, and  
\begin{equation}\label{eq_spec_graf_filt_normL}
  \Hm_i = h_i(\Lcb),  \quad \Gm_i = g_i(\Lcb), \quad i \in \{ 0,1 \}.
\end{equation}
The diagonal matrix $\Jm = \Sm^{\top}_{\Ac}\Sm_{\Ac}^{\phantom{\top}}  - \Sm^{\top}_{\Bc}\Sm_{\Bc}^{\phantom{\top}}$,  
with entries 
\begin{equation}\label{eq_f_setindicator}
 \Jm_{i,i} = \left\{ \begin{array}{cc}
		1  & \mbox{if } i \in \Ac \\
		-1 & \mbox{if } i \in \Bc,
	\end{array} \right.
\end{equation}
is used to establish the following property of the normalized Laplacian of bipartite graphs. 
\begin{proposition}[Spectral folding \cite{chung1997spectral}]
\label{prop_folding_bipartite} Let $\Lcb$ be the normalized Laplacian of a bipartite graph,  we have that
$\Lcb \uv = \lambda \uv $ if and only if $\Jm \uv$ is also an eigenvector with eigenvalue $2 - \lambda$.
\end{proposition}
\NEW{Thus, eigenvalues come in pairs ($\lambda$, $2 -\lambda$) mirrored around the middle frequency $\lambda=1$. For a given $\lambda$, the eigenvector $\uv$ and its folded version  $\Jm\uv$ have the same values for entries in $\Ac$, while signs are changed for the entries in $\Bc$. }
\autoref{prop_folding_bipartite} is the key property used by \cite{narang2012perfect,narang2013compact}  to design perfect reconstruction and $\Id$-orthogonal BFBs.  
\begin{theorem}[Perfect Reconstruction \cite{narang2012perfect}]\label{th_narang_pr}
A two-channel filter bank on a bipartite graph with spectral graph filters given by \eqref{eq_spec_graf_filt_normL} is PR if and only if,  $ \forall \lambda \in \sigma(\Lcb,\Id)$
\begin{align}\label{eq_pr1}
    g_0(\lambda)h_0(\lambda) +  g_1(\lambda)h_1(\lambda) &= 2, \\
    h_1(\lambda)g_1(2-\lambda) - h_0(\lambda)g_0(2-\lambda) &= 0. \label{eq_pr2}
\end{align}
\end{theorem}
The proof  follows from using     $ \Sm_\Ac^{\top} \Sm_\Ac^{\phantom{\top}} = \frac{1}{2}(\Id +\Jm)$ and $\Sm_\Bc^{\top} \Sm_\Bc^{\phantom{\top}} = \frac{1}{2}(\Id -\Jm)$ in the PR condition  \eqref{eq_PR_condition_sampling}, in combination with \autoref{prop_folding_bipartite} (see \cite{narang2012perfect}).
\autoref{prop_folding_bipartite} can also be used to design  $\Id$-orthogonal  filter banks.
\begin{theorem}[Orthogonality \cite{narang2012perfect}]\label{th_narang_ortho}
Under the same conditions of \autoref{th_narang_pr},  a  filter bank is $\Id$-orthogonal if and only if, 
$ \forall \lambda \in \sigma(\Lcb,\Id)$
\begin{align}\label{eq_orthogonal_filters1}
    h_0^2(\lambda) + h_1^2(\lambda) &= 2,\\
    h_1(\lambda)h_1(2-\lambda) - h_0(\lambda)h_0(2-\lambda) &= 0.\label{eq_orthogonal_filters2}
\end{align}
\end{theorem}

\NEW{The main advantage of $\Id$-orthogonal over non $\Id$-orthogonal PR filter banks is the energy preservation property (\autoref{def_Qorth}), essential for compression and de-noising applications. 
Because no polynomial solutions to \eqref{eq_orthogonal_filters1} and \eqref{eq_orthogonal_filters2} exist \cite{narang2013compact},  $\Id$-orthogonal filter banks  cannot be implemented with polynomial graph filters.
Instead, polynomial biorthogonal filter banks  with near $\Id$-orthogonality can be designed \cite{narang2013compact}.
\begin{proposition}\cite{narang2013compact,tay2015techniques}
	Biorthogonal filters  defined by
	\begin{align}\label{eq_bior}
		h_0(\lambda) = g_1(2-\lambda), \quad 
		h_1(\lambda) = g_0(2-\lambda), \\
  h_0(\lambda)h_1(2-\lambda) + h_0(2-\lambda)h_1(\lambda) = 2
	\end{align}
	are perfect reconstruction. 
\end{proposition}
In what follows we will show that the PR, $\Id$-orthogonality and biorthogonal conditions are not limited to bipartite graphs and the normalized Laplacian. The key insight is a generalization of  \autoref{prop_folding_bipartite}. 
 } 
%
%
%
%
%
\subsection{From bipartite to arbitrary graphs: Lazy filter bank}
\label{ssec:lazy_example}
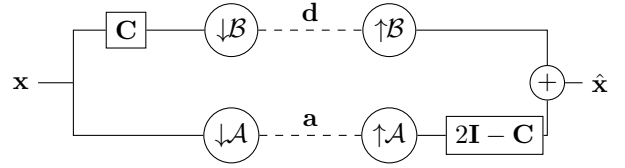
\begin{figure}[t]
    \centering
\def\hpdeltay{1}
\def\lpdeltay{-1}
\def\inputtosplitx{1}
\def\splittofilterx{1}
\def\filtertosamplingx{2}
\def\minspacedownupx{3}
\def\filtertooutputsplitx{1}
\begin{tikzpicture}[scale=0.7]
    \node (input) at (0, 0) {$\xv$};
    \coordinate (input_split) at (\inputtosplitx,0);

	\node[draw,rectangle] (H1) at ($(\inputtosplitx+\splittofilterx,\hpdeltay)$) {$\Cm$};
	\node[draw,circle,inner sep=2pt] (downB) at ($(H1)+(\filtertosamplingx,0)$) {$\downarrow\mkern-6mu \Bc$};

	\node[draw,circle,inner sep=2pt] (downA) at ($(H0)+(\filtertosamplingx,0)$) {$\downarrow\mkern-6mu\Ac$};

	\node[draw,circle,inner sep=2pt] (upB) at ($(downB)+(\minspacedownupx,0)$) {$\uparrow\mkern-6mu\Bc$};

	\node[draw,circle,inner sep=2pt] (upA) at ($(downA)+(\minspacedownupx,0)$) {$\uparrow\mkern-6mu\Ac$};
	\node[draw,rectangle] (G0) at ($(upA)+(\filtertosamplingx,0)$) {$2\Id - \Cm$};

    \node[draw,circle, inner sep=1pt](add) at ($(G1)+(\filtertooutputsplitx,\lpdeltay)$) {$+$};


    \node (output) at ($(add)+(\inputtosplitx,0)$) {$\hat{\xv}$};

    \draw (input) -- (input_split);
	\draw (input_split) |- (H1) -- (downB);
	\draw (input_split) |- (downA);

	\draw[dashed] (downB) -- node[above] {$\dv$} (upB);
	\draw[dashed] (downA) -- node[above] {$\av$} (upA);

    \draw (upB)  -| (add);
    \draw (upA) -- (G0) -| (add);

	\draw (add) -- (output);
\end{tikzpicture}
\caption{Lazy two-channel filter bank}
\label{fig_lazy_2chan}
\end{figure}
\NEW{Before stating our main results, we  show how our extension from bipartite to arbitrary graphs would work for the ``lazy'' filter bank  of \autoref{fig_lazy_2chan}, where $\Cm$ is given by} 
\begin{equation}\label{eq_Cmatrix}
    \Cm = \begin{bmatrix}
        \Id_{\Ac} & \Cm_{\Ac \Bc} \\
        \Cm_{\Bc \Ac} & \Id_{\Bc}
    \end{bmatrix}.
\end{equation}
$\Cm_{\Ac \Bc}$  and $\Cm_{\Bc \Ac}$ are arbitrary rectangular matrices of dimensions $\vert \Ac \vert \times \vert \Bc \vert$ and $\vert \Bc \vert \times \vert \Ac \vert$, respectively. \NEW{ 
It is easy to verify that  the lazy filter bank is  \emph{perfect reconstruction} when  $\Cm$ is given by \eqref{eq_Cmatrix}.
Since $\dv = \Sm_{\Bc} \Cm \xv$ and $\av = \Sm_{\Ac}\xv$, we have that
\begin{align}\nonumber
    \hat{\xv} &=\Sm_{\Bc}^{\top}\dv  + (2\Id - \Cm)\Sm_{\Ac}^{\top}\av=(\Sm_{\Bc}^{\top}\Sm_{\Bc}\Cm  + (2\Id - \Cm)\Sm_{\Ac}^{\top}\Sm_{\Ac})\xv \\ 
    &=\left(\begin{bmatrix}
        \mathbf{0} & \mathbf{0} \\
        \Cm_{\Bc \Ac} & \Id_{\Bc}
    \end{bmatrix}
    + \begin{bmatrix}
        \Id_{\Ac} & \mathbf{0} \\
        -\Cm_{\Bc \Ac} & \mathbf{0}
    \end{bmatrix} \right)\xv = \xv.
\end{align}
If we take $\Cm = \Lcb$ for a bipartite graph,  the lazy filter bank   is  a special case of a \emph{perfect reconstruction} BFB (\autoref{fig:2chanfb})  with filters $\Hm_0 = \Id$, $\Hm_1 = \Lcb$, $\Gm_0 = 2\Id - \Lcb$, and $\Hm_1 = \Id$, 
corresponding to 
SGFs of the $(\Lcb,\Id)$-GFT with \emph{biorthogonal} filter kernels  $h_0(\lambda) = g_1(\lambda) = 1$,  $h_1(\lambda) = \lambda$, and $g_0(\lambda) = h_1(2-\lambda)$. 
}
In the following subsections we will show that  it is also possible to choose $\Cm$ as a function of the variation operator $\Mm$ of a non-bipartite  graph,   
 which allows us to generalize the theorems from \autoref{subsec_BFB} to arbitrary graphs.
\subsection{Spectral folding on arbitrary graphs}
\label{ssec_folding}
We generalize \autoref{prop_folding_bipartite} to other graphs and variation operators by using  the  $\Qm$ inner product. Using this result, we propose a new $(\Mm,\Qm)$-GFT and show that it can be used to  obtain \emph{Generalized Filter Banks} on arbitrary graphs.
First we define  the  spectral folding property for a $(\Mm,\Qm)$-GFT.
\begin{definition}[Spectral folding]\label{def_spectral_folding}
Given a graph $\Gc$ with variation operator $\Mm\succeq 0$, inner product $\Qm \succ 0$ and a partition $\Ac$, $\Bc$. The $(\Mm,\Qm)$-GFT has the {\bf spectral folding property}, 
if for all generalized eigenpairs $(\uv,\lambda)$    then  $(\Jm\uv,(2-\lambda))$ is also a generalized eigenpair, that is:   
\begin{equation}
    \Mm \uv = \lambda \Qm \uv \Leftrightarrow  \Mm \Jm\uv = (2-\lambda) \Qm \Jm\uv,
\end{equation}
where $\Jm = \Sm_\Ac^{\top} \Sm_\Ac^{\phantom{\top}}  - \Sm_\Bc^{\top} \Sm_\Bc^{\phantom{\top}}$, as in \eqref{eq_f_setindicator}. 
\end{definition}
%
The following theorem  completely characterizes the spectral folding property.  
The proof  can be found in   \autoref{app_proof_folding_variation}.
\begin{theorem}[Spectral folding]
\label{prop:folding-variation}
Given a graph $\Gc$ with variation operator $\Mm\succeq 0$, inner product $\Qm$ and a partition $\Ac = \lbrace  1, \cdots, \vert \Ac \vert \rbrace$, $\Bc = \Vc \setminus \Ac$, for which $\Mm_{\Ac \Ac}$ and $\Mm_{\Bc\Bc}$ are invertible. The $(\Mm,\Qm)$-GFT has the spectral folding property if and only if  $\Qm$ is chosen as 
\begin{equation}
\Qm=\begin{bmatrix}\label{eq_M_Q}
    \Mm_{\Ac\Ac} & \mathbf{0} \\ \mathbf{0} & \Mm_{\Bc\Bc}
    \end{bmatrix}.
\end{equation}
\end{theorem}
\NEW{The condition that $\Mm_{\Ac \Ac}$ and $\Mm_{\Bc \Bc}$ are non singular is satisfied for any vertex partition  if $\Mm \succ 0$. This condition also holds  if $\Mm$ is the combinatorial or normalized Laplacian of a connected graph. }
\autoref{prop:folding-variation}  generalizes \autoref{prop_folding_bipartite},  demonstrating that the spectral folding property is not unique to the  $(\Lcb,\Id)$-GFT of bipartite graphs, and in fact,  it is always satisfied if the inner product matrix  $\Qm$ is chosen as in \eqref{eq_M_Q}.  

Using \eqref{eq_M_Q},  the fundamental matrix $\Zm=\Um \Lambdam \Um^{\top}\Qm$ is 
\begin{equation}\label{eq_fundamental_matrix_folding}
    \Zm = \Qm^{-1}\Mm = \begin{bmatrix}
    \Id_{\Ac} & \Mm_{\Ac\Ac}^{-1} \Mm_{\Ac \Bc} \\
    \Mm_{\Bc\Bc}^{-1} \Mm_{\Bc\Ac} &\Id_{\Bc}
    \end{bmatrix},
\end{equation}
so that $\Zm$ has the same structure as $\Cm$ in \eqref{eq_Cmatrix} and thus $\Zm$ can be used to construct  ``lazy'' filter banks (see \autoref{fig_lazy_2chan}). In \autoref{ssec_filterbanks_arbitrary} we apply \autoref{prop:folding-variation} with the fundamental matrix $\Zm$ in \eqref{eq_fundamental_matrix_folding}  to design GFBs on arbitrary graphs. 
\NEW{To gain some intuition about   $(\Mm,\Qm)$-GFTs with spectral folding,  
we study some of their properties from vertex, spectral and probabilistic perspectives (\autoref{sec_properties_examples}) and propose     criteria to choose vertex partitions with favorable properties (\autoref{sec_vertex_partitioning}).  }
\subsection{Proof of Theorem \ref{prop:folding-variation} (Spectral folding)}
\label{app_proof_folding_variation}
First we  prove that  \eqref{eq_M_Q} implies the spectral folding property (\autoref{def_spectral_folding}).
Since $\Mm$ and $\Qm$ are positive semidefinite, and $\Qm$ is non singular, there is a full set of generalized eigenvectors \cite{horn2012matrixbook}. Let $\uv = [\uv_\Ac^{\top}, \uv_\Bc^{\top} ]^{\top}$ and  $\Mm \uv = \lambda \Qm \uv$, then
\begin{align}
    \Mm_{\Ac\Ac} \uv_\Ac + \Mm_{\Ac\Bc} \uv_\Bc &= \lambda \Qm_\Ac \uv_\Ac,\label{eq_block_genEV1}  \\
    \Mm_{\Bc\Ac} \uv_\Ac + \Mm_{\Bc\Bc} \uv_\Bc &= \lambda \Qm_\Bc \uv_\Bc.\label{eq_block_genEV2}
\end{align}
Set $\mathbf{v} = \Jm\uv=[\uv_\Ac^{\top}, -\uv_\Bc^{\top} ]^{\top}$, and using  \eqref{eq_block_genEV1} and \eqref{eq_block_genEV2} we get
\begin{align}
\Mm \vv &= 
    \begin{bmatrix}
    \Mm_{\Ac\Ac} \uv_\Ac  -\Mm_{\Ac\Bc} \uv_\Bc \\
    \Mm_{\Bc\Ac} \uv_\Ac  -\Mm_{\Bc\Bc} \uv_\Bc
    \end{bmatrix} \\
    &=
    \begin{bmatrix}
     2 \Mm_{\Ac\Ac} \uv_\Ac  -\lambda \Qm_{\Ac} \uv_\Ac \\
    \lambda \Qm_\Bc \uv_\Bc  - 2\Mm_{\Bc\Bc} \uv_\Bc
    \end{bmatrix}=(2-\lambda) \Qm \vv.
\end{align}
The second equality is implied by $\Mm_{\Ac\Ac} = \Qm_\Ac$, and $\Mm_{\Bc\Bc} = \Qm_\Bc$. To prove the other direction of the equivalence in \autoref{def_spectral_folding} we  set $\gamma=  2-\lambda$ and repeat the same steps.

Now, we prove that spectral folding implies \autoref{eq_M_Q}.
Let
\begin{equation}
\Qm =
        \begin{bmatrix}
            \Qm_{\Ac}    & \Qm_{\Ac\Bc} \\
            \Qm_{\Bc\Ac} & \Qm_{\Bc}
        \end{bmatrix}.    
\end{equation}
We will first show that $\mathbf{Q}$ is  block diagonal.
Using the spectral folding assumption and given two generalized eigenvectors $\mathbf{u}$ and $\mathbf{v}$ with unit $\Qm$-norm, we have that $\Jm\uv$ and $\Jm\vv$ are also generalized eigenvectors of unit $\Qm$-norm. In addition,   $\Jm\uv$ is $\Qm$-orthogonal to $\Jm\vv$, and $\uv$ is $\Qm$-orthogonal to $\vv$.
Therefore, given the matrix of generalized eigenvectors $\Um$,
\begin{equation}\label{eq_orthoplusorth}
 \Um^\top\left(\Qm+\Jm\Qm\Jm\right)\Um=2\Id
    \text{.}   
\end{equation}
We replace  the right hand side of \eqref{eq_orthoplusorth} with $2 \Um^{\top}\Qm \Um = 2\Id$.  After simplification  we obtain $\Jm\Qm\Jm=\Qm$, meaning that $\Qm_{\Ac\Bc}=\mathbf{0}$, and $\Qm_{\Bc\Ac}=\mathbf{0}$.

Next, we prove that $\Qm_\Ac=\Mm_{\Ac \Ac}$ and $\Qm_\Bc=\Mm_{\Bc \Bc}$.
Using the spectral folding property, and given $\uv$, a generalized eigenvector with eigenvalue $\lambda$, we obtain:
\begin{align}
    \Mm(\Id+\Jm)\uv &= (\lambda\Qm+(2-\lambda)\Qm\Jm)\uv \label{eq_folding_MAUA}\\
    \Mm(\Id-\Jm)\uv &= (\lambda\Qm-(2-\lambda)\Qm\Jm)\uv \label{eq_folding_MBUB}.
\end{align}
\eqref{eq_folding_MAUA} and \eqref{eq_folding_MBUB} imply 
  $2\Mm_{\Ac \Ac}\uv_\Ac = 2\Qm_\Ac\uv_\Ac$ and $2\Mm_{\Bc\Bc}\uv_\Bc = 2\Qm_\Bc\uv_\Bc$, respectively.
Gathering all these equations for all eigenvectors into matrix form, we obtain $\begin{bmatrix}
            \Mm_{\Ac \Ac}    & \mathbf{0} \\
            \mathbf{0} & \Mm_{\Bc \Bc}
        \end{bmatrix}\Um=\begin{bmatrix}
            \Qm_{\Ac}    & \mathbf{0} \\
            \mathbf{0} & \Qm_{\Bc}
        \end{bmatrix}\Um$.
Since $\Um$ is invertible, we obtain   that $\Mm_{\Ac \Ac}=\Qm_\Ac$ and $\Mm_{\Bc \Bc}=\Qm_\Bc$. 
\subsection{Generalized filter banks on arbitrary graphs}
\label{ssec_filterbanks_arbitrary}
We use the spectral folding property (\autoref{def_spectral_folding}) and   \autoref{prop:folding-variation}  to construct perfect reconstruction (\autoref{th_PR_arbitrary})  and $\Qm$-orthogonal (\autoref{th_orthogonality_arbitrary})   filter banks. Our proofs (Appendices \ref{app_PR} and  \ref{app_Orth}) follow closely the proofs of \autoref{th_narang_pr} and \autoref{th_narang_ortho} for the bipartite case \cite{narang2012perfect}, with the main difference being the use of spectral graph filters of the $(\Mm,\Qm)$-GFT. 
We state the conditions for perfect reconstruction and $\Qm$-orthogonal  filter banks (see \autoref{def_Qorth}) next.
\begin{theorem}\label{th_PR_arbitrary}
Consider a positive semi-definite variation operator $\Mm$  and a vertex partition $\Ac$, $\Bc$.  $\Qm$  is  
chosen according to \autoref{prop:folding-variation}  so that the $(\Mm,\Qm)$-GFT has the spectral folding property. 
A  two-channel filter bank with SGFs
 \begin{align}\label{eq_spec_graf_filt_Z1}
  \Hm_i &= h_i(\Zm) = \Um h_i(\Lam)\Um^{\top}\Qm \textnormal{ and }\\ \label{eq_spec_graf_filt_Z2}
  \Gm_i &= g_i(\Zm) = \Um g_i(\Lam)\Um^{\top}\Qm
\end{align}
 is perfect reconstruction, if and only if  for all $\lambda \in \sigma(\Mm,\Qm)$
\begin{align}\label{eq_pr1_Q}
    g_0(\lambda)h_0(\lambda) +  g_1(\lambda)h_1(\lambda) &= 2, \\
    h_1(\lambda)g_1(2-\lambda) - h_0(\lambda)g_0(2-\lambda) &= 0. \label{eq_pr2_Q}
\end{align}
\end{theorem}
%
%
%

\begin{theorem}\label{th_orthogonality_arbitrary}
Under the  conditions of  \autoref{th_PR_arbitrary},   a two-channel filter bank is $\Qm$-orthogonal if and only if,
\begin{align}\label{eq_orthogonal_filters1_Q}
    h_0^2(\lambda) + h_1^2(\lambda) &= 2,\\
    h_1(\lambda)h_1(2-\lambda) - h_0(\lambda)h_0(2-\lambda) &= 0,\label{eq_orthogonal_filters2_Q}
\end{align}
for all $\lambda \in \sigma(\Mm,\Qm)$.
\end{theorem}
%
%
%
\NEW{Note that the PR and $\Qm$-orthogonality conditions on  $h_i$ and $g_i$ are the same for BFBs and GFBs.}
\NEW{
Thus filter designs for BFBs can be used to construct GFBs. 
From a computational perspective, spectral graph filters that have 
polynomial filter implementations  are more desirable. As in the BFB case, there are no exactly polynomial solutions to the  $\Qm$-orthogonality conditions  \eqref{eq_orthogonal_filters1_Q} and  \eqref{eq_orthogonal_filters2_Q} \cite{narang2013compact}.  However,  the biorthogonal filters from   \eqref{eq_bior} also satisfy the PR conditions of \autoref{th_PR_arbitrary} and have polynomial solutions. 
Since the biorthogonal designs from \cite{narang2013compact} are designed to approximately satisfy conditions \eqref{eq_orthogonal_filters1_Q} and  \eqref{eq_orthogonal_filters2_Q},  they are also approximately $\Qm$-orthogonal.  }

\NEW{
\begin{remark}
While $\Id$-orthogonal filter banks  are preferable in many scenarios (e.g., for coding applications),  nearly $\Id$-orthogonal solutions are often used in practice if they provide other useful properties. As an example,  $\Id$-orthogonal  filter banks cannot be constructed with finite impulse response linear phase filters (other than the Haar filters) \cite{vetterli1995wavelets}, thus  nearly $\Id$-orthogonal biorthogonal filters have been used because of the advantages of symmetry for image coding applications  \cite{Taubman2002JPEG2000I}. 
For GFBs,  we will show in \autoref{sec_vertex_partitioning} that by optimizing the downsampling sets based on a numerical stability criteria, the  matrix $\Qm$ can be close to a diagonal matrix.
\end{remark}
}
\begin{remark} The zeroDC filter bank\cite{narang2013compact} was  proposed so that DC (constant) signals are mapped to the lowest graph frequency ($\lambda =0$). This is achieved by multiplying the input signal by $\Dm^{1/2}$ before applying the analysis filter bank (based on the $(\Lm,\Id)-GFT$), and multiplying by $\Dm^{-1/2}$ at the output of the synthesis filter bank. This ensures that a constant input signal has zero response in the high pass channel. \cite{narang2013compact} showed that biorthogonal zeroDC filter banks can be implemented with polynomials of the  random walk Laplacian of a bipartite graph. The zeroDC filter banks can be derived as a special case of our framework with the $(\Lm,\Dm)$-GFT, by noticing that for bipartite graphs with Laplacian $\Lm$, \autoref{prop:folding-variation}  leads to choosing $\Qm = \Dm$.  
\end{remark}
\subsection{Tree structured generalized filter banks}
\label{ssec_treeGFB}
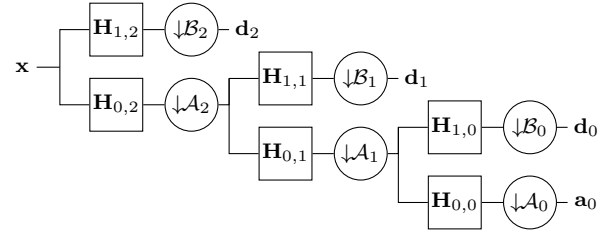
\begin{figure}[tb]
\centering
\def\hpdeltay{1}
\def\lpdeltay{-1}
\def\inputtosplitx{1}
\def\splittofilterx{1.5}
\def\filtertosamplingx{2}
\def\minspacedownupx{3}
\def\filtertooutputsplitx{1}
\def\splittooutput{0.5}

\footnotesize
\def\minsize{20}
\def\upfrac{0.7}

\colorlet{input}{black!80!black}
\colorlet{detail}{black!80!black}
\colorlet{approx}{black!80!black}
\begin{tikzpicture}[scale=0.5]
    \node (input) at (0, 0) {\color{input} $\xv$};
    \coordinate (input_split) at (\inputtosplitx,0);

	\node[draw,rectangle,inner sep=0,minimum size=\minsize] (H1) at ($(\inputtosplitx+\splittofilterx,\hpdeltay)$) {$\Hm_{1,2}$};
	\node[draw,circle,inner sep=0,minimum size=\minsize] (downB) at ($(H1)+(\filtertosamplingx,0)$) {$\downarrow\mkern-6mu \Bc_2$};

	\node[draw,rectangle,inner sep=0,minimum size=\minsize] (H0) at ($(\inputtosplitx+\splittofilterx,\lpdeltay)$) {$\Hm_{0,2}$};
	\node[draw,circle,inner sep=0,minimum size=\minsize] (downA) at ($(H0)+(\filtertosamplingx,0)$) {$\downarrow\mkern-6mu\Ac_2$};

	\coordinate (output1_split1) at ($(downB) + (\inputtosplitx,0)$);
	\node (d2) at ($(output1_split1) + (\splittooutput,0)$){\color{detail} $\dv_2$};

    \draw (input) -- (input_split);
	\draw (input_split) |- (H1) -- (downB);
	\draw (input_split) |- (H0) -- (downA);
    \draw(downB)--(output1_split1);



	\coordinate (output0_split1) at ($(downA) + (\inputtosplitx,\upfrac)$);
	\coordinate (output0_split2) at ($(output0_split1) + (0, -\hpdeltay + \lpdeltay)$);

	\node[draw,rectangle,inner sep=0,minimum size=\minsize] (H01) at ($(output0_split1)+(\splittofilterx,0)$) {$\Hm_{1, 1}$};
	\node[draw,rectangle,inner sep=0,minimum size=\minsize] (H00) at ($(H01)+(0, -\hpdeltay + \lpdeltay)$) {$\Hm_{0, 1}$};
	\node[draw,circle,inner sep=0,minimum size=\minsize] (downAA) at ($(H00)+(\filtertosamplingx,0)$) {$\downarrow\mkern-6mu\Ac_{1}$};
	\node[draw,circle,inner sep=0,minimum size=\minsize] (downAB) at ($(H01)+(\filtertosamplingx,0)$) {$\downarrow\mkern-6mu\Bc_{1}$};
	
	\coordinate (output01_split1) at ($(downAB) + (\inputtosplitx,0)$);
	\node (d1) at ($(output01_split1) + (\splittooutput,0)$){\color{detail} $\dv_1$};
	
	\draw (downA) -| (output0_split1) -- (H01) -- (downAB);
	\draw  (output0_split1) -- (output0_split2) -- (H00) -- (downAA);
	\draw(downAB)--(output01_split1);

 	\coordinate (output00_split1) at ($(downAA) + (\inputtosplitx,\upfrac)$);
 	\coordinate (output00_split2) at ($(output00_split1) + (0, -\hpdeltay + \lpdeltay)$);

 	\node[draw,rectangle,inner sep=0,minimum size=\minsize] (H001) at ($(output00_split1)+(\splittofilterx,0)$) {$\Hm_{1, 0}$};
 	\node[draw,rectangle,inner sep=0,minimum size=\minsize] (H000) at ($(H001)+(0, -\hpdeltay + \lpdeltay)$) {$\Hm_{0, 0}$};
 	\node[draw,circle,inner sep=0,minimum size=\minsize] (downAAA) at ($(H000)+(\filtertosamplingx,0)$) {$\downarrow\mkern-6mu\Ac_{0}$};
 	\node[draw,circle,inner sep=0,minimum size=\minsize] (downAAB) at ($(H001)+(\filtertosamplingx,0)$) {$\downarrow\mkern-6mu\Bc_{0}$};

	\coordinate (output001_split1) at ($(downAAB) + (\inputtosplitx,0)$);
	\node (d0) at ($(output001_split1) + (\splittooutput,0)$){\color{detail} $\dv_0$};

	\coordinate (output000_split1) at ($(downAAA) + (\inputtosplitx,0)$);
	\node (a0) at ($(output000_split1) + (\splittooutput,0)$){\color{approx} $\av_0$};
	
 	\draw  (downAA) -| (output00_split1) -- (H001) -- (downAAB);
 	\draw  (output00_split1) -- (output00_split2) -- (H000) -- (downAAA);

	\draw(downAAB)--(output001_split1);
	\draw(downAAA)--(output000_split1);
\end{tikzpicture}
\caption{Tree structured analysis  filter bank}
    \label{fig:iterated_fb_1}
\end{figure}

%
Tree structured GFBs are formed by  concatenating  two-channel GFBs.  For all resolution levels, the  graphs, variation operators and sampling sets are given and fixed, but otherwise arbitrary.  See \autoref{fig:iterated_fb_1} for an example with $L=3$ levels. 
%

 We assume the input signal is at  resolution $L$, thus $\av_L = \xv$. The outputs of the low and high  pass channels at resolution $\ell<L$ are called approximation and detail coefficients, and are denoted by    $\av_{\ell}$, and $\dv_{\ell}$,    respectively. 
 The sampling sets obey, $\Vc = \Ac_{L}$, and  for $\ell<L$,  $\Vc_{\ell} = \Ac_{\ell}$,   $\Ac_{\ell+1} = \Ac_{\ell}\cup\Bc_{\ell}$, and $\Bc_{\ell} = \Ac_{\ell+1}\setminus \Ac_{\ell}$. The graph at resolution $\ell$  is denoted by $\Gc_{\ell} = (\Vc_{\ell},\Ec_{\ell})$, and has variation operator $\Mm_{\ell}$ with corresponding inner product matrix $\Qm_{\ell}$, chosen so that the $(\Mm_{\ell}, \Qm_{\ell})$-GFT  has the spectral folding property.
 We will consider a family of PR filter banks, with analysis  and synthesis operators at resolution $\ell$ denoted by $\Tm_{a,\ell}$ and $\Tm_{s,\ell} = \Tm_{a,\ell}^{-1}$, respectively. The analysis equation at resolution $\ell$ is given by
\begin{equation}
    \Tm_{a,\ell} \av_{\ell+1} = \begin{bmatrix}
    \av_{\ell}^{\top} &
    \dv_{\ell}^{\top}
    \end{bmatrix}^{\top},
\end{equation}
where $\av_{\ell} = \Sm_{\Ac_{\ell}}\Hm_{0,\ell} \av_{\ell+1}$, and $\dv_{\ell} = \Sm_{\Bc_{\ell}}\Hm_{1,\ell} \av_{\ell+1}$.
The synthesis operator implements
\begin{equation}\label{eq_synthesis_iterated_1level}
   \av_{\ell+1} = \Tm_{s,\ell}\begin{bmatrix}
   	\av_{\ell}^{\top} &
   	\dv_{\ell}^{\top}
   \end{bmatrix}^{\top}= \Gm_{0,\ell} \Sm^{\top}_{\Ac_{\ell}} \av_{\ell} + \Gm_{1,\ell} \Sm^{\top}_{\Bc_{\ell}} \dv_{\ell}.
\end{equation}
After applying (\ref{eq_synthesis_iterated_1level}) recursively we can represent $\xv$ as a linear combination of coefficients at various resolutions $\cv = [\av^{\top}_0,  \dv^{\top}_0,    \cdots,   \dv^{\top}_{L-1}]^{\top}$, and define the synthesis operator of the tree structured filter bank $\Tcb_s$ via the equation
\begin{equation}\label{eq_synthesis_tree1}
    \xv = \Tcb_s \cv.
\end{equation}
Because $\Tcb_s$ is the composition of the synthesis operators $\Tm_{s,\ell}$ at multiple resolutions, there is a corresponding analysis operator $\Tcb_a = \Tcb_s^{-1}$, that can be implemented as a composition of analysis operators $\Tm_{a,\ell}$.

\section{Properties, interpretation and examples}
\label{sec_properties_examples}
\NEW{The term \emph{graph Fourier transform} (GFT) associated to the eigenvectors of a graph operator is often justified by the fact that the discrete Fourier transform (DFT) diagonalizes the adjacency matrix of circulant graphs. 
More generally, the frequency interpretation of the $(\Mm,\Id)$-GFT is justified from its  variational definition, i.e., \eqref{eq_MQGFT_def1} and \eqref{eq_MQGFT_def2}. 
In this section we motivate the use of the proposed $(\Mm,\Qm)$-GFTs for graph signal representation, provide insights about its frequency interpretation, and through examples help further  understand
 the role of $\Qm$. }
\subsection{Spectral properties of the \texorpdfstring{$(\Mm,\Qm)$}{(M,Q)}-GFT}
\label{app_properties_Z}
%
\subsubsection{Eigenvalue bounds} \NEW{For any graph with  normalized Laplacian  $\Lcb = \Dm^{-1/2}\Lm \Dm^{-1/2}$, its eigenvalues belong to the $[0,2]$ interval, moreover $\sigma(\Lm,\Dm) = \sigma(\Lcb,\Id) \subset [0,2]$ \cite{chung1997spectral}. Also,  for the $(\Lm,\Dm)$-GFT and $(\Lcb,\Id)$-GFT,     $\lambda_n=2$ if and only if the graph is bipartite \cite{chung1997spectral}. For the $(\Mm,\Qm)$-GFT there is a similar result. }
\begin{proposition}\label{prop_eigenvalue_bound}
If the $(\Mm,\Qm)$-GFT has the spectral folding property then $\sigma(\Mm,\Qm) \subset [0,2]$, and  $\lambda_n=2$ if and only if $\lambda_1=0$. 
\end{proposition}
\begin{proof}
For all $\xv$, $\xv^{\top}\Mm \xv \geq 0$ and $\xv^{\top}\Jm \Mm \Jm \xv \geq 0$, therefore
\begin{equation}\label{eq_bound_xMx}
    0 \leq \xv^{\top}\Mm \xv \leq \xv^{\top}\Mm \xv + \xv^{\top}\Jm \Mm \Jm \xv = 2 \xv^{\top}\Qm \xv, 
\end{equation}
where we use the identity $\Mm + \Jm\Mm\Jm = 2\Qm$. 
Now for any generalized  eigenvector $\uv$ with unit $\Qm$-norm and eigenvalue $\lambda$, we have $\uv^{\top} \Mm \uv = \lambda \uv^{\top} \Qm \uv = \lambda$, and using  \eqref{eq_bound_xMx} we obtain 
\begin{equation}
     0 \leq   \frac{\uv^{\top}\Mm \uv}{\uv^{\top}\Qm \uv} = \lambda  \leq 2.
\end{equation}
The  statement, $\lambda_n = 2$ if and only if $\lambda_1=0$, follows directly from the spectral folding property.
\end{proof}

\subsubsection{Middle frequency $\lambda=1$} \NEW{In the bipartite case, the subspace associated to the eigenvectors corresponding to $\lambda =1$ has the least frequency discrimination because energies at this frequency contribute  equally to the low pass and high pass channels. This can be seen from  \eqref{eq_orthogonal_filters2} evaluated at $\lambda=1$, which shows that $h_0(1)^2=h_1(1)^2$.
Since the  subspace associated to $\lambda=1$ has  dimension at least $\vert \vert \Ac \vert - \vert \Bc \vert  \vert$,  bipartite graphs with more balanced partitions are usually preferred  \cite{zeng2017bipartite}.  
There is a similar result for the $(\Mm,\Qm)$-GFT.}
\begin{proposition}\label{prop_middle_lambda}
If the $(\Mm,\Qm)$-GFT has the spectral folding property with vertex partition partition $\Ac, \Bc$, then the multiplicity of $\lambda =1$ is at least $\vert \vert \Ac \vert - \vert \Bc \vert  \vert$.
\end{proposition}
\begin{proof}
We will show that
\begin{equation}\label{eq_lambda1_dimension}
   \vert \{ i: \lambda_i=1 \} \vert \geq  \vert \vert \Ac \vert - \vert \Bc \vert   \vert.
\end{equation}
If $\lambda=1$, and $\uv$ is the corresponding generalized eigenvector, then $\Zm \uv = \uv$, hence the dimension of the null space of  $\Id - \Zm$  is equal to the multiplicity of $\lambda=1$. Note that $\Id - \Zm$  has a the sparsity pattern of a bipartite adjacency matrix, thus $(\Id - \Zm)^2$ is block diagonal, and  
\begin{align}
    \vert \{ i: \lambda_i=1 \} \vert &= \dim\ker(\Id - \Zm ) \\
    &= n - \rank(\Zm_{\Ac\Bc}) - \rank(\Zm_{\Bc\Ac})\\
    &= n-2\rank(\Mm_{\Ac\Bc})\\
    &\geq n - 2 \min(\vert \Ac \vert, \vert \Bc \vert) \\
    &= \vert \vert \Ac \vert - \vert \Bc \vert \vert.
\end{align}
 The rank of $\Id - \Zm$ is the sum of the ranks  of the blocks (this can be proven using the fact that t$(\Id - \Zm)^2$ is block diagonal).  Then the rank of each individual term is equal to the rank of $\Mm_{\Ac\Bc}$, because $\Qm$ is non-singular. The rank of a matrix is upper bounded by the minimum between the number of rows and columns. Finally we use the fact that $n = \vert \Ac \vert + \vert \Bc \vert$.
\end{proof}

\subsubsection{Multiplicity of $\lambda_1$} \NEW{When $\Mm$ is a generalized Laplacian the multiplicity of the smallest eigenvalue is equal to the number of connected components of the graph \cite{biyikoglu2007laplacian}.  The smallest generalized eigenvalue also has this property.}
\begin{proposition}
If $\Mm \succeq 0$ is a generalized Laplacian, that is, $\Mm_{ij} \leq 0$ for all $i \neq j$,  the $(\Mm,\Qm)$-GFT has the spectral folding property, and $n \geq 2$, then the multiplicity of $\lambda_1$ is equal to the number of connected components of the graph. 
\end{proposition}
\begin{proof}
Let $\lambda_1$ be the smallest generalized eigenvalue. The operation $\hat{\Mm} = \Mm - \lambda_1\Qm$ preserves generalized eigenvectors, and shifts all generalized eigenvalues by $-\lambda_1$. Then the multiplicity of $\lambda_1$ is equal to dimension of the null space of $\hat{\Mm}$, thus the desired result is proven if  we can show that $\hat{\Mm}$ is a positive semi-definite generalized Laplacian. To show that $\hat{\Mm} \succeq 0$, note \eqref{eq_MQGFT_gen_eigen} implies that $\hat{\Mm} = \sum_{i=1}^n (\lambda_i - \lambda_1) \Qm \uv_i \uv_i^{\top}\Qm$, thus  $\xv^{\top} \hat{\Mm} \xv \geq 0$ for any $\xv$. To see that $\hat{\Mm}$ is also a generalized Laplacian we use the identity
\begin{equation}
    \hat{\Mm} = (1-\lambda_1)\Qm + \Mm - \Qm.
\end{equation}
Because $n \geq 2$ and  the spectral folding property, $\lambda_1 \leq 1$, and $(1-\lambda_1)\Qm$ is a generalized Laplacian. Since $\Mm - \Qm$ has zero diagonal and  non positive off-diagonal entries, then $\hat{\Mm}$ must be a generalized Laplacian.
\end{proof}
\subsubsection{$(\Mm,\Qm)$-GFT examples}
\begin{figure}[t]
     \centering
\begin{subfigure}[b]{0.44\textwidth}
\centering
		\includegraphics[width=0.75\textwidth]{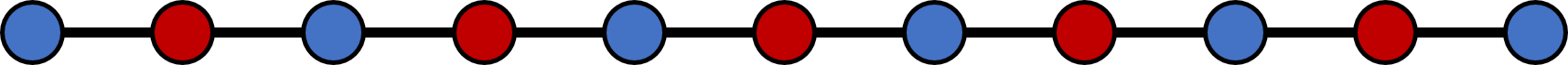}
		\caption{$\Gc_1$: Bipartite path graph}
\label{fig:path}
	\end{subfigure}%
	
\begin{subfigure}[b]{0.44\textwidth}
\centering
		\includegraphics[width=0.75\textwidth]{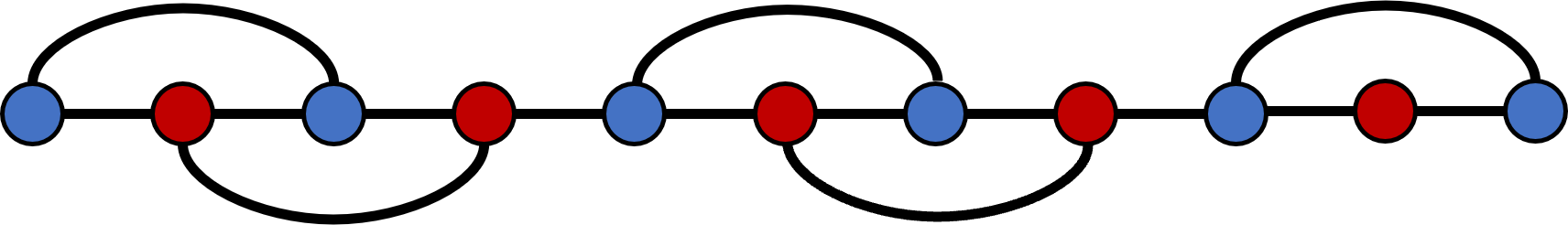}
		\caption{$\Gc_2$: Non-bipartite path graph}
\label{fig:path2}
	\end{subfigure}%
     \caption{Examples of two similar path graphs with $n=11$ nodes. Vertex partition is marked with blue (nodes in $\Ac$) and red (nodes in $\Bc$).}
     \label{fig:path_graphs}
 \end{figure}
 \begin{figure}[t]
     \centering
\begin{subfigure}[b]{0.48\textwidth}
\centering
\includegraphics[width=1\textwidth]{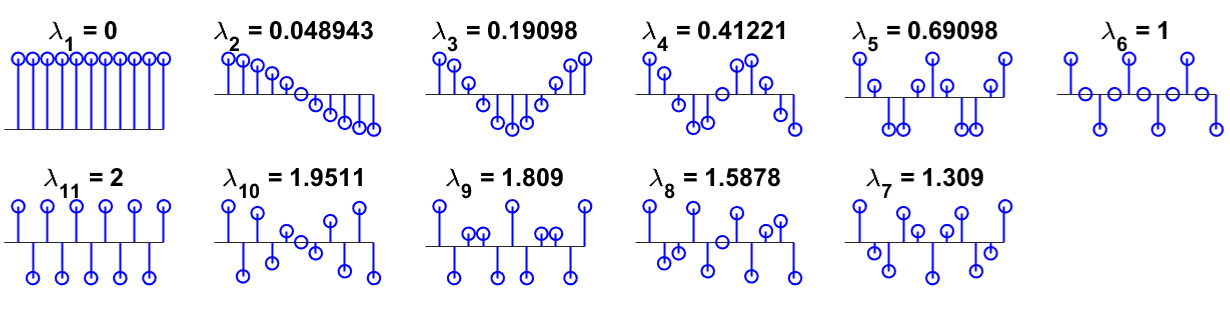}
		\caption{Basis functions  of the $(\Lm_1,\Dm_1)$-GFT of graph $\Gc_1$}
\label{fig:path_LD}
	\end{subfigure}%
 \vspace{0.5cm}
\begin{subfigure}[b]{0.48\textwidth}
\centering	\includegraphics[width=1\textwidth]{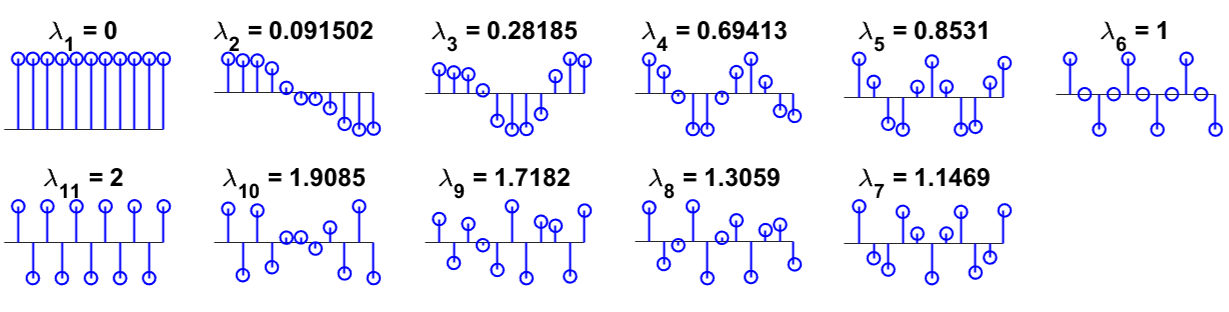}
		\caption{Basis functions  of the $(\Lm_2,\Qm_2)$-GFT of graph $\Gc_2$}
\label{fig:path2_LQ}
	\end{subfigure}%
 \vspace{0.5cm}
	\begin{subfigure}[b]{0.48\textwidth}
\centering
		\includegraphics[width=1\textwidth]{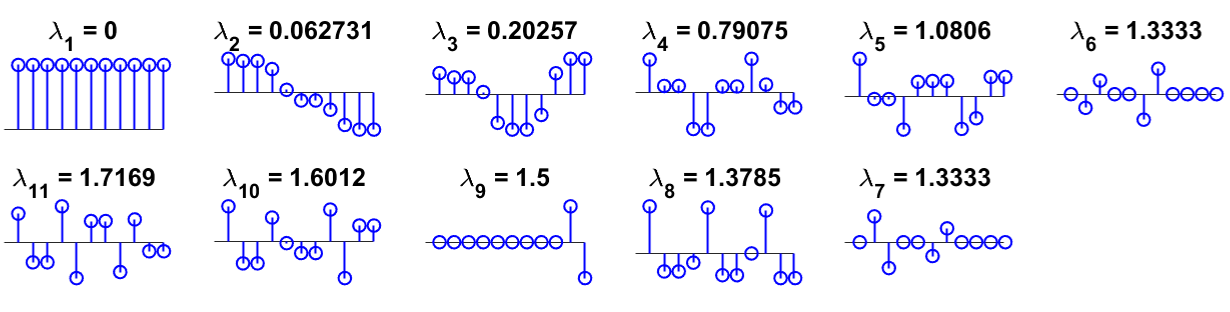}
		\caption{Basis functions  of the $(\Lm_2,\Dm_2)$-GFT of graph $\Gc_2$}
\label{fig:path2_LD}
	\end{subfigure}%
     \caption{Graph Fourier transforms for graphs of \autoref{fig:path_graphs}}
     \label{fig:path_graphs_MQ}
 \end{figure}
 \NEW{
$\Gc_1$ in \autoref{fig:path} is a bipartite path graph, while $\Gc_2$ in \autoref{fig:path2} is a non bipartite graph, formed by adding a few edges to $\Gc_1$. All edge weights in $\Gc_1$ and $\Gc_2$ are equal to $1$.  For each graph we use their combinatorial Laplacian $\Lm$ as variation operator. The vertex partition is marked by colored vertices, where blue and red correspond to the sets $\Ac$ and $\Bc$, respectively. The Laplacian and degree matrices of $\Gc_i$ are denoted by $\Lm_i$ and $\Dm_i$, respectively. 
In \autoref{fig:path_LD} and \autoref{fig:path2_LQ} we plot the $(\Mm,\Qm)$-GFTs for graphs $\Gc_1$ and $\Gc_2$ that have the spectral folding property. Since  $\Gc_1$ is bipartite, we plot $(\Lm_1,\Dm_1)$-GFT, while for $\Gc_2$ we plot the $(\Lm_2,\Qm_2)$-GFT where $\Qm_2$ is chosen according to \autoref{prop:folding-variation}. 
As predicted by \autoref{prop_eigenvalue_bound}, the generalized eigenvalues lie in $[0,2]$ and $\lambda_{n}=2$ ($n=11$) for both graphs. The generalized eigenvalue/eigenvector pairs $(\lambda, \uv)$ and $(2-\lambda, \Jm \uv)$ can be easily observed, as predicted by \autoref{prop:folding-variation}. Note that also  for both graphs the generalized eigenvalue $\lambda=1$ is simple, thus agreeing with the lower bound $\vert \vert \Ac \vert - \vert \Bc \vert \vert = 1$ from \autoref{prop_middle_lambda}.
In \autoref{fig:path2_LD} we plot the $(\Lm_2, \Dm_2)$-GFT of $\Gc_2$. The graph frequencies belong the interval $[0,2]$, but they are not symmetric around $1$.While both $(\Lm_i,\Dm_i)$-GFTs are the eigenvectors of the random walk Laplacian, their basis functions  are very different, specially at the higher frequencies. 
 Some of the basis functions of $(\Lm_2,\Dm_2)$-GFTs present some undesirable features such as repeated eigenvalues ($\lambda_6=\lambda_7$) and  localized basis functions $\uv_6, \uv_7, \uv_9$.   In this example the  $(\Mm,\Qm)$-GFTs that have the spectral folding produce better bases for signal representation. 
 }
\subsection{Vertex domain interpretation} 
\label{ssec_vertex_Z}
Traditionally,  implementations of SGFs using polynomials of the variation operator $\Mm$ (i.e., the fundamental matrix of the $(\Mm,\Id)$-GFT) have been preferred for their  
efficiency 
(sparse matrix vector products)  and interpretability 
(localized vertex domain operations). 
%
Since $\Qm$ is not diagonal unless the graph is bipartite,   $\Zm=\Qm^{-1}\Mm$  can be much denser than $\Mm$, and thus the product $\Zm\xv$ may  no longer be implemented with localized vertex domain operations.  

In this section we show that $\Qm^{-1}$, and  thus the fundamental matrix $\Zm$,  can be  approximated by polynomials of sparse matrices. In fact,  for certain choices of  vertex partitions, $\Zm$ may be approximately sparse and thus the operation $\Zm \xv$ is localized in the vertex domain. We show that this product can be described in terms of vertex domain operations involving  bipartite and disconnected graphs. The complexity of SGFs as a function of the vertex partition is studied in  \autoref{sec_vertex_partitioning}.

For simplicity we only consider the combinatorial Laplacian, so that  $\Mm = \Lm$. 
%
Given  a vertex partition $\Ac, \Bc$,  any graph can be decomposed as the sum of a bipartite graph and a disconnected graph with $2$ or more connected components (the adjacency matrix of a disconnected graph is block diagonal if vertices in the same connected component are labeled consecutively). An example is depicted in \autoref{fig:bip_block_graph_decomp}.
These graphs obey the following identities
\begin{equation}
	\Lm = \Lbip + \Lblock, \quad \Wm = \Wbip + \Wblock, \quad \Dm = \Dbip + \Dblock,
\end{equation}
where the super indices \emph{bi} and \emph{bd} refer to bipartite and block diagonal, respectively (see  
\autoref{fig:bip_block_graph_decomp}). In addition, we have that
\begin{equation}
	\Lblock = \Dblock - \Wblock,\quad \Lbip = \Dbip - \Wbip.
\end{equation}
%
%
%
%
%
%
%
%
%
We have the following factorization for $\Zm$.
\begin{proposition}\label{prop_bip_block_smoothing}
	If $\Mm = \Lm$  and   $(\Lm,\Qm)$-GFT has the spectral folding property, then  the fundamental matrix   is equal to
	\begin{equation}
		\Zm = \Qm^{-1}\Lm= \Id - \Pblock \Pbip,
	\end{equation}
	where   $\Pbip = (\Dbip)^{-1}\Wbip$, and $\Pblock =\Qm^{-1}\Dbip$, are both right stochastic non negative matrices, with bipartite and block diagonal structure, respectively.
\end{proposition}
The proof can be found in Appendix \ref{app_bip_block_smoothing}.
The matrix $\Pblock \Pbip$ is a two step smoothing operator, since 
\begin{equation}
	y_i = (\Pbip \xv)_i = \left\{ \begin{array}{cc}
		\frac{1}{(\Dbip)_{ii}}\sum_{j \in \Bc}w_{ij}x_j & \mbox{if } i \in \Ac \\
		\frac{1}{(\Dbip)_{ii}}\sum_{j \in \Ac}w_{ij}x_j & \mbox{if } i \in \Bc.
	\end{array} \right., 
\end{equation}
where a smoothed signal is obtained through linear combinations of neighbors on the complement set. 
Since $\Pblock$ is block diagonal and non negative, the second filtering step is also a low pass filter that   only uses connections within $\Ac$ or $\Bc$. 
 \begin{figure}[t]
     \centering
    \begin{subfigure}[b]{0.16\textwidth}
         \centering
		\includegraphics[width=0.9\textwidth]{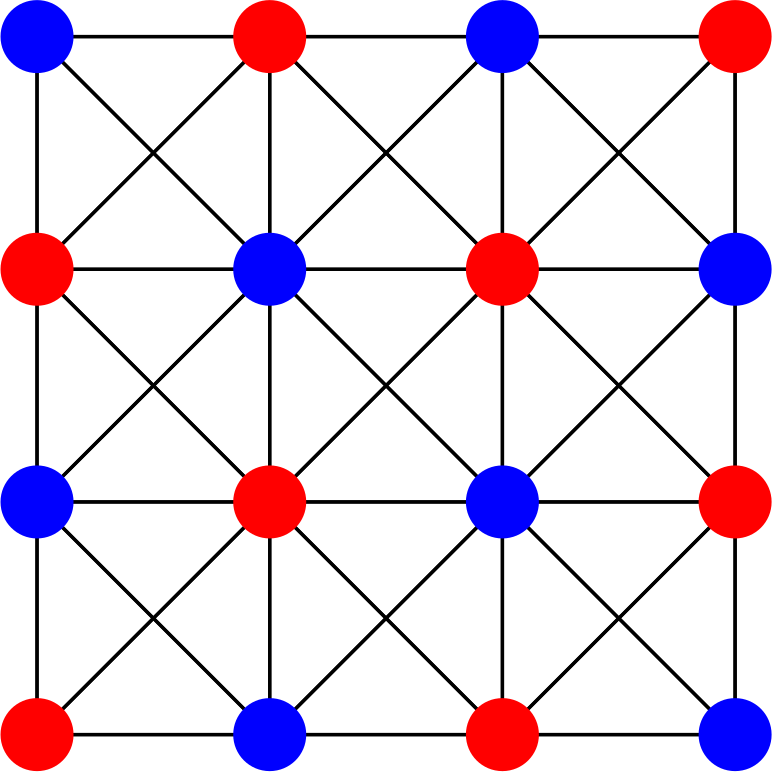}
		\caption{$\Wm = \Wbip + \Wblock$}
\label{fig:image_graph_8con}
	\end{subfigure}%
\begin{subfigure}[b]{0.16\textwidth}
\centering
		\includegraphics[width=0.9\textwidth]{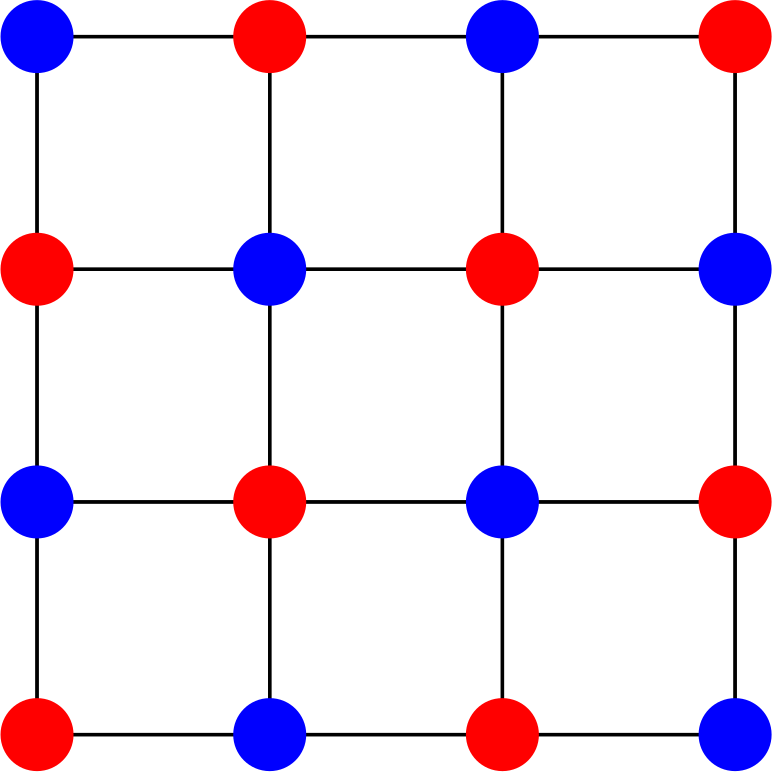}
		\caption{$\Wbip$}
\label{fig:image_graph_4con}
	\end{subfigure}%
\begin{subfigure}[b]{0.16\textwidth}
\centering
		\includegraphics[width=0.9\textwidth]{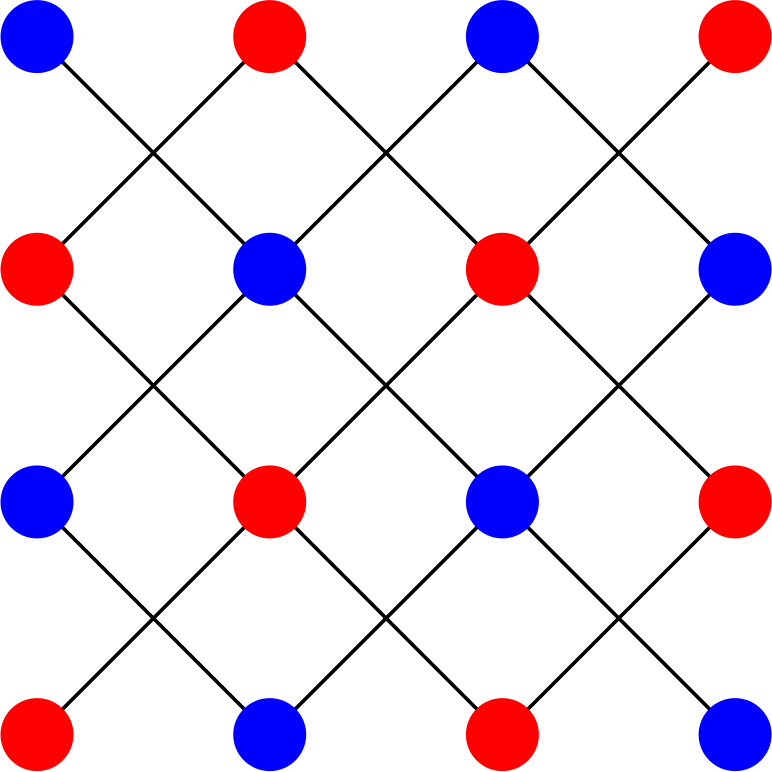}
		\caption{ $\Wblock$}
\label{fig:image_graph_block}
	\end{subfigure}
	\begin{subfigure}[b]{0.23\textwidth}
	\centering
		\includegraphics[width=0.9\textwidth]{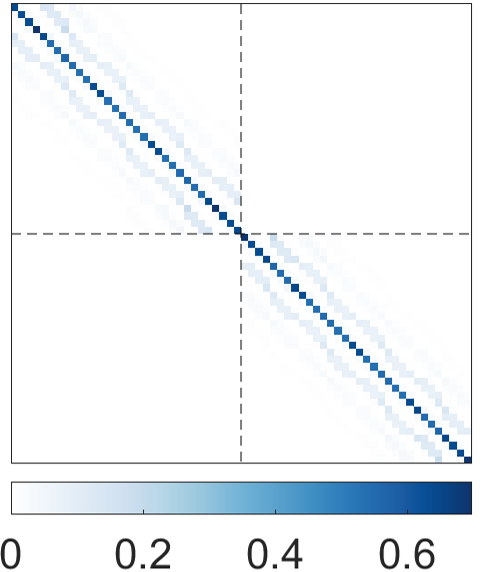}
		\caption{$\Pblock=\Qm^{-1}\Dbip$}
\label{fig:fundamental_Pbd}
	\end{subfigure}%
\begin{subfigure}[b]{0.23\textwidth}
\centering
		\includegraphics[width=0.9\textwidth]{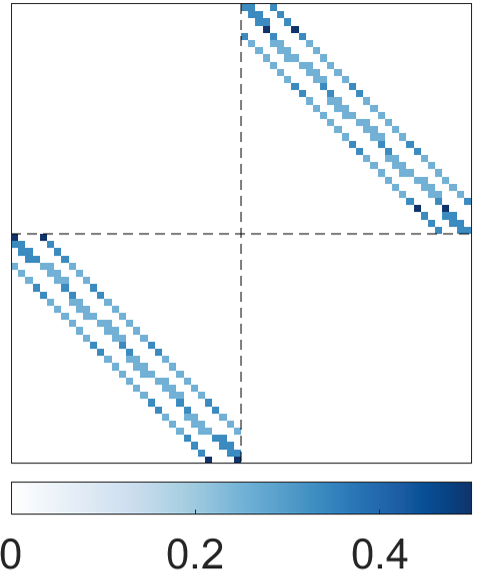}
		\caption{$\Pbip=(\Dbip)^{-1}\Wbip$}
\label{fig:fundamental_Pbi}
	\end{subfigure}  
\begin{subfigure}[b]{0.23\textwidth}
\centering
		\includegraphics[width=0.9\textwidth]{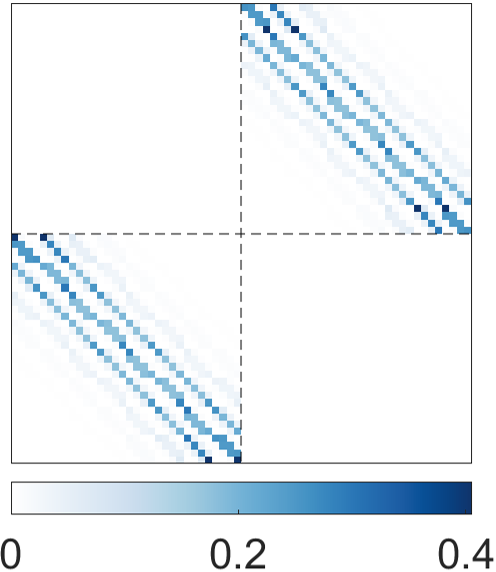}
		\caption{$\Pblock\Pbip$}
\label{fig:fundamental_P}
	\end{subfigure}%
	\begin{subfigure}[b]{0.23\textwidth}
\centering
		\includegraphics[width=0.9\textwidth]{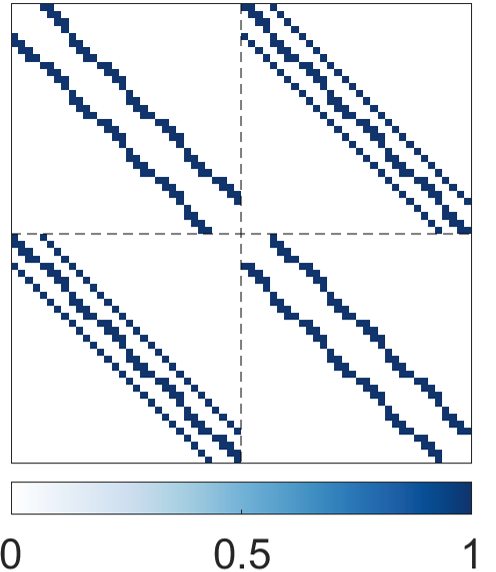}
		\caption{$\Wm$}
\label{fig:fundamental_W}
	\end{subfigure}
     \caption{Decomposition of $8$ connected $4 \times 4$ grid graph (\ref{fig:image_graph_8con}) as sum of bipartite (\ref{fig:image_graph_4con}) and block diagonal graphs (\ref{fig:image_graph_block}). Graph matrices from \autoref{prop_bip_block_smoothing} are shown in Figures \ref{fig:fundamental_Pbd}- \ref{fig:fundamental_W}. }
     \label{fig:image_graphs1}
 \end{figure}
 
As an example we consider the 
$8$-connected grid graph in \autoref{fig:image_graph_8con}, commonly used in image processing \cite{cheung2018graph}. 
The vertex partition (red or blue) can be used to decompose the graph into (i) a bipartite $4$-connected grid (\autoref{fig:image_graph_4con}), containing all vertical and horizontal connections,  and (ii) a block diagonal graph  containing  the diagonal connections (see \autoref{fig:image_graph_block}). The adjacency matrix of the $8$-connected grid is depicted in \autoref{fig:fundamental_W}, where all the edge weights have been set to $1$, and the vertices are labeled so that $\Ac = \lbrace 1, 2, \cdots, \vert \Ac \vert \rbrace$.
 $\Pblock$ and $\Pbip$ are depicted in \autoref{fig:fundamental_Pbd} and \autoref{fig:fundamental_Pbi}, respectively, and their product is depicted in \autoref{fig:fundamental_P}. While  $\Pbip$ and $\Wm$ are sparse, the matrices $\Pblock$ and $\Pblock\Pbip$ are dense. However, \autoref{fig:fundamental_Pbd} and \autoref{fig:fundamental_P} show that most of the entries of $\Pblock$ and $\Pblock\Pbip$ are close to zero. The approximate sparsity of $\Zm$ and $\Pblock\Pbip$ can  be explained by the following result.
 \begin{proposition}\label{prop_neuman_Qinv}
 Let $\Lm$ be the combinatorial Laplacian of a connected graph. If $\Qm$ is chosen according to \autoref{prop:folding-variation}, then
 \begin{equation} \label{eq_neuman_Qinv}
     \Qm^{-1} = \Dm^{-\frac{1}{2}}\left(\sum_{k=0}^{\infty}(\Dm^{-\frac{1}{2}}\Wblock \Dm^{-\frac{1}{2}})^{k}\right)\Dm^{-\frac{1}{2}}.
 \end{equation}
 \end{proposition}
\begin{proof}
Because the graph is connected, $\Qm \succ 0$. Since $\Qm = \Dm - \Wblock$, we have that $\Id - \Dm^{-1/2}\Wblock\Dm^{-1/2} \succ 0$, which implies that $\Vert  \Dm^{-1/2}\Wblock\Dm^{-1/2} \Vert <1$. Then $(\Id - \Dm^{-1/2}\Wblock\Dm^{-1/2})^{-1} = \sum_{k=1}^{\infty}(\Dm^{-1/2}\Wblock\Dm^{-1/2})^k$, and $\Qm^{-1} = \Dm^{-1/2}(\Id - \Dm^{-1/2}\Wblock\Dm^{-1/2})^{-1}\Dm^{-1/2}$. 
\end{proof}
 If we only keep the first $m$ terms in \eqref{eq_neuman_Qinv}, the norm of the remaining terms decays exponentially  as  $\Oc(\Vert  \Dm^{-1/2}\Wblock\Dm^{-1/2} \Vert^{m+1})$. If the vertex partition is designed so that $\Wblock$ is sparse and has small weights, this error decays faster.  \NEW{ While in this work we do not approximate $\Qm^{-1}$, we propose a vertex partitioning algoritm (\autoref{sec_vertex_partitioning}) that  minimizes the $\ell_1$ norm of $\Dm^{-1/2}\Wblock\Dm^{-1/2}$ that   leads to a sparse $\Qm$, and  an approximately sparse fundamental matrix $\Zm$. }
%
\subsection{A probabilistic interpretation of GFBs}
Our analysis is based on the lazy filter bank from \autoref{fig_lazy_2chan}. 
\label{subsec_lazy_2chan}
Assume that $\xv$ is a zero mean Gaussian vector with covariance matrix $\Sigmam$, and select $\Mm =\Sigmam^{-1}$  as the variation operator.   We will show that 
$\Zm$ given by \eqref{eq_fundamental_matrix_folding} is the matrix that produces high pass coefficients $\Zm \xv$ with the least norm, hence  $\Zm$ solves
\begin{equation}\label{eq_optimality_Z}
    \min_{\Cm} \E[\Vert \Cm \xv \Vert^2_{\Id}], \textnormal{ s.t. } \Cm = \begin{bmatrix}
        \Id_{\Ac} & \Cm_{\Ac \Bc} \\
        \Cm_{\Bc \Ac} & \Id_{\Bc}
    \end{bmatrix},
\end{equation}
where $\Ac$ and $\Bc = \Vc\setminus \Ac$ are fixed, and  $\Cm$ is the high pass filter of the lazy two channel  filter bank from \autoref{fig_lazy_2chan}. 
Later we study properties of the tree structured iterated lazy filter bank, and provide an alternative probabilistic perspective on its orthogonality. 
%
We can reduce \eqref{eq_optimality_Z} to the equivalent problem
\begin{align}
    &\min_{\Cm_{\Ac \Bc}, \Cm_{\Bc \Ac}} \E[\Vert \xv_{\Ac} + \Cm_{\Ac \Bc}\xv_{\Bc} \Vert_{\Id}^2] + \E[\Vert \xv_{\Bc} + \Cm_{\Bc \Ac}\xv_{\Ac} \Vert_{\Id}^2]  \\
    &= \min_{\Cm_{\Ac \Bc}} \E[\Vert \xv_{\Ac} + \Cm_{\Ac \Bc}\xv_{\Bc} \Vert_{\Id}^2] + \min_{ \Cm_{\Bc \Ac}} \E[\Vert \xv_{\Bc} + \Cm_{\Bc \Ac}\xv_{\Ac} \Vert_{\Id}^2] \nonumber.
\end{align}
Minimization of  $\E[\Vert \xv_{\Ac} + \Cm_{\Ac \Bc}\xv_{\Bc} \Vert_{\Id}^2]$ corresponds to optimal linear prediction of $\xv_{\Ac}$ from $\xv_{\Bc}$, which fortunately, has a closed form solution for Gaussian distributions.

\begin{proposition}\label{prop_Z_gaussian}
Given a vertex partition $\Ac, \Bc$,  if $\Mm = \Sigmam^{-1}$, then $\Zm=\Qm^{-1}\Mm$ from \eqref{eq_fundamental_matrix_folding} is the minimizer of \eqref{eq_optimality_Z}, and 
\begin{equation}
\Zm \xv
= \xv - \begin{bmatrix}
\E[\xv_{\Ac} | \xv_{\Bc}] \\
\E[\xv_{\Bc} | \xv_{\Ac}]
\end{bmatrix}
= \xv - \begin{bmatrix}
	\Sigmam_{\Ac \Bc} \Sigmam_{\Bc \Bc}^{-1}\xv_{\Bc} \\
	\Sigmam_{\Bc \Ac} \Sigmam_{\Ac \Ac}^{-1}\xv_{\Ac}
\end{bmatrix}.
\end{equation}
\end{proposition}
 The proof of this result is a direct consequence of
\begin{equation}
 \Sigmam_{\Ac \Bc} \Sigmam_{\Bc \Bc}^{-1} = - \Mm_{\Ac \Ac}^{-1}    \Mm_{\Ac \Bc}, 
 \Sigmam_{\Bc \Ac} \Sigmam_{\Ac \Ac}^{-1} = - \Mm_{\Bc \Bc}^{-1}    \Mm_{\Bc \Ac},
\end{equation}
which follows  from  $\Mm \Sigmam = \Id$.
From the  $(\Mm,\Qm)$-GFT perspective, $\Zm$ is a high pass filter (see \autoref{ssec_vertex_Z}).  \autoref{prop_Z_gaussian} illustrates that the filtering operation $\Zm \xv$ is the prediction error of an optimal  linear predictor  for the Gaussian distribution.
%
%
%

Now we take the lazy filter bank from \autoref{fig_lazy_2chan} and form a tree structured filter bank as in \autoref{fig:iterated_fb_1}. Filters at resolution $\ell$ are given by $\Hm_{0,\ell} = \Id$, $\Hm_{1,\ell} = \Zm_{\ell}$, $\Gm_{0,\ell} = 2\Id - \Zm_\ell$, and $\Gm_{1,\ell} = \Id$. We will show that  \autoref{prop_Z_gaussian} implies that this tree structured  filter bank  produces sub-bands with uncorrelated coefficients. 
Assume that the input signal  $\xv = \av_L$ is a zero mean Gaussian with covariance $\Sigmam = \Mm^{-1}$. Since the low pass channel corresponds to down-sampling $\av_{\ell+1}$ on the set $\Ac_\ell$, we have that for  $\ell<L$
\begin{equation}\label{eq_lazy_a_ell}
\av_{\ell} = \xv_{\Ac_\ell}, 
\end{equation}
  is a $\vert \Ac_\ell \vert$-dimensional zero mean Gaussian vector with covariance matrix $\Sigmam_{\Ac_\ell, \Ac_\ell}$. Therefore, at resolution $\ell$ we will use the variation operator $\Mm_\ell =\Sigmam_{\Ac_\ell, \Ac_\ell}^{-1} $, which can be computed using Schur complements  \cite{dorfler2012kron}
\begin{equation}
\Mm_{\ell-1} = (\Mm_{\ell})_{\Ac_\ell \Ac_\ell} - (\Mm_{\ell})_{\Ac_\ell \Bc_\ell}((\Mm_{\ell})_{\Bc_\ell \Bc_\ell})^{-1} (\Mm_{\ell})_{\Bc_\ell \Ac_\ell}.
\end{equation}
 The detail coefficients   are
\begin{equation}\label{eq_lazy_d_ell}
    \dv_{\ell} = \xv_{\Bc_\ell} - {\Sigmam}_{\Bc_\ell \Ac_\ell}\Sigmam^{-1}_{\Ac_\ell \Ac_\ell} \xv_{\Ac_\ell}.
\end{equation}
The coefficient vector obtained after iterating the filter bank $L$ times is given by $\cv = [\av_0^{\top}, \dv_0^{\top}, \cdots, \dv_{L-1}^{\top}]^{\top} = \Tcb_a \xv$.   These coefficients are uncorrelated, more precisely:
\begin{proposition}\label{prop_orth_lazy}
The inverse covariance matrix of $\cv$ is equal to
\begin{equation}
(\E[\cv \cv^{\top}])^{-1} = \begin{bmatrix}
\Mm_{0}  & \mathbf{0} & \cdots & \mathbf{0} \\
\mathbf{0} & \Qm_{1,0} & \ddots & \mathbf{0}\\
\vdots & \ddots &  \ddots &  \vdots \\
\mathbf{0} & \mathbf{0} & \mathbf{0} & \Qm_{1,L-1}
\end{bmatrix},
\end{equation}
where   $\Qm_{1,\ell}$ is the submatrix given by
\begin{equation}\label{eq_sub_QEll}
\Qm_{1,\ell} = \Qm_{\ell+1}(\Bc_{\ell}, \Bc_{\ell}).
\end{equation}
\end{proposition}
The proof  follows from \autoref{prop_Z_gaussian} (see Appendix \ref{app_prop_orth_lazy}).

\section{Vertex partitioning}
\label{sec_vertex_partitioning}
In this section 
we  assume  $\Mm$ is given, and we wish to optimize  $\Ac$ and $\Bc=\Vc \setminus \Ac$, as a function of $\Mm$.
We start by formulating this problem as an optimization of the condition number of $\Qm$. We provide some theoretical justifications for this choice of objective based on complexity and stability of GFBs. We end the section with our proposed solution.
\subsection{Problem formulation}
\NEW{$\Jm$ defined in \eqref{eq_f_setindicator} completely characterizes the vertex partition since $\Jm_{ii} = 1$ if $ i \in \Ac$, and $\Jm_{ii} = -1$ if $i \in \Bc$. Consequently, we will use a vector $\fv$ containing the diagonal terms of $\Jm$ as an optimization variable. 
We propose to minimize  $\kappa(\Qm)= \Vert \Qm \Vert \Vert \Qm^{-1}\Vert$, the condition number of $\Qm$, subject to  spectral folding  and balanced partition constraints,  namely, 
\begin{align}\label{eq_min_condition}
	\min_{\fv: \fv_i^2=1} \kappa(\Qm)\quad\textnormal{ s.t. } &\Qm = \frac{1}{2}(\Mm + \diag(\fv) \Mm\diag(\fv))\\
	&\left\vert \sum_{i \in \Vc} \fv_i \right\vert \leq 1. \nonumber
	\end{align}
The constraint $\fv_i^2 = 1$ ensures that the vector $\fv$ has entries equal to $1$ or $-1$. The downsampling sets are recovered as $\Ac = \lbrace i: \fv_i>0\rbrace$, and $\Bc = \Ac^c$.  To ensure the $(\Mm,\Qm)$-GFT has the  spectral folding property we  incorporate the constraint   $\Qm = \frac{1}{2}(\Mm + \diag(\fv) \Mm\diag(\fv))$.  When there is an even number of nodes,  the constraint $\vert \sum_{i \in \Vc} \fv_i \vert \leq 1$ becomes $ \sum_{i \in \Vc} \fv_i = 0$ so that  $\vert \Ac \vert = \vert \Bc \vert$. When  the number of nodes is odd, the constraint $\vert \sum_{i \in \Vc} \fv_i \vert \leq 1$ becomes $\vert \sum_{i \in \Vc} \fv_i \vert= 1$, so that  $\vert \vert \Ac \vert - \vert \Bc \vert \vert =1$. These constraints guarantee a balanced partition, which is a necessary condition for good frequency selectivity  (see  \autoref{app_properties_Z} for more details).
}
\subsection{Justification for condition number minimization}
\NEW{The reason for seeking vertex partitions so that the  $\Qm$ has a small condition number is due to numerical stability and complexity of GFBs.  }
%
For large graphs, performing eigendecomposition to implement a SGF is infeasible due to high computational complexity, thus   polynomial graph filters are used instead. For BFBs, a polynomial graph filter  of degree $d$ can be implemented by sparse matrix vector products (of the normalized Laplacian) with complexity $\Oc(d\vert \Ec \vert)$  \cite{narang2012perfect,sakiyama2016spectral}.
For arbitrary graphs,   polynomials of $\Zm$ require the computation of products of the form $\Zm \xv$. However, although $\Mm$ may be sparse, the fundamental matrix $\Zm = \Qm^{-1}\Mm$ is not.  Hence,  a naive implementation of the product $\Zm\xv$ has complexity $\Oc(n^2)$, while direct computation of $\Zm = \Qm^{-1}\Mm$ has complexity $\Oc(n^3)$ due to matrix inversion. 
A more efficient approach is to decompose the product $\yv = \Zm \xv$   into: 1) computing a sparse matrix vector product $\wv = \Mm \xv$, 
and 2) solving  a   system of equations $\Qm \yv = \wv$. 
The second step is the computation bottleneck, since it requires solving a  linear system. 

Large scale positive semi-definite linear system are solved by iterative methods such as the Conjugate Gradient (CG), whose complexity per iteration is that of  a sparse matrix vector products with $\Qm$. The condition number of $\Qm$ is directly responsible for: 1)   the convergence rate of CG algorithms (and thus the total number of iterations), and 2) stability to perturbations of the system $\Qm \yv = \wv$.   From a computational perspective, the best case scenario occurs when  $\Qm$ is sparse and has a small condition number. In practice this is attained when $\Qm$ is  diagonal, and the graph is bipartite, as will be shown in the next subsection.

%
\begin{figure*}[ht]
	\centering
	\begin{subfigure}[b]{0.31\textwidth}
		\includegraphics[width=1\textwidth]{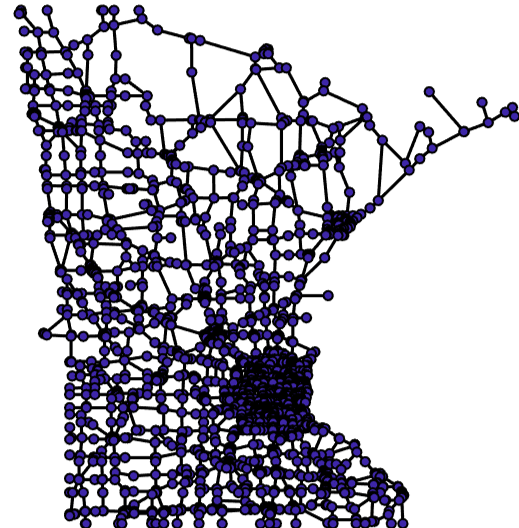}
		\caption{$\Wm$}
		\label{fig:minn}
	\end{subfigure} 
	\begin{subfigure}[b]{0.31\textwidth}
		\includegraphics[width=1\textwidth]{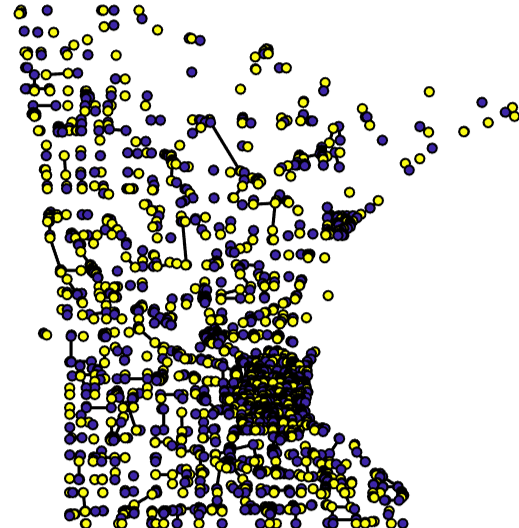}
		\caption{$\Wblock$ with max-cut  partitioning}
				\label{fig:minnQmax-cut}
	\end{subfigure}
	\begin{subfigure}[b]{0.31\textwidth}
		\includegraphics[width=1\textwidth]{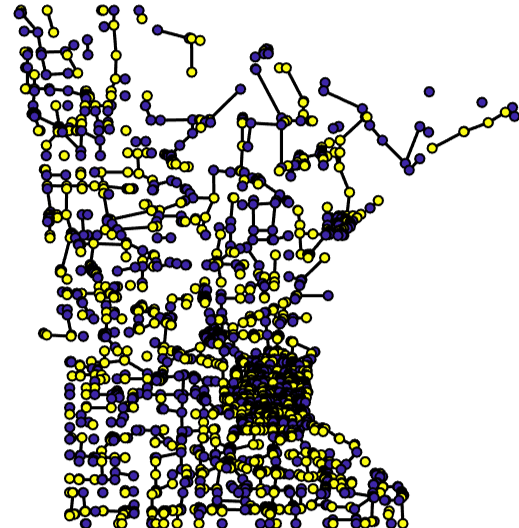}
		\caption{$\Wblock$ with random  partitioning}
				\label{fig:minnQrand}
	\end{subfigure}
	\caption{Comparison of matrices  $\Wblock = \frac{1}{2}(\Wm + \diag(\fv)\Wm\diag(\fv))$  using different vertex partitioning algorithms for the Minnesota graph \cite{gleich2015matlabbgl}. Vertex partitions are marked by blue/yellow circles. Figures were generated using the  toolbox \cite{girault2017grasp}.}
	\label{fig:minnesota}
\end{figure*}
\subsection{Approximate solution for generalized Laplacians}
Because of the non convex constraint  $\fv_i^2=1$,  \eqref{eq_min_condition} is a   non convex minimization problem. In this section we obtain an upper bound for $\kappa(\Qm)$, which allows us to pose an alternative, more tractable optimization problem. 
For the rest of this section we  assume that $\Mm$ is a positive semi definite generalized graph Laplacian, that is,  $\Mm = \Vm - \Wm$, where $\Vm$ is a diagonal matrix with positive entries and $\Wm$ is a non negative matrix with zero diagonal.
Since $\Qm$ is given by \autoref{prop:folding-variation}, we have that  $\Qm = \Vm - \Wblock \succ 0$,  where $\Wblock = (1/2)(\Wm + \diag(\fv)\Wm\diag(\fv))$  is the block diagonal part of $\Wm$. We can   bound  $\kappa(\Qm)$ with
\begin{equation}\label{eq_bound_kappa1}
	\kappa(\Qm) \leq \kappa(\Vm) ({1+\rho(\Ac)})/({1-\rho(\Ac)}),
\end{equation}
where $\kappa(\Vm)$ is the condition number of $\Vm$, and $\rho(\Ac) = \Vert \Vm^{-1/2}\Wblock \Vm^{-1/2} \Vert$. A derivation of this bound is given in Appendix \ref{app_bound_kappa1}.
 When the graph is bipartite, $\fv$ can be chosen so that $\Qm$ is diagonal resulting in $\rho(\Ac) = 0$, making  the bound  tight. 
Since the right side of \eqref{eq_bound_kappa1} is decreasing with $\rho(\Ac)$, we can minimize  $\rho(\Ac)$ as a function of the vertex partition. 
As a simplification,  we use the  sequence of bounds for the operator norm of a matrix, $\Vert\Am \Vert \leq \Vert \Am \Vert_F \leq \Vert \Am \Vert_1 = \sum_{i,j}\vert a_{ij} \vert$, which results in $\rho(\Ac) \leq \sum_{i,j}(1+\fv_i \fv_j) w_{ij}/\sqrt{v_i v_j}$. We propose solving instead:  
\begin{equation}\label{eq_min_L1_norm}
\min_{\fv: \fv_i^2=1}  \sum_{i,j}(1+\fv_i \fv_j)\tilde{w}_{ij}   \textnormal{ s.t. }   \left\vert \sum_{i \in \Vc} \fv_i \right\vert \leq 1
\end{equation}
where $\tilde{w}_{ij}$ is the $ij$ entry of  $\tilde{\Wm} = \Vm^{-1/2}\Wm\Vm^{-1/2} $.
Using  the identity $\sum_{i,j}\fv_i \fv_j\tilde{w}_{ij} = \fv^{\top} \tilde{\Wm} \fv = \sum_{i,j}\tilde{w}_{ij} - \fv^{\top}\tilde{\Lm}\fv$, we have that 
  \eqref{eq_min_L1_norm} is equivalent to 
\begin{equation}\label{eq_max-cut_f}
\max_{\fv: \fv_i^2=1} \fv^{\top} \tilde{\Lm} \fv  \text{ s.t. }   \left\vert \sum_{i \in \Vc} \fv_i \right\vert \leq 1,
\end{equation}
where  $\tilde{\Dm} = \diag(\tilde{\Wm} \mathbf{1})$,  and $\tilde{\Lm} = \tilde{\Dm} - \tilde{\Wm}$.  \eqref{eq_max-cut_f} is an instance of weighted maximum cut (WMC) \cite{goemans1995improved}, with an additional balanced partition constraint.
It is well known that WMC is NP hard, thus we consider a  spectral partitioning approximation \cite{shuman2020localized,aspvall1984graph}, that computes $\tilde{\uv}_n = \argmax_{\Vert \uv \Vert=1} \uv^{\top}\tilde{\Lm}\uv$, and sets  $\fv = \sign(\tilde{\uv}_n - \tau \mathbf{1})$, and $\Ac = \lbrace i \in \Vc: \fv_i =1 \rbrace$. 
The parameter $\tau \in \R$ can be tuned to ensure $\vert \sum_{i \in \Vc} \fv_i \vert \leq 1$.
\subsection{Minnesota graph example}
\begin{figure}[ht]
    \centering
    \includegraphics[width=0.35\textwidth]{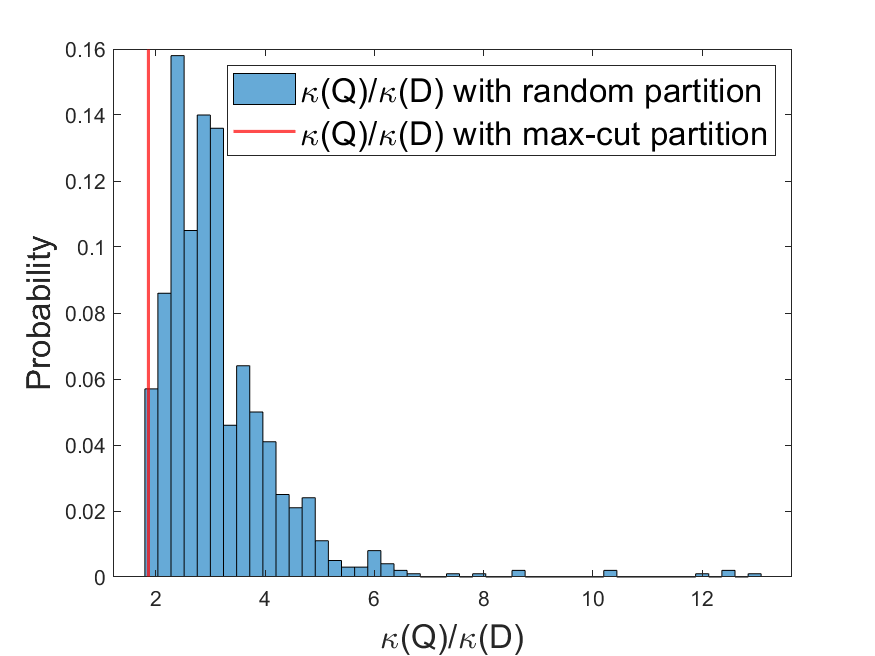}
    \caption{Distribution of the normalized condition number of $\Qm$ when using random vertex partitioning. }
    \label{fig:minnesota_histo}
\end{figure}

\NEW{ We implement the proposed max-cut  vertex partitioning algorithm and a random partitioning algorithm that assigns nodes to $\Ac$ with probability $1/2$. We consider the Minnesota road graph \cite{gleich2015matlabbgl}, which has $n=2640$ nodes and $s=3302$ edges with unit  weights. Its adjacency matrix $\Wm$ is depicted in \autoref{fig:minn}. We use the combinatorial Laplacian $\Lm$ as variation operator. 
In \autoref{fig:minnesota} we display the sparsity patterns of  $\Wblock = \Dm - \Qm =  (1/2)(\Wm + \diag(\fv)\Wm\diag(\fv))$, where $\fv$ has been obtained via the proposed max-cut sampling (\autoref{fig:minnQmax-cut}) and via random partitioning (\autoref{fig:minnQrand}).  Random vertex partition produces a less sparse matrix $\Qm$, which retains  $49.6\%$ of the edges from $\Wm$. In contrast, the proposed max-cut vertex partitioning, produces a matrix $\Qm$ which only has $14.56\%$ of the edges of $\Wm$, and is much closer to $\Dm$. 
We also compare the condition numbers of $\Qm$ as a function of the vertex partition. We generate $1000$ realizations of random partitions, and display the distribution of $\kappa(\Qm)/\kappa(\Dm)$ in  \autoref{fig:minnesota_histo}. The proposed max-cut partition achieves $\kappa(\Qm)/\kappa(\Dm) = 1.863$, and  our simulation shows that  the normalized condition number is always greater than  $1.863$ for the the random partitions. \textcolor{black}{ Because the Minnesota graph is very sparse ($s=3302 \approx 1.25 n$), the sub-graph associated to $\Pblock$  (and $\Qm$)  have many more than $2$ connected components, resulting in both $\Zm$ and $\Qm^{-1}$ also being exactly sparse. When using max-cut partitioning $\Zm$  has $1.2806 s$ non-zero off diagonal entries, but  when  random partitioning is used the number of non-zero off diagonal entries in $\Zm$ ranges from $2.7733 s$ to $4.4170 s$, with average $3.3559s$.}
}
In the next section, we show that this algorithm can be efficiently implemented  for large graphs (since it only requires computing an eigenvector), while  improving the signal representation. 
\begin{figure}[t]
	\centering
	\begin{subfigure}{.2\textwidth}
		\centering
		\includegraphics[width=0.9\textwidth]{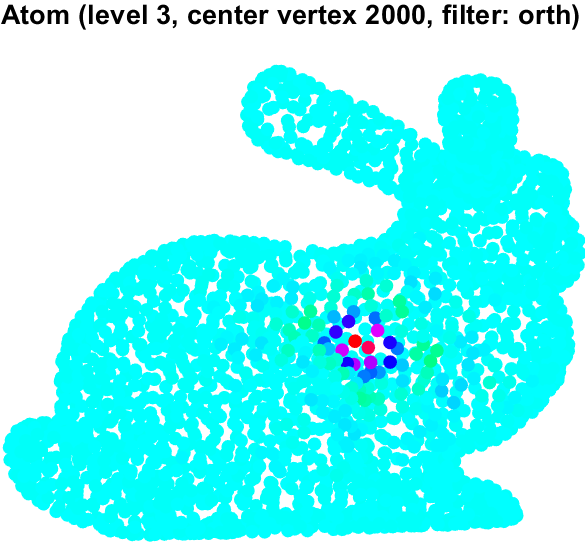}
		\caption{GFB Meyer L=3}
		\label{fig:bunny_gfb_orth_L3}
	\end{subfigure}
	\begin{subfigure}{.2\textwidth}
		\centering
		\includegraphics[width=0.9\textwidth]{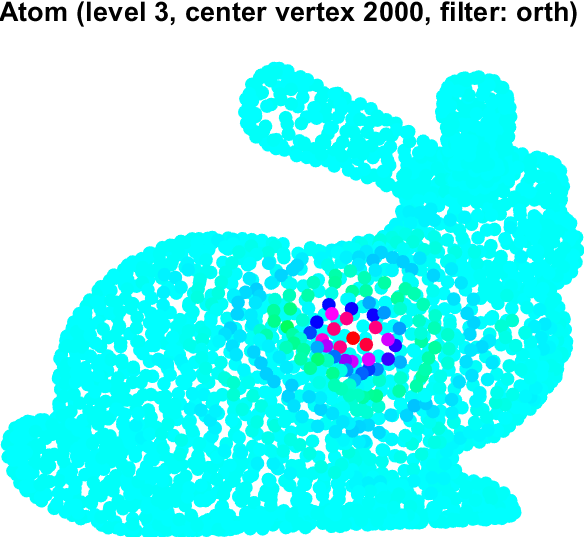}
		\caption{SDS Meyer L=3}
		\label{fig:bunny_spec_orth_L3}
	\end{subfigure}
	\begin{subfigure}{.2\textwidth}
		\centering
		\includegraphics[width=0.9\textwidth]{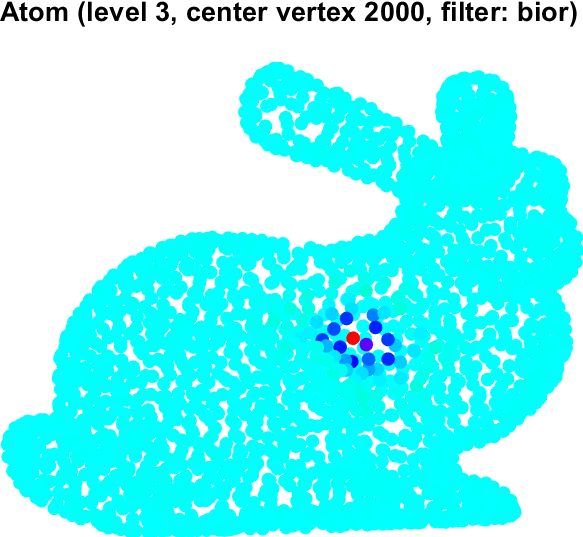}
		\caption{GFB CDF-9/7 L=3}
		\label{fig:bunny_gfb_bior_L3}
	\end{subfigure}
	\begin{subfigure}{.2\textwidth}
		\centering
		\includegraphics[width=0.9\textwidth]{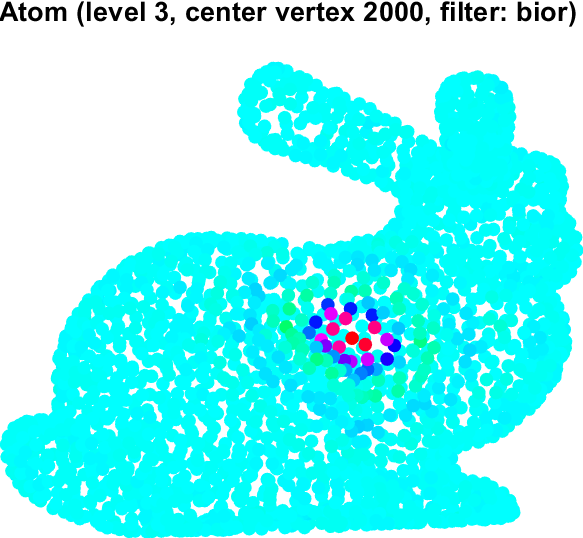}
		\caption{SDS CDF-9/7 L=3}
		\label{fig:bunny_spec_bior_L3}
	\end{subfigure}
	\caption{Low frequency atoms  on Bunny point cloud centered at node $2000$. (left) Proposed  generalized filter banks (GFB), and (right) filter banks based on spectral domain sampling (SDS). }
	\label{fig:bunny_atom}
\end{figure}
\begin{figure}[t]
	\centering
	\begin{subfigure}[b]{0.24\textwidth}
		\includegraphics[width=1\textwidth]{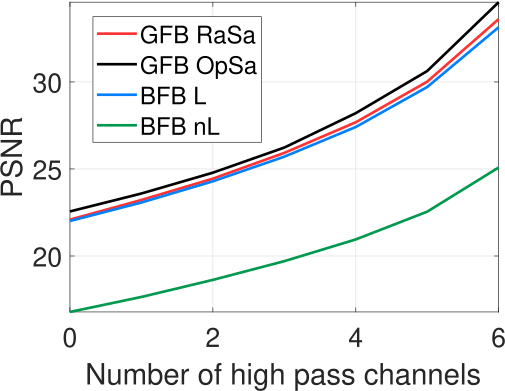}
		\caption{longdress}
	\end{subfigure}
	\begin{subfigure}[b]{0.24\textwidth}
		\includegraphics[width=1\textwidth]{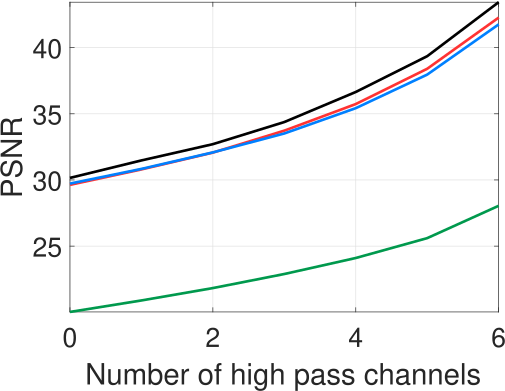}
		\caption{loot}
	\end{subfigure}%
	\caption{3D point cloud attribute approximation.   GFBs  with random partitioning (GFB RaSa) and  max-cut partitioning from \autoref{sec_vertex_partitioning} (GFB OpSa). BFBs using random partitioning with the combinatorial Laplacian (BFB L), and the normalized Laplacian (BFB nL).  }
	\label{fig:approximation_8i}
\end{figure}
\section{Experiments on 3D point clouds}
 \label{sec_exp}
In this section we present  numerical results showing that: 1) GFB can be implemented efficiently on graphs with hundreds of thousands of nodes, 2) the basis functions of GFBs are localized in the vertex domain and 3) for point clouds, GFBs with optimized vertex partitions can provide better signal representations than BFBs.
%
\subsection{Vertex domain localization}
We start by visualizing the vertex domain behavior of the GFB basis functions, i.e.,  the columns of the synthesis operator $\Tcb_s$ of a tree structured filter bank    (see \autoref{ssec_treeGFB} and \autoref{fig:iterated_fb_1}).  We compare them  to those of  spectral domain sampling (SDS) filter banks \cite{sakiyama2019two}, which are  close to the proposed GFB in terms of graph properties  (see \autoref{tab_relatedwork}). We show that although   $\Zm$ is (numerically)  dense, for point clouds,  $\Zm$ is approximately sparse and   polynomial filters of $\Zm$ are localized in the vertex domain. 
%
%

 Since SDS filter banks require full eigendecomposition, for complexity reasons  we use a small \emph{Bunny} point cloud, which has $n = 2503$ nodes. For both filter banks we fix the  number of levels to $L=3$ and consider two designs: biorthogonal CDF-9/7 filters, and orthogonal Meyer filters.  Note that these filters are exactly those developed for traditional 1D wavelets.
For graph construction, we use  nonnegative kernel regression (NNK)  \cite{shekkizhar2020graph}  initialized with k-nearest-neighbors (KNN) using  $K=20$  and inverse distance as edge weights. For GFBs, we use the spectral partitioning algorithm from \autoref{sec_vertex_partitioning}.
\autoref{fig:bunny_atom} displays low frequency basis functions, centered at a single node. For visualization purposes, we normalize the    basis functions, so  their  entries have magnitude at most $1$. For both filter designs, the basis functions of the GFBs   are  more localized than their SDS   counterparts. The  GFB with CDF-9/7 filters is implemented using polynomials of $\Zm$, thus producing the most localized basis functions. 
\subsection{Representation of  3D point cloud attributes}
3D point clouds  consist of list of point coordinates $\Vm = [\vv_i] \in \R^{n \times 3}$, and color  attributes $\Am \in \R^{n\times 3}$.  A graph is constructed where each point is assigned to a node and the edge weight between nodes $i$ and $j$ is $w_{ij} = 1/\Vert \vv_i - \vv_j\Vert$. A sparse edge set is obtained using the KNN and NNK \cite{shekkizhar2020graph} graph construction algorithms.
We consider the first frame of the \emph{loot} and \emph{longdress} sequences of the 8iVFBv2 dataset \cite{d20178i}, which have $n=784,142$ and $n=765,821$ points respectively.  We implement the tree structured GFB of \autoref{fig:iterated_fb_1} with $L=7$ levels,   using biorthogonal   analysis filters $h_0(\lambda) = \frac{1}{2a_0}(2-\lambda)(1+\lambda)$, and   $h_1(\lambda) = a_0 \lambda$, with  gain $a_0 = 0.735$, and synthesis filters computed with \eqref{eq_bior}.

We apply the GFB to each column of the color matrix $\Am$ independently, and obtain the approximation by only keeping the low-pass coefficients $\av_0$  and a subset of the  
high-pass coefficients    $\dv_i$ for $0 \leq i \leq m-1$, with $m \in {1,\cdots, L}$, and zeroing out the rest. \NEW{
For each filter bank, we tested various graph constructions and chose the one giving the best performance. We tested KNN graphs with $K \in \lbrace 5, 10, 15, 20 \rbrace$ and NNK graphs initialized with KNN using $K=20$. To construct bipartite graphs, we take a KNN or NNK graph, a random vertex partition $\Ac,\Bc$, and use the subgraph  consisting of only the edges from $\Ac$ to $\Bc$. }
For GFBs, the best results (higher PSNR) are obtained with NNK graphs, while for BFB the best results are obtained with KNN graphs.  \autoref{fig:approximation_8i} shows the PSNR between the color signal and its approximation as a function of the number of high pass channels ($m$).
 As expected, the BFB with the normalized Laplacian has the worst performance, which  can be attributed to the non constant DC signal \cite{narang2013compact,tzamarias2021graphBior}. By using the BFB with the $(\Lm,\Dm)$-GFT, which results in the random walk Laplacian, we recover the zeroDC filter bank from \cite{narang2013compact}, which has a much improved performance, consistent with previous studies \cite{tzamarias2021graphBior,narang2013compact}. The proposed GFB with random sampling always outperforms the best BFB, although by a small margin. When using optimal sampling with GFBs, performance is again improved consistently across datasets\footnote{\NEW{The superiority of GFBs is consistent for other frames, and other sequences of the 8i dataset. Those results are omitted due to space.} }. 
\begin{figure}[t]
    \centering
    \includegraphics[width=0.35\textwidth]{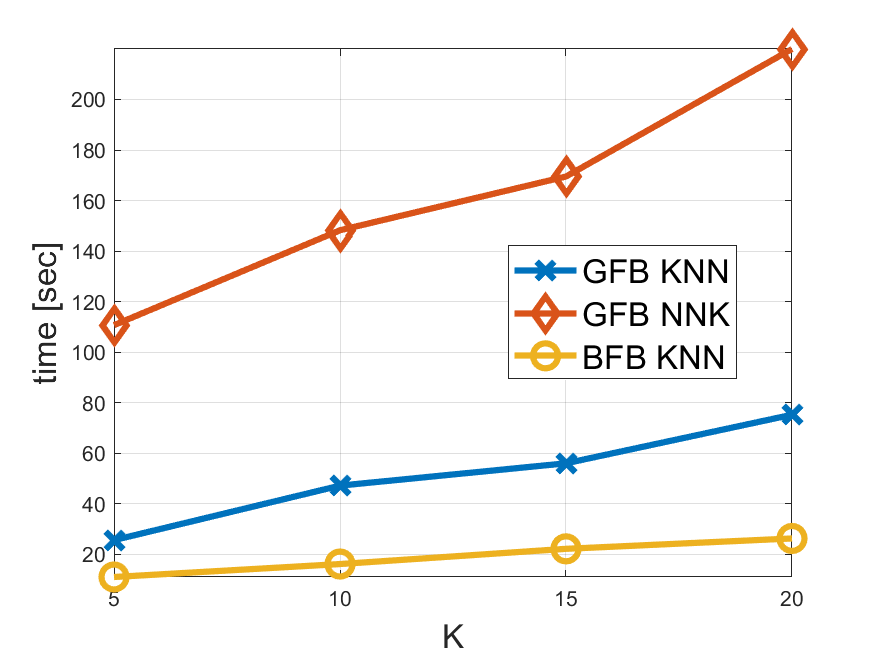}
    \caption{Run time of BFBs implemented with random vertex partitioning and KNN graphs, and  GFBs implemented with max-cut vertex partitioning on KNN and NNK graphs. The first $20$ frames of the ''longdress`` sequence of 3D point clouds are considered.}
    \label{fig:complexity}
\end{figure}
\subsection{Complexity}

We compute the run time of the tree structured analysis and synthesis filter banks with $L = 7$ levels using the biorthogonal filters  from the previous section,  applied to the first 20 frames of the “longdress” sequence. 
These point clouds have an average of approximately $795,000$ points per frame. We compare BFBs constructed with KNN graphs and random balanced partitioning (BFF KNN), GFB constructed with KNN graphs and max-cut based vertex partitioning (GFB KNN), and GFB constructed with NNK graphs and max-cut based vertex partitioning (GFB NNK). NNK graphs are initialized with KNN graphs with varying parameters $K$. 
The complexity of our implementation is dominated by graph construction (KNN and NNK), vertex partitioning, and solving the sparse linear system of the form $\Qm \zv = \uv$. The complexity of each of these components is proportional to the graph sparsity, and the parameter $K$ in KNN. In \autoref{fig:complexity} we plot the average run time over $20$ trials as a function of $K$. Note that   $K$ serves as a proxy for graph sparsity since the number of edges in a KNN and NNK graphs is $\Oc(n K)$.  The figure shows that our implementations of GFB  has a run time higher than their BFB counterparts, but still  scales approximately linearly with $K$.
%
\section{Conclusion}
\label{sec_conc}
This paper proposed two-channel filter banks   on arbitrary graphs with positive semi definite variation operators $\Mm$ and arbitrary vertex partitions $\Ac, \Bc$ for downsampling. Because of the spectral folding property, previous designs were only valid for the normalized Laplacian of bipartite graphs. Our main contribution is showing that the spectral folding property is satisfied by the generalized eigenvalues and eigenvectors of arbitrary graphs, if the inner product matrix is properly chosen. Based on this, we proposed generalized filter banks (GFB) implemented with spectral graph filters of the generalized eigenvectors. \NEW{We also studied other theoretical aspects of  GFBs including: properties of the generalized eigenvalues, vertex domain localization and  probabilistic interpretations. We showed that even though GFBs can use arbitrary vertex partitions for downsampling, partitions that have a large cut (the sum of edge weights from $\Ac$ to $\Bc$ is large) lead to computationally efficient and stable implementations. Our numerical results show that indeed, GFBs have localized basis functions and they can be efficiently implemented on large graphs with hundreds of thousands of nodes,  while   outperforming bipartite filter banks on a signal representation tasks. }
 \NEW{Some important directions for future work include:}
 
\NEW{
     \emph{Tree structured filter banks:} We constructed tree structured GFBs by concatenating two channel GFBs. There is ample room for study of properties of these tree structured filter banks from both probabilistic perspectives (beyond the lazy filter bank case),  and the design of  variation and downsampling operators. In the supplementary material we obtain frame bounds that reveal some of these multi resolution dependencies. 
    }
    
    \NEW{
     \emph{Vertex partitioning:}  Better vertex partitions could be obtained using different approximations to  \eqref{eq_min_condition} or other objective functions, e.g., by using probabilistic models as in \autoref{sec_properties_examples}, or  techniques from the graph signal sampling literature \cite{tanaka2020sampling}. 
}
%
 %
\bibliographystyle{IEEEtran}

\bibliography{refs}

\begin{thebibliography}{10}
\providecommand{\url}[1]{#1}
\csname url@samestyle\endcsname
\providecommand{\newblock}{\relax}
\providecommand{\bibinfo}[2]{#2}
\providecommand{\BIBentrySTDinterwordspacing}{\spaceskip=0pt\relax}
\providecommand{\BIBentryALTinterwordstretchfactor}{4}
\providecommand{\BIBentryALTinterwordspacing}{\spaceskip=\fontdimen2\font plus
\BIBentryALTinterwordstretchfactor\fontdimen3\font minus
  \fontdimen4\font\relax}
\providecommand{\BIBforeignlanguage}[2]{{%
\expandafter\ifx\csname l@#1\endcsname\relax
\typeout{** WARNING: IEEEtran.bst: No hyphenation pattern has been}%
\typeout{** loaded for the language `#1'. Using the pattern for}%
\typeout{** the default language instead.}%
\else
\language=\csname l@#1\endcsname
\fi
#2}}
\providecommand{\BIBdecl}{\relax}
\BIBdecl

\bibitem{ortega2018graph}
A.~Ortega, P.~Frossard, J.~Kova{\v{c}}evi{\'c}, J.~M. Moura, and
  P.~Vandergheynst, ``Graph signal processing: Overview, challenges, and
  applications,'' \emph{Proceedings of the IEEE}, vol. 106, no.~5, pp.
  808--828, 2018.

\bibitem{shuman2013emerging}
D.~I. Shuman, S.~K. Narang, P.~Frossard, A.~Ortega, and P.~Vandergheynst, ``The
  emerging field of signal processing on graphs: Extending high-dimensional
  data analysis to networks and other irregular domains,'' \emph{IEEE signal
  processing magazine}, vol.~30, no.~3, pp. 83--98, 2013.

\bibitem{sandryhaila2013discrete}
A.~Sandryhaila and J.~M. Moura, ``Discrete signal processing on graphs,''
  \emph{IEEE transactions on signal processing}, vol.~61, no.~7, pp.
  1644--1656, 2013.

\bibitem{ortega2021book}
A.~Ortega, \emph{Introduction to Graph Signal Processing}.\hskip 1em plus 0.5em
  minus 0.4em\relax Cambridge University Press, 2022.

\bibitem{vetterli1995wavelets}
M.~Vetterli and J.~Kovacevic, \emph{Wavelets and subband coding}.\hskip 1em
  plus 0.5em minus 0.4em\relax Prentice-hall, 1995.

\bibitem{daubechies1992ten}
I.~Daubechies, \emph{Ten lectures on wavelets}.\hskip 1em plus 0.5em minus
  0.4em\relax Siam, 1992, vol.~61.

\bibitem{vaidyanathan2006multirate}
P.~P. Vaidyanathan, \emph{Multirate systems and filter banks}.\hskip 1em plus
  0.5em minus 0.4em\relax Pearson Education India, 2006.

\bibitem{mallat1999wavelet}
S.~Mallat, \emph{A wavelet tour of signal processing}.\hskip 1em plus 0.5em
  minus 0.4em\relax Elsevier, 1999.

\bibitem{shuman2020localized}
D.~I. Shuman, ``Localized spectral graph filter frames: A unifying framework,
  survey of design considerations, and numerical comparison,'' \emph{IEEE
  Signal Processing Magazine}, vol.~37, no.~6, pp. 43--63, 2020.

\bibitem{narang2012perfect}
S.~K. Narang and A.~Ortega, ``Perfect reconstruction two-channel wavelet filter
  banks for graph structured data,'' \emph{IEEE Transactions on Signal
  Processing}, vol.~60, no.~6, pp. 2786--2799, 2012.

\bibitem{narang2013compact}
------, ``Compact support biorthogonal wavelet filterbanks for arbitrary
  undirected graphs,'' \emph{IEEE transactions on signal processing}, vol.~61,
  no.~19, pp. 4673--4685, 2013.

\bibitem{sakiyama2016spectral}
A.~Sakiyama, K.~Watanabe, and Y.~Tanaka, ``Spectral graph wavelets and filter
  banks with low approximation error,'' \emph{IEEE Transactions on Signal and
  Information Processing over Networks}, vol.~2, no.~3, pp. 230--245, 2016.

\bibitem{tay2015techniques}
D.~B. Tay and J.~Zhang, ``Techniques for constructing biorthogonal bipartite
  graph filter banks,'' \emph{IEEE Transactions on Signal Processing}, vol.~63,
  no.~21, pp. 5772--5783, 2015.

\bibitem{nguyen2014compression}
H.~Q. Nguyen, P.~A. Chou, and Y.~Chen, ``Compression of human body sequences
  using graph wavelet filter banks,'' in \emph{2014 IEEE International
  Conference on Acoustics, Speech and Signal Processing (ICASSP)}.\hskip 1em
  plus 0.5em minus 0.4em\relax IEEE, 2014, pp. 6152--6156.

\bibitem{anis2016compression}
A.~Anis, P.~A. Chou, and A.~Ortega, ``Compression of dynamic 3d point clouds
  using subdivisional meshes and graph wavelet transforms,'' in \emph{2016 IEEE
  International Conference on Acoustics, Speech and Signal Processing
  (ICASSP)}.\hskip 1em plus 0.5em minus 0.4em\relax IEEE, 2016, pp. 6360--6364.

\bibitem{zeng2017hyperspectral}
J.~Zeng, G.~Cheung, Y.-H. Chao, I.~Blanes, J.~Serra-Sagrist{\`a}, and
  A.~Ortega, ``Hyperspectral image coding using graph wavelets,'' in \emph{2017
  IEEE Intl. Conf. on Image Proc. (ICIP)}.\hskip 1em plus 0.5em minus
  0.4em\relax IEEE, 2017, pp. 1672--1676.

\bibitem{tzamarias2019compression}
D.~E.~O. Tzamarias, K.~Chow, I.~Blanes, and J.~Serra-Sagrist{\`a},
  ``Compression of hyperspectral scenes through integer-to-integer spectral
  graph transforms,'' \emph{Remote Sensing}, vol.~11, no.~19, p. 2290, 2019.

\bibitem{iizuka2014depth}
Y.~Iizuka and Y.~Tanaka, ``Depth map denoising using collaborative graph
  wavelet shrinkage on connected image patches,'' in \emph{2014 IEEE Intl.
  Conf. on Image Proc. (ICIP)}.\hskip 1em plus 0.5em minus 0.4em\relax IEEE,
  2014, pp. 828--832.

\bibitem{levorato2012reduced}
M.~Levorato, S.~Narang, U.~Mitra, and A.~Ortega, ``Reduced dimension policy
  iteration for wireless network control via multiscale analysis,'' in
  \emph{2012 IEEE Global Communications Conference (GLOBECOM)}.\hskip 1em plus
  0.5em minus 0.4em\relax IEEE, 2012, pp. 3886--3892.

\bibitem{sharma2018efficient}
V.~K. Sharma, D.~K. Srivastava, and P.~Mathur, ``Efficient image steganography
  using graph signal processing,'' \emph{IET Image Processing}, vol.~12, no.~6,
  pp. 1065--1071, 2018.

\bibitem{qiao2019target}
Y.-l. Qiao, Y.~Zhao, and X.-y. Men, ``Target recognition in sar images via
  graph wavelet transform and {2DPCA},'' in \emph{Proceedings of the 2nd Intl.
  Conf. on Image and Graphics Processing}, 2019, pp. 3--7.

\bibitem{tay2017bipartite}
D.~B. Tay and A.~Ortega, ``Bipartite graph filter banks: Polyphase analysis and
  generalization,'' \emph{IEEE Transactions on Signal Processing}, vol.~65,
  no.~18, pp. 4833--4846, 2017.

\bibitem{narang2010local}
S.~K. Narang and A.~Ortega, ``Local two-channel critically sampled filter-banks
  on graphs,'' in \emph{2010 IEEE Intl. Conf. on Image Proc.}\hskip 1em plus
  0.5em minus 0.4em\relax IEEE, 2010, pp. 333--336.

\bibitem{zeng2017bipartite}
J.~Zeng, G.~Cheung, and A.~Ortega, ``Bipartite approximation for graph wavelet
  signal decomposition,'' \emph{IEEE Transactions on Signal Processing},
  vol.~65, no.~20, pp. 5466--5480, 2017.

\bibitem{jiang2019admm}
A.~Jiang, J.~Wan, Y.~Tang, B.~Ni, and Y.~Zhu, ``Admm-based bipartite graph
  approximation,'' in \emph{ICASSP 2019-2019 IEEE International Conference on
  Acoustics, Speech and Signal Processing (ICASSP)}.\hskip 1em plus 0.5em minus
  0.4em\relax IEEE, 2019, pp. 5421--5425.

\bibitem{pavez2018learning}
E.~Pavez, H.~E. Egilmez, and A.~Ortega, ``Learning graphs with monotone
  topology properties and multiple connected components,'' \emph{IEEE
  Transactions on Signal Processing}, vol.~66, no.~9, pp. 2399--2413, 2018.

\bibitem{kumar2020unified}
S.~Kumar, J.~Ying, J.~d.~M. Cardoso, and D.~P. Palomar, ``A unified framework
  for structured graph learning via spectral constraints,'' \emph{Journal of
  Machine Learning Research}, vol.~21, no.~22, pp. 1--60, 2020.

\bibitem{tzamarias2021graphBior}
D.~E. Tzamarias, E.~Pavez, B.~Girault, A.~Ortega, I.~Blanes, and
  J.~Serra-Sagrist{\`a}, ``Orthogonality and zero dc tradeoffs in biorthogonal
  graph filterbanks,'' in \emph{ICASSP 2021-2021 IEEE International Conference
  on Acoustics, Speech and Signal Processing (ICASSP)}.\hskip 1em plus 0.5em
  minus 0.4em\relax IEEE, 2021, pp. 5509--5513.

\bibitem{girault2018irregularity}
B.~Girault, A.~Ortega, and S.~S. Narayanan, ``Irregularity-aware graph fourier
  transforms,'' \emph{IEEE Transactions on Signal Processing}, vol.~66, no.~21,
  pp. 5746--5761, 2018.

\bibitem{Taubman2002JPEG2000I}
D.~S. Taubman and M.~W. Marcellin, \emph{JPEG2000 - image compression
  fundamentals, standards and practice}.\hskip 1em plus 0.5em minus 0.4em\relax
  Kluwer, 2002.

\bibitem{lu2019fast}
K.-S. Lu and A.~Ortega, ``Fast graph fourier transforms based on graph symmetry
  and bipartition,'' \emph{IEEE Transactions on Signal Processing}, vol.~67,
  no.~18, pp. 4855--4869, 2019.

\bibitem{d20178i}
E.~d’Eon, B.~Harrison, T.~Myers, and P.~A. Chou, ``8i voxelized full bodies-a
  voxelized point cloud dataset,'' \emph{ISO/IEC JTC1/SC29 Joint WG11/WG1
  (MPEG/JPEG) input document WG11M40059/WG1M74006}, 2017.

\bibitem{spielman2011graph}
D.~A. Spielman and N.~Srivastava, ``Graph sparsification by effective
  resistances,'' \emph{SIAM Journal on Computing}, vol.~40, no.~6, pp.
  1913--1926, 2011.

\bibitem{hein2007graph}
M.~Hein, J.-Y. Audibert, and U.~v. Luxburg, ``Graph laplacians and their
  convergence on random neighborhood graphs,'' \emph{Journal of Machine
  Learning Research}, vol.~8, no. Jun, pp. 1325--1368, 2007.

\bibitem{chou2019volumetric}
P.~A. Chou, M.~Koroteev, and M.~Krivoku{\'c}a, ``A volumetric approach to point
  cloud compression—part i: Attribute compression,'' \emph{IEEE Transactions
  on Image Processing}, vol.~29, pp. 2203--2216, 2019.

\bibitem{krivokuca2019volumetric}
M.~Krivoku{\'c}a, P.~A. Chou, and M.~Koroteev, ``A volumetric approach to point
  cloud compression--part ii: Geometry compression,'' \emph{IEEE Transactions
  on Image Processing}, vol.~29, pp. 2217--2229, 2019.

\bibitem{girault2020graph}
B.~Girault, A.~Ortega, and S.~S. Narayayan, ``Graph vertex sampling with
  arbitrary graph signal hilbert spaces,'' in \emph{2020 IEEE International
  Conference on Acoustics, Speech and Signal Processing (ICASSP)}.\hskip 1em
  plus 0.5em minus 0.4em\relax IEEE, 2020, pp. 5670--5674.

\bibitem{lu2020perceptually}
K.-S. Lu, A.~Ortega, D.~Mukherjee, and Y.~Chen, ``Perceptually inspired
  weighted mse optimization using irregularity-aware graph fourier transform,''
  in \emph{2020 IEEE Intl. Conf. on Image Proc. (ICIP)}.\hskip 1em plus 0.5em
  minus 0.4em\relax IEEE, 2020, pp. 3384--3388.

\bibitem{chung1997spectral}
F.~R. Chung and F.~C. Graham, \emph{Spectral graph theory}.\hskip 1em plus
  0.5em minus 0.4em\relax American Mathematical Soc., 1997, no.~92.

\bibitem{pavez2020spectral}
E.~Pavez, B.~Girault, A.~Ortega, and P.~A. Chou, ``Spectral folding and
  two-channel filter-banks on arbitrary graphs,'' in \emph{2021 IEEE
  International Conference on Acoustics, Speech, and Signal Processing
  (ICASSP)}.\hskip 1em plus 0.5em minus 0.4em\relax IEEE, 2021.

\bibitem{teke2016extending_1}
O.~Teke and P.~P. Vaidyanathan, ``Extending classical multirate signal
  processing theory to graphs—part i: Fundamentals,'' \emph{IEEE Transactions
  on Signal Processing}, vol.~65, no.~2, pp. 409--422, 2016.

\bibitem{teke2016extending_2}
------, ``Extending classical multirate signal processing theory to
  graphs—part ii: M-channel filter banks,'' \emph{IEEE Transactions on Signal
  Processing}, vol.~65, no.~2, pp. 423--437, 2016.

\bibitem{sakiyama2014oversampled}
A.~Sakiyama and Y.~Tanaka, ``Oversampled graph laplacian matrix for graph
  filter banks,'' \emph{IEEE Transactions on Signal Processing}, vol.~62,
  no.~24, pp. 6425--6437, 2014.

\bibitem{shuman2015multiscale}
D.~I. Shuman, M.~J. Faraji, and P.~Vandergheynst, ``A multiscale pyramid
  transform for graph signals,'' \emph{IEEE Transactions on Signal Processing},
  vol.~64, no.~8, pp. 2119--2134, 2015.

\bibitem{chen2015discrete}
S.~Chen, R.~Varma, A.~Sandryhaila, and J.~Kova{\v{c}}evi{\'c}, ``Discrete
  signal processing on graphs: Sampling theory,'' \emph{IEEE transactions on
  signal processing}, vol.~63, no.~24, pp. 6510--6523, 2015.

\bibitem{li2019scalable}
S.~Li, Y.~Jin, and D.~I. Shuman, ``Scalable $ m $-channel critically sampled
  filter banks for graph signals,'' \emph{IEEE Transactions on Signal
  Processing}, vol.~67, no.~15, pp. 3954--3969, 2019.

\bibitem{sakiyama2019two}
A.~Sakiyama, K.~Watanabe, Y.~Tanaka, and A.~Ortega, ``Two-channel critically
  sampled graph filter banks with spectral domain sampling,'' \emph{IEEE
  Transactions on Signal Processing}, vol.~67, no.~6, pp. 1447--1460, 2019.

\bibitem{kotzagiannidis2019splines}
M.~S. Kotzagiannidis and P.~L. Dragotti, ``Splines and wavelets on circulant
  graphs,'' \emph{Applied and Computational Harmonic Analysis}, vol.~47, no.~2,
  pp. 481--515, 2019.

\bibitem{ekambaram2015spline}
V.~N. Ekambaram, G.~C. Fanti, B.~Ayazifar, and K.~Ramchandran, ``Spline-like
  wavelet filterbanks for multiresolution analysis of graph-structured data,''
  \emph{IEEE Transactions on Signal and Information Processing over Networks},
  vol.~1, no.~4, pp. 268--278, 2015.

\bibitem{cheung2018graph}
G.~Cheung, E.~Magli, Y.~Tanaka, and M.~K. Ng, ``Graph spectral image
  processing,'' \emph{Proceedings of the IEEE}, vol. 106, no.~5, pp. 907--930,
  2018.

\bibitem{kao2014graph}
J.-Y. Kao, A.~Ortega, and S.~S. Narayanan, ``Graph-based approach for motion
  capture data representation and analysis,'' in \emph{2014 IEEE Intl. Conf. on
  Image Proc. (ICIP)}.\hskip 1em plus 0.5em minus 0.4em\relax IEEE, 2014, pp.
  2061--2065.

\bibitem{narang2012graph}
S.~K. Narang, Y.~H. Chao, and A.~Ortega, ``Graph-wavelet filterbanks for
  edge-aware image processing,'' in \emph{2012 IEEE Statistical Signal
  Processing Workshop (SSP)}.\hskip 1em plus 0.5em minus 0.4em\relax IEEE,
  2012, pp. 141--144.

\bibitem{lu2018learning}
K.-S. Lu, E.~Pavez, and A.~Ortega, ``On learning laplacians of tree structured
  graphs,'' in \emph{2018 IEEE Data Science Workshop (DSW)}.\hskip 1em plus
  0.5em minus 0.4em\relax IEEE, 2018, pp. 205--209.

\bibitem{anis2017critical}
A.~Anis and A.~Ortega, ``Critical sampling for wavelet filterbanks on arbitrary
  graphs,'' in \emph{2017 IEEE International Conference on Acoustics, Speech
  and Signal Processing (ICASSP)}.\hskip 1em plus 0.5em minus 0.4em\relax IEEE,
  2017, pp. 3889--3893.

\bibitem{crovella2003graph}
M.~Crovella and E.~Kolaczyk, ``Graph wavelets for spatial traffic analysis,''
  in \emph{IEEE INFOCOM 2003. Twenty-second Annual Joint Conference of the IEEE
  Computer and Communications Societies (IEEE Cat. No. 03CH37428)},
  vol.~3.\hskip 1em plus 0.5em minus 0.4em\relax IEEE, 2003, pp. 1848--1857.

\bibitem{coifman2006diffusion}
R.~R. Coifman and M.~Maggioni, ``Diffusion wavelets,'' \emph{Applied and
  Computational Harmonic Analysis}, vol.~21, no.~1, pp. 53--94, 2006.

\bibitem{hammond2011wavelets}
D.~K. Hammond, P.~Vandergheynst, and R.~Gribonval, ``Wavelets on graphs via
  spectral graph theory,'' \emph{Applied and Computational Harmonic Analysis},
  vol.~30, no.~2, pp. 129--150, 2011.

\bibitem{cloninger2020natural}
A.~Cloninger, H.~Li, and N.~Saito, ``Natural graph wavelet packet
  dictionaries,'' \emph{Journal of Fourier Analysis and Applications}, vol.~27,
  no.~3, pp. 1--33, 2021.

\bibitem{dong2019learning}
X.~Dong, D.~Thanou, M.~Rabbat, and P.~Frossard, ``Learning graphs from data: A
  signal representation perspective,'' \emph{IEEE Signal Processing Magazine},
  vol.~36, no.~3, pp. 44--63, 2019.

\bibitem{mateos2019connecting}
G.~Mateos, S.~Segarra, A.~G. Marques, and A.~Ribeiro, ``Connecting the dots:
  Identifying network structure via graph signal processing,'' \emph{IEEE
  Signal Processing Magazine}, vol.~36, no.~3, pp. 16--43, 2019.

\bibitem{miraki2021spectral}
A.~Miraki, H.~Saeedi-Sourck, N.~Marchetti, and A.~Farhang, ``Spectral domain
  spline graph filter bank,'' \emph{IEEE Signal Processing Letters}, vol.~28,
  pp. 469--473, 2021.

\bibitem{horn2012matrixbook}
R.~A. Horn and C.~R. Johnson, \emph{Matrix analysis}.\hskip 1em plus 0.5em
  minus 0.4em\relax Cambridge university press, 2012.

\bibitem{biyikoglu2007laplacian}
T.~Biyikoglu, J.~Leydold, and P.~F. Stadler, \emph{Laplacian eigenvectors of
  graphs: Perron-Frobenius and Faber-Krahn type theorems}.\hskip 1em plus 0.5em
  minus 0.4em\relax Springer, 2007.

\bibitem{Kurras.ICML.2014}
S.~Kurras, U.~Luxburg, and G.~Blanchard, ``{The f-Adjusted Graph Laplacian: a
  Diagonal Modification with a Geometric Interpretation},'' in
  \emph{{Proceedings of the 31st International Conference on Machine
  Learning}}, vol.~32, no.~2, Bejing, China, 22--24 Jun 2014, pp. 1530--1538.

\bibitem{egilmez2017graph}
H.~E. Egilmez, E.~Pavez, and A.~Ortega, ``Graph learning from data under
  laplacian and structural constraints,'' \emph{IEEE Journal of Selected Topics
  in Signal Processing}, vol.~11, no.~6, pp. 825--841, 2017.

\bibitem{pavez2020ragft}
E.~Pavez, B.~Girault, A.~Ortega, and P.~A. Chou, ``Region adaptive graph
  {F}ourier transform for {3D} point clouds,'' in \emph{2020 IEEE Intl. Conf.
  on Image Proc. (ICIP)}.\hskip 1em plus 0.5em minus 0.4em\relax IEEE, 2020.

\bibitem{anis2016efficient}
A.~Anis, A.~Gadde, and A.~Ortega, ``Efficient sampling set selection for
  bandlimited graph signals using graph spectral proxies,'' \emph{IEEE
  Transactions on Signal Processing}, vol.~64, no.~14, pp. 3775--3789, 2016.

\bibitem{dorfler2012kron}
F.~Dorfler and F.~Bullo, ``Kron reduction of graphs with applications to
  electrical networks,'' \emph{IEEE Transactions on Circuits and Systems I:
  Regular Papers}, vol.~60, no.~1, pp. 150--163, 2012.

\bibitem{gleich2015matlabbgl}
D.~Gleich, ``The matlabbgl matlab library,'' 2015.

\bibitem{girault2017grasp}
B.~Girault, S.~S. Narayanan, A.~Ortega, P.~Gon{\c{c}}alves, and E.~Fleury,
  ``Grasp: A matlab toolbox for graph signal processing,'' in \emph{2017 IEEE
  International Conference on Acoustics, Speech and Signal Processing
  (ICASSP)}.\hskip 1em plus 0.5em minus 0.4em\relax IEEE, 2017, pp. 6574--6575.

\bibitem{goemans1995improved}
M.~X. Goemans and D.~P. Williamson, ``Improved approximation algorithms for
  maximum cut and satisfiability problems using semidefinite programming,''
  \emph{Journal of the ACM (JACM)}, vol.~42, no.~6, pp. 1115--1145, 1995.

\bibitem{aspvall1984graph}
B.~Aspvall and J.~R. Gilbert, ``Graph coloring using eigenvalue
  decomposition,'' \emph{SIAM Journal on Algebraic Discrete Methods}, vol.~5,
  no.~4, pp. 526--538, 1984.

\bibitem{shekkizhar2020graph}
S.~Shekkizhar and A.~Ortega, ``Graph construction from data by non-negative
  kernel regression,'' in \emph{ICASSP 2020-2020 IEEE International Conference
  on Acoustics, Speech and Signal Processing (ICASSP)}.\hskip 1em plus 0.5em
  minus 0.4em\relax IEEE, 2020, pp. 3892--3896.

\bibitem{tanaka2020sampling}
Y.~Tanaka, Y.~C. Eldar, A.~Ortega, and G.~Cheung, ``Sampling signals on graphs:
  From theory to applications,'' \emph{IEEE Signal Processing Magazine},
  vol.~37, no.~6, pp. 14--30, 2020.

\end{thebibliography}


\vfill
\section{Properties and proofs}

\subsection{Proof of \autoref{prop_bip_block_smoothing}}
\label{app_bip_block_smoothing}
Recall that $\Qm$ from \eqref{eq_M_Q} obeys
\begin{equation}
    \Qm = \Dbip + \Lblock = \Dm - \Wblock.
\end{equation}
Using this notation, the fundamental matrix can be written 
\begin{equation}
    \Zm = \Id - (\Dm - \Wblock)^{-1}\Wbip.
\end{equation}
Because $\Qm^{-1}$   is non negative, and 
 $\Zm \mathbf{1=0}$,   the matrix $(\Dm - \Wblock)^{-1}\Wbip$ is  right stochastic. 
In fact, we can write it as follows
\begin{equation}\label{eq_normalized_WQ}
(\Dm - \Wblock)^{-1}\Wbip =  \Qm^{-1}\Wbip =   \Pblock \Pbip,
\end{equation}
where $\Pbip = (\Dbip)^{-1}\Wbip$, and $\Pblock = ((\Dbip)^{-1}(\Dbip + \Lblock))^{-1}$, are right stochastic matrices. We can verify this by computing
\begin{equation}
    (\Dbip)^{-1}\Wbip \mathbf{1} = (\Dbip)^{-1}\Dbip \mathbf{1} = \mathbf{1}.
\end{equation}
Note that $(\Pblock)^{-1}$ is right stochastic since
\begin{align*}
 (\Dbip)^{-1}(\Dbip + \Lblock) \mathbf{1} &=    (\Dbip)^{-1}(\Dbip\mathbf{1} + \Lblock\mathbf{1})\\
 &=(\Dbip)^{-1}(\Dbip\mathbf{1}) = \mathbf{1}.
\end{align*}
We have proven that $(\Pblock)^{-1} \mathbf{1}=\mathbf{1}$, and conclude by right multiplying by $\Pblock$.

\subsection{Proof of Theorem \ref{th_PR_arbitrary} (Perfect Reconstruction GFB)}
\label{app_PR}
The projector  onto the eigenspace with eigenvalue $\lambda$ is
\begin{equation}
\Pm_{\lambda} = \sum_{i: \lambda_i = \lambda} \uv_i \uv_i^{\top} \Qm,
\end{equation}
It is easy to verity that $\Pm_{\lambda}^2 = \Pm_{\lambda}$, and $\Pm_{\lambda} \Pm_{\gamma} = \mathbf{0}$ when $\lambda \neq \gamma$.
Spectral graph filters can be written using the projectors,  
\begin{equation}
\Hm_i = \sum_{\lambda \in \sigma(\Mm,\Qm)} h_i(\lambda) \Pm_{\lambda}, \quad \Gm_i = \sum_{\lambda \in \sigma(\Mm,\Qm)} g_i(\lambda) \Pm_{\lambda}.
\end{equation}
Then we have that
\begin{align}
&\mathbf{G}_0 \mathbf{H}_0 + \mathbf{G}_1 \mathbf{H}_1 = \sum_{\mathclap{\lambda \in \sigma(\Mm,\Qm)}} (g_0(\lambda)h_0(\lambda) + g_1(\lambda)h_1(\lambda))\Pm_{\lambda}, \\
&\mathbf{G}_0 \Jm \mathbf{H}_0 - \mathbf{G}_1\Jm \mathbf{H}_1= 
\sum_{\mathclap{\lambda,\gamma \in \sigma(\Mm,\Qm)}}(g_0(\lambda) h_0(\gamma) -g_1(\lambda) h_1(\gamma) )\Pm_{\lambda}\Jm \Pm_{\gamma}.
\end{align}
\autoref{prop:folding-variation} implies $\Jm \Pm_{\gamma} = \Pm_{2-\gamma}\Jm$. Using  orthogonality,  the only terms that survive are those for which $\lambda = 2 -\gamma$, hence
\begin{align}
&\mathbf{G}_0 \Jm \mathbf{H}_0 - \mathbf{G}_1\Jm \mathbf{H}_1=\nonumber \\
&\sum_{\lambda \in \sigma(\Mm,\Qm)}(g_0(\lambda) h_0(2-\lambda) -g_1(\lambda) h_1(2-\lambda) )\Pm_{\lambda}\Jm.
\end{align}
The proof follows directly from these identities. Note that if (\ref{eq_pr1}) and (\ref{eq_pr2}) are true, then $\mathbf{G}_0 \Jm \mathbf{H}_0 - \mathbf{G}_1\Jm \mathbf{H}_1 = \mathbf{0}$, and $\mathbf{G}_0 \mathbf{H}_0 + \mathbf{G}_1 \mathbf{H}_1 = 2\sum_{\lambda \in \sigma(\Mm,\Qm)}  \Pm_{\lambda} = 2\Id$. For the converse, assume that $\mathbf{G}_0 \mathbf{H}_0 + \mathbf{G}_1 \mathbf{H}_1 + \mathbf{G}_0 \Jm \mathbf{H}_0 - \mathbf{G}_1\Jm \mathbf{H}_1 = 2\Id$. Now let $\gamma \in \sigma(\Mm,\Qm)$. Right multiplication by $\Pm_{\gamma}$ implies (\ref{eq_pr1}), while right multiplication by $\Pm_{2-\gamma}$ leads to   (\ref{eq_pr2}).

\subsection{Proof of  \autoref{th_orthogonality_arbitrary} ($\Qm$-orthogonal GFB)}
\label{app_Orth}
We need to show that $\Tm_a^{\top} \Qm \Tm_a = \Qm$. We will use the following properties.
\begin{itemize}
    \item Since $\Qm$ is block diagonal, we have that $\Qm\Jm = \Jm \Qm$.
    \item If $\Hm$ is a spectral filter, then $\Hm^{\top}\Qm= \Qm \Hm$.  Indeed $\Hm =\Um h(\Lam)\Um^{\top}\Qm $, then
    \begin{equation}
     \Hm^{\top}\Qm   = \Qm \Um h(\Lam)\Um^{\top}\Qm = \Qm \Hm.
    \end{equation}
\end{itemize}
Since $\Id + \Jm$ and $\Id - \Jm$ have disjoint support,
\begin{align}
    \Tm_a^{\top} \Qm \Tm_a &=\frac{1}{2} \Hm_0^{\top}\Qm (\Id + \Jm)\Hm_0 +  \frac{1}{2} \Hm_1^{\top} \Qm (\Id - \Jm)\Hm_1 \\
    &=\frac{1}{2}\Qm  \Hm_0 (\Id + \Jm)\Hm_0 +  \frac{1}{2} \Qm \Hm_1 (\Id - \Jm)\Hm_1.
\end{align}
Using the projectors $\Pm_{\lambda}$, we  can write
\begin{multline}
2\Tm_a^{\top} \Qm \Tm_a =\sum_{\mathclap{\lambda \in \sigma(\Mm,\Qm)}} (h_0^2(\lambda) + h_1^2(\lambda))\Qm \Pm_{\lambda} + \\
\sum_{\mathclap{\lambda \in \sigma(\Mm,\Qm)}} (h_0(\lambda)h_0(2-\lambda) - h_1(\lambda)h_1(2-\lambda))\Qm \Pm_{\lambda}\Jm.
\end{multline}
If conditions (\ref{eq_orthogonal_filters1_Q}) and (\ref{eq_orthogonal_filters2_Q}) are true, then $\Tm_a^{\top} \Qm \Tm_a = \Qm$. To show the  converse, we use   the same strategy used to prove Theorem \ref{th_PR_arbitrary}, where we right multiply by $\Pm_{\gamma}$ and $\Pm_{2-\gamma}$.

\subsection{Proof of  \autoref{prop_orth_lazy}}
\label{app_prop_orth_lazy}
First, observe that   (\ref{eq_lazy_a_ell}) and (\ref{eq_lazy_d_ell}) imply that for each resolution $\ell$,  $\E[\av_\ell \dv_\ell^{\top}] = \mathbf{0}$. Linearity of the expectation implies that any linear transformation of $\av_\ell$ is also uncorrelated with $\dv_\ell$, then $\E[\av_0 \dv_\ell^{\top}] = \mathbf{0}$ and $\E[\dv_\tau \dv_\ell^{\top}] = \mathbf{0}$, 
for all $\ell \geq 0$, and for all $\tau \neq \ell$. Finally,  the covariance of the detail coefficients is
\begin{equation*}
 \E[\dv_\ell \dv_\ell^{\top}] = \Sigmam_{\Bc_\ell \Bc_\ell} - \Sigmam_{\Bc_\ell \Ac_\ell} \Sigmam_{\Ac_\ell \Ac_\ell}^{-1}\Sigmam_{\Ac_\ell \Bc_\ell} = ((\Mm_{\ell+1})_{\Bc_\ell \Bc_\ell})^{-1}.
\end{equation*}
We conclude by using the spectral folding property and  \eqref{eq_sub_QEll}, leading to $(\Mm_{\ell+1})_{\Bc_\ell \Bc_\ell} = (\Qm_{\ell+1})_{\Bc_\ell \Bc_\ell} = \Qm_{1,\ell}$.
\subsection{Derivation of \eqref{eq_bound_kappa1}}
\label{app_bound_kappa1}
Since $\Qm = \Vm^{1/2}(\Id - \Vm^{-1/2} \Wblock \Vm^{-1/2})\Vm^{1/2}\succ 0$, we have that $\Vm^{-1/2} \Wblock \Vm^{-1/2} \prec 1$, and   $\rho(\Ac) = \Vert \Vm^{-1/2}\Wblock \Vm^{-1/2} \Vert <1$. Using triangular inequality we have the bound $\Vert \Qm \Vert \leq \Vert \Vm \Vert(1+\rho(\Ac))$. The fact that $\rho(\Ac)<1$ also implies that
\begin{equation}\label{eq_inverse_Q}
	\Qm^{-1} =  \Vm^{-1/2}\left( \Id -\Vm^{-1/2}\Wblock \Vm^{-1/2}   \right)^{-1} \Vm^{-1/2}.
\end{equation}
The norm of $\Qm^{-1}$ can be bounded as $\Vert \Qm^{-1} \Vert \leq \Vert \Vm^{-1} \Vert (1-\rho(\Ac))^{-1}$. Combining the upper bounds for $\Vert \Qm \Vert$ and $\Vert \Qm^{-1}\Vert$ results in the bound for the condition number
\begin{equation}
	\kappa(\Qm) \leq \kappa(\Vm) ({1+\rho(\Ac)})/({1-\rho(\Ac)}).
\end{equation}
%
%

\section{Multi-channel filter bank}
\label{sec_multires_filterbank}
In this section we study properties of  tree structured filter banks formed by  concatenating  two-channel filter banks.  For all resolution levels, the  graphs, variation operators and sampling sets are given and fixed.  An example with $L=3$ levels is depicted in   \autoref{fig:iterated_fb_1}. 
%
\subsection{Analysis and synthesis operators}
 We assume the input signal is at  resolution $L$, thus $\av_L = \xv$. The outputs of the low and high  pass channels at resolution $\ell<L$ are called approximation and detail coefficients, and are denoted by    $\av_{\ell}$, and $\dv_{\ell}$,    respectively. 
 The sampling sets obey $\Vc_{\ell} = \Ac_{\ell}$, and for $\ell<L$, $\Ac_{\ell+1} = \Ac_{\ell}\cup\Bc_{\ell}$. The graph at resolution $\ell$  is denoted by $\Gc_{\ell} = (\Vc_{\ell},\Ec_{\ell})$, and has variation operator $\Mm_{\ell}$ with corresponding inner product matrix $\Qm_{\ell}$, chosen so that the $(\Mm_{\ell}, \Qm_{\ell})$-GFT  has the spectral folding property.
 The analysis equation at resolution $\ell$ is given by
\begin{equation}
    \begin{bmatrix}
    \av_{\ell}^{\top} &
    \dv_{\ell}^{\top}
    \end{bmatrix}^{\top} = 
    \Tm_{a,\ell} \av_{\ell+1}.
\end{equation}
The synthesis operator is  $\Tm_{s,\ell}$, and satisfies 
\begin{equation}\label{eq_synthesis_iterated_1levelv2}
   \av_{\ell+1} = \Tm_{s,\ell}\begin{bmatrix}
   	\av_{\ell}^{\top} &
   	\dv_{\ell}^{\top}
   \end{bmatrix}^{\top}=\Thetam_{0,\ell} \av_{\ell} + \Thetam_{1,\ell} \dv_{\ell},
\end{equation}
where we have defined the matrices
\begin{equation}
    \Thetam_{0,\ell} =  \Gm_{0,\ell} \Sm^{\top}_{\Ac_{\ell}}, \quad \Thetam_{1,\ell}= \Gm_{1,\ell} \Sm^{\top}_{\Bc_{\ell}}, 
\end{equation}
with dimensions $\vert \Ac_{\ell+1} \vert \times \vert \Ac_{\ell} \vert$,  and $\vert \Ac_{\ell+1} \vert \times \vert \Bc_{\ell} \vert $,  respectively.
After applying (\ref{eq_synthesis_iterated_1levelv2}) recursively we get the synthesis equation
\begin{equation}
    \xv=\Thetam_{0,L-1} \av_{L-1} + \Thetam_{1,L-1} \dv_{L-1}
       = \Psim_{0}\av_0 + \sum_{k=0}^{L-1} \Phim_k\dv_k.
\end{equation}
The matrices $\Psim_{k}$ and $\Phim_{k}$ are given by
\begin{equation}
    \Psim_k = \Thetam_{0,L-1} \Thetam_{0,L-2}\cdots \Thetam_{0,k}, \quad 
    \Phim_k = \Psim_{k+1} \Thetam_{1,k},
\end{equation}
for $k \in [L-1]$, and $\Psim_{L}=\Id$. By collecting the coefficients into a vector $\cv = [\av^{\top}_0,  \dv^{\top}_0,    \cdots,   \dv^{\top}_{L-1}]^{\top}$,  we can write 
\begin{equation}
    \xv = 
    \begin{bmatrix}
    \Psim_{0} & \Phim_{0} &  \cdots & \Phim_{L-1}
    \end{bmatrix}
	\cv= \Tcb_s \cv,
\end{equation}
where $\Tcb_s$ and $\Tcb_a = \Tcb_s^{-1}$ are the synthesis and analysis operators of the tree structured filter bank, respectively.
%
\subsection{Frame bounds}
Consider a compression application, where $\cv$ and $\hat{\cv}$ are the unquantized and quantized coefficients, respectively. We would like to write the reconstruction error 
\begin{equation}\label{eq_compression_error}
 \xv - \hat{\xv} = \Tcb_s(\cv - \hat{\cv})   
\end{equation}
as a function of the quantization error $\cv - \hat{\cv}$. The challenge, is that the   operators $\Tcb_a$ and $\Tcb_s$ are not orthogonal (under any norm). This is true even if each two channel filter bank at resolution $\ell$ is $\Qm_{\ell}$ orthogonal, in fact
\begin{equation}\label{eq_parseval_1level}
\Vert \av_{\ell+1}  \Vert_{\Qm_{\ell+1}}^2 = \Vert \av_{\ell}  \Vert_{\Qm_{0,\ell}}^2 + \Vert \dv_{\ell}  \Vert_{\Qm_{1,\ell}}^2,
\end{equation}
where we have  defined  
\begin{equation}\label{eq_sub_QEllv2}
\Qm_{0,\ell} = \Qm_{\ell+1}(\Ac_{\ell}, \Ac_{\ell}),\quad 
\Qm_{1,\ell} = \Qm_{\ell+1}(\Bc_{\ell}, \Bc_{\ell}).
\end{equation}
Note that if we wanted to write $\Vert \av_{\ell}  \Vert_{\Qm_{0,\ell}}^2$ as the sum of the energies of $\av_{\ell-1}$ and $\dv_{\ell-1}$, in some norm, and apply recursively \eqref{eq_parseval_1level},  we would need that 
\begin{equation}\label{eq_condition_parseval_ver1}
\Qm_{0,\ell} = \Qm_{\ell}, 
\end{equation}
%
which is satisfied by the normalized Laplacians of bipartite graphs at all resolution levels (since $\Qm_{0,\ell} = \Qm_{\ell}=\Id$). However, \eqref{eq_condition_parseval_ver1} does not  
hold for other variation operators on arbitrary graphs. Fortunately, we can still establish a relation between the norms of the coefficients at different resolutions. We will present a frame bound that relates the norms of $\xv$ and its coefficients $\cv$, when the two channel filter bank at level $\ell$ is $\Qm_{\ell}$ orthogonal. The frame constants depend on the amount by which \eqref{eq_condition_parseval_ver1} is violated.  First we recall a basic result that relates different  vector norms.
\begin{lemma}\cite{horn2012matrixbook}\label{lemma_equi_norms}
	Let $\Rm_1$ and $\Rm_2$ be positive definite matrices, then 
	\begin{equation}\label{eq_equivalent_norms}
		\alpha \Vert \xv \Vert^2_{\Rm_2} \leq  \Vert \xv \Vert^2_{\Rm_1} \leq \beta \Vert \xv \Vert^2_{\Rm_2},
	\end{equation}
	for any $\xv$, where
	\begin{equation}
		\alpha = \inf_{\xv \neq 0} \frac{\xv^{\top}\Rm_1 \xv}{\xv^{\top}\Rm_2 \xv},\quad  \beta = \sup_{\xv \neq 0} \frac{\xv^{\top}\Rm_1 \xv}{\xv^{\top}\Rm_2 \xv}.
	\end{equation}
\end{lemma}
Equalities in \eqref{eq_equivalent_norms} are attained if and only if $\Rm_1 = c \Rm_2$ for a given constant $c>0$, in which case $\alpha = \beta=c$. 
For each $\ell \geq 0$, we define
\begin{equation}\label{eq_alpha_ell_beta_ell}
    \alpha_{\ell} = \inf_{\xv \neq 0} \frac{\xv^{\top} \Qm_{0,\ell}  \xv}{\xv^{\top} \Qm_{\ell}  \xv}, \quad \beta_{\ell}=\sup_{\xv \neq 0} \frac{\xv^{\top} \Qm_{0,\ell}  \xv}{\xv^{\top} \Qm_{\ell}  \xv}.
\end{equation}
Also,  we define, $A_{L-1}=B_{L-1}=1$, and for $0 \leq \ell < L-1$,
\begin{equation}
    A_{\ell} = \prod_{k=\ell+1}^{L-1} \alpha_k, \quad  B_{\ell} = \prod_{k=\ell+1}^{L-1} \beta_k.
\end{equation}
We also consider the following inner product matrix
\begin{equation}\label{eq_Q_iterated_filterbank}
{\Qm} = \begin{bmatrix}
\Qm_{0,0} & \mathbf{0} & \cdots & \mathbf{0} \\
\mathbf{0} & \Qm_{1,0} & \ddots & \mathbf{0}\\
\vdots & \ddots &  \ddots &  \vdots \\
\mathbf{0} & \mathbf{0} & \mathbf{0} & \Qm_{1,L-1}
\end{bmatrix}.
\end{equation}
Now we present or main result of this section for the tree structured filter bank of \autoref{fig:iterated_fb_1}. The proof is given in Appendix \ref{ap_proof_th_frame}.
\begin{theorem}\label{th_frame} For a  $L$-level  tree structured filter bank,  composed of $\Qm_{\ell}$ orthogonal two channel filter banks, the synthesis operator $\Tcb_s$ satisfies
\begin{equation}\label{eq_frame}
    C \Vert \cv \Vert_{\Qm}^2 \leq \Vert \Tcb_s \cv \Vert^2_{\Qm_L} \leq D  \Vert \cv\Vert_{\Qm}^2, 
\end{equation}
for all $\xv = \Tcb_s \cv$, 
with constants given by
\begin{equation}
    C = \min_{0 \leq \ell \leq L-1} A_{\ell}, \quad D = \max_{0 \leq \ell \leq L-1} B_{\ell}.
\end{equation}
\end{theorem} 
Now we can revisit   \eqref{eq_compression_error} and bound the reconstruction error using \autoref{th_frame}, leading to $\Vert \xv - \hat{\xv} \Vert^2_{\Qm_L} \leq D\Vert \cv - \hat{\cv}\Vert^2_{\Qm}$, where $\Qm$ is given by  \eqref{eq_Q_iterated_filterbank}.
\begin{remark}
We have established that the synthesis operator $\Tcb_s$ is a frame with constants $C$ and $D$ \cite{vetterli1995wavelets,mallat1999wavelet}.   A frame is called tight, if the constants in the upper and lower bound are equal. In this case, $C=D$ when \eqref{eq_condition_parseval_ver1} is true for all $\ell \geq 0$. GFBs are tight frames when all the graphs $\Gc_{\ell}$ are bipartite, and  BFBs are used  for each $\ell \geq 0$, since we have that $A = B =1$, $\Qm_L = \Qm = \Id$, and $\Vert \Tcb_s \cv \Vert_{\Id} = \Vert \cv\Vert_{\Id}$. 
\end{remark}
The constants $C$, $D$, and the inner product $\Qm_L$ and $\Qm$ that appear in \autoref{th_frame} depend on the graphs, variation operators and downsampling sets chosen at each resolution level.   

\subsection{Proof of \autoref{th_frame} (frame bound of tree structured GFBs)}
\label{ap_proof_th_frame}
Let   $ \cv =[\av_0^{\top}, \dv_0^{\top}, \cdots, \dv_{L-1}^{\top}]$, and $\xv = \Tcb_s \cv$. 
The reason we cannot apply \eqref{eq_parseval_1level} recursively, is because \eqref{eq_condition_parseval_ver1} does not hold. We can instead use Lemma \ref{lemma_equi_norms}, and  the inequalities
\begin{align} \label{eq_bound_norms_iteratedFB}
   \alpha_{\ell} \Vert \av_{\ell} \Vert_{\Qm_{\ell}}^2 \leq  \Vert \av_{\ell} \Vert_{\Qm_{0,\ell}}^2 \leq \beta_{\ell} \Vert \av_{\ell} \Vert_{\Qm_{\ell}}^2.
\end{align}
Indeed, using $\Qm_L$ orthogonality   and \eqref{eq_bound_norms_iteratedFB} implies
\begin{align}
    \Vert \xv \Vert_{\Qm_L}^2 &= \Vert \av_{L-1} \Vert_{\Qm_{0,L-1}}^2  + \Vert \dv_{L-1} \Vert_{\Qm_{0,L-1}}^2\\
    &\leq \beta_{L-1}\Vert \av_{L-1} \Vert_{\Qm_{L-1}}^2 + \Vert \dv_{L-1} \Vert_{\Qm_{0,L-1}}^2.
\end{align}
We  apply  $\Qm_{\ell}$ orthogonality and \eqref{eq_bound_norms_iteratedFB} for each $\ell \leq L-1$ and, 
\begin{equation}
 \Vert \xv \Vert_{\Qm_L}^2 \leq B_0 \Vert \av_0\Vert^2_{\Qm_{0,0}} + \sum_{\ell = 0}^{L-1}B_{\ell}  \Vert \dv_{\ell}\Vert^2_{\Qm_{1,\ell}} \leq D \Vert \cv \Vert^2_{\Qm}.
\end{equation}
We can use the lower bounds in \eqref{eq_bound_norms_iteratedFB}, to get the lower bound
\begin{equation}
C \Vert \cv \Vert^2_{\Qm} \leq A_0 \Vert \av_0\Vert^2_{\Qm_{0,0}} + \sum_{\ell = 0}^{L-1}A_{\ell}  \Vert \dv_{\ell}\Vert^2_{\Qm_{1,\ell}} \leq \Vert \xv \Vert_{\Qm_L}^2.
\end{equation}

\end{document}